\pdfoutput=1 
\documentclass[apj]{emulateapj}
\usepackage{epsfig}
\usepackage[figuresright]{rotating}

\submitted{Draft version \today, \  accepted by the Astronomical Journal}

\def\eg{{e.g.}}

\shortauthors{Charlot et al.}
\shorttitle{The Celestial Reference Frame at 24 and 43 GHz}

\begin{document}

\title{The Celestial Reference Frame at 24 and 43 GHz. II. Imaging} 
\author{
P. Charlot\altaffilmark{1,2},
D. A. Boboltz\altaffilmark{3},
A. L. Fey\altaffilmark{3},
E. B. Fomalont\altaffilmark{4},
B. J. Geldzahler\altaffilmark{5},\\
D. Gordon\altaffilmark{6},
C. S. Jacobs\altaffilmark{7},
G. E. Lanyi\altaffilmark{7},
C. Ma\altaffilmark{8},
C. J. Naudet\altaffilmark{7},\\
J. D. Romney\altaffilmark{9},
O. J. Sovers\altaffilmark{10},
and
L. D. Zhang\altaffilmark{7}}

\altaffiltext{1}{Universit\'e de Bordeaux, Observatoire Aquitain des Sciences de l'Univers, 
BP~89, 33271 Floirac Cedex, France}
\altaffiltext{2}{CNRS, Laboratoire d'Astrophysique de Bordeaux -- UMR~5804, 
BP~89, 33271 Floirac Cedex, France}
\altaffiltext{3}{U.S. Naval Observatory,
3450 Massachusetts Ave., NW, Washington, DC 20392-5420}
\altaffiltext{4}{National Radio Astronomy Observatory,
520 Edgemont Rd., Charlottesville, VA  22903}
\altaffiltext{5}{National Aeronautics and Space Administration,
300 E. St., SW, Washington, DC 20546}
\altaffiltext{6}{NVI Inc./NASA Goddard Space Flight Center,
Greenbelt, MD  20771}
\altaffiltext{7}{Jet Propulsion Laboratory, Caltech,
4800 Oak Grove Dr., Pasadena, CA 91109}
\altaffiltext{8}{NASA Goddard Space Flight Center,
Greenbelt, MD  20771}
\altaffiltext{9}{National Radio Astronomy Observatory,
P.O. Box O, Socorro, NM 87801}
\altaffiltext{10}{Remote Sensing Analysis Systems,
2092 Sinaloa Ave., Altadena, CA 91001}

\begin{abstract}  

We have measured the sub-milli-arcsecond structure of 274 extragalactic
sources at 24 and 43 GHz in order to assess their astrometric suitability
for use in a high frequency celestial reference frame (CRF).   Ten sessions of
observations with the Very Long Baseline Array have been conducted over the
course of $\sim$5 years, with a total of 1339 images produced for the 274 sources.
There are several quantities that can be used to characterize the
impact of intrinsic source structure on astrometric observations including
the source flux density, the flux density variability, the source structure index,
the source compactness, and the compactness variability.  
A detailed analysis of these imaging quantities shows that
(1) our selection of compact sources from 8.4 GHz catalogs yielded 
sources with flux densities, averaged over the sessions in which each source
was observed, of about 1 Jy at both 24 and 
43 GHz, (2) on average the source flux densities at 24 GHz varied by 
20\%--25\% relative to their mean values, with variations in the 
session-to-session flux density scale being less than 10\%,
(3) sources were found to be more compact with less intrinsic structure 
at higher frequencies, and (4) variations of the core radio emission relative to the 
total flux density of the source are less than 8\% on average at 24 GHz.  
We conclude that the reduction in the effects due to 
source structure gained by observing at higher frequencies will result 
in an improved CRF and a pool of high-quality 
fiducial reference points for use in spacecraft navigation over the 
next decade.

\end{abstract}

\keywords{astrometry --- quasars: general --- radio continuum: galaxies ---
surveys}

\section{INTRODUCTION}

The International Celestial Reference Frame (ICRF) was formally 
adopted as the fundamental celestial reference frame (CRF) by the International 
Astronomical Union (IAU) in 1997.  The catalog includes precise 
astrometric positions of over 600 extragalactic compact radio sources 
distributed uniformly over the sky.  These positions were determined 
from the analysis of thousands of dual-frequency S/X-band (2.3/8.4 GHz) 
Very Long Baseline Interferometry (VLBI) observational sessions recorded 
between 1979 and 1995.  The frame itself was defined by the 212 
highest-quality ``defining" sources with typical position accuracies of 
0.25 milli-arcseconds (mas), while the axes of the frame are accurate 
to 0.02 mas \citep{MA:98}.   In addition to the 212 defining sources, 
positions for 294 less observed ``candidate'' sources along with 102 
``other'' sources with excessive position variation, were also given to 
increase the density of the frame \citep{MA:98}.  Since
its adoption, incremental updates to the catalog of sources have 
occurred in the form of two extensions to the ICRF \citep{FEY:04}
using hundreds of additional sessions along with improved 
analysis and modeling techniques.  However, the positions of the 
original 212 defining sources have remained constant through these
extensions.  

The compact extragalactic radio sources that comprise the ICRF 
have been the subject of extensive study since the inception of 
VLBI techniques.  In the standard theory, \citep[e.g.][]{BK:79}, 
the jet-like emission from quasars and active galactic nuclei is assumed to be 
powered by a central engine where energetic phenomena occur.   
The observed, frequency dependent, intrinsic structure of extragalactic radio 
sources typically consists of a flat spectrum ($S\propto\nu^\alpha, \alpha\approx0$) 
unresolved core at the base of the jet where the optical depth is near unity 
($\tau \approx 1$) and extended emission in the form of multiple steep spectrum 
($\alpha\approx-0.5$ to $-$1.5) jet components.  These components 
often move away from the core along the direction of the jet, sometimes 
at apparent superluminal speeds.  The emission outside of the 
core has been shown to be extended on scales larger than the accuracy of the 
astrometric position measurements \citep{FCF:96}.  This extended emission, or
intrinsic source structure, contributes to the uncertainty in the measured
astrometric positions of sources that comprise the ICRF.
In addition to extended emission, frequency dependent opacity effects 
can contribute to variations in the measured astrometric positions of
extragalactic radio sources.  In particular, opacity conditions in the region 
near the base of the jet can cause the measured core position to move 
inward along the jet direction as a function of increasing radio frequency 
of the observations.  Such core shifts have been measured for some 
compact sources \citep[e.g.][]{LOBANOV:98, KOVALEV:08}. 

Based upon the discussion above, there are potential advantages to 
observing astrometric sources at radio frequencies higher than the 
typical S/X band observations.  First, the resolution of the observations 
is increased, thus improving the astrometric accuracy.  Second, the 
effects due to intrinsic source structure may be reduced since the core
and jet components have different spectral characteristics
and the core is expected to be more dominant at high
frequencies. We have undertaken a program to observe a 
number of extragalactic sources at K band (24 GHz) and Q band (43 GHz) 
using the 10 stations of the Very Long Baseline Array (VLBA).   At these higher 
frequencies, only the VLBA provides the stability, imaging capabilities and 
frequency coverage to enable such a program.  The long term goals 
of the program include:  1) developing a high-frequency CRF with a 
variety of applications including improved deep space 
navigation,  2) providing the astronomical community with an extended 
catalog of calibrator sources for VLBI observations at 24 and 43 GHz,  and 
3) studying the effects of the intrinsic structure of extragalactic sources
to improve the astrometric accuracy of future high-frequency reference 
frames.   A detailed discussion of the program and the astrometric 
results is contained in a companion paper 
\citep[][hereafter Paper I]{LANYI:09}.   In this paper, we concentrate on the 
imaging aspects of the program and the effects of observed source structure 
and variability on astrometric accuracy.   

Theoretically, an optimal CRF would be composed of sources with
strong, non-variable, point-like emission.   In reality, however, virtually 
all sources possess some structure and intrinsic variability in the emission 
over time.  In addition, there is the potential for sudden flaring events even 
for normally quiescent sources.   Thus it is highly desirable to characterize
and monitor the nature of the sources used in a CRF through periodic 
VLBI imaging. The VLBA has previously been used to make simultaneous 
dual-frequency S/X-band (2.3/8.4 GHz) observations of a total of 389 ICRF sources 
\citep{FCF:96,FC:97,FC:00}.  To date, approximately $90\%$ of the ICRF sources
north of $-20^\circ$ declination have been imaged at least once
at both 2.3 and 8.4~GHz.  Based on the initial work of \cite{CHARLOT:90}, 
the database of VLBA X-band and S-band images was analyzed by \cite{FC:97,FC:00} 
in order to quantitatively improve our understanding of the relationship between 
extended source structure and the astrometric positions determined from 
VLBI.  Here we discuss our growing database of high-frequency
images of potential extragalactic reference frame sources.  We apply 
similar analysis techniques in an attempt to characterize the impact of 
extended source structure on the astrometric accuracy of the catalog 
of source positions obtained from our VLBA high-frequency data (Paper I).

\section{OBSERVATIONS AND DATA REDUCTION \label{SEC:OBS}}

We observed a total of 351 extragalactic radio sources using 
the 10 antennas of the VLBA \citep{NAPIER:94} of
the National Radio Astronomy Observatory (NRAO)\footnote{The National 
Radio Astronomy Observatory is a facility of the National Science Foundation 
operated under cooperative agreement by Associated Universities, Inc.}
over the course of 10 sessions spanning five years from 2002--2007.   In all 
10 of the sessions, observations were recorded at K-band (24 GHz).  
Additionally, in four of the 10 sessions, observations were also recorded 
at Q-band (43 GHz).  In two of the sessions, dual S/X-band (2.3/8.4 GHz) 
observations were interspersed with the K-band observations 
to investigate potential ionospheric correction methods.   A summary of 
the VLBA K- and Q-band observations relevant to the imaging program
is presented in Table~\ref{TAB:OBS}.

The frequency information listed in Table~\ref{TAB:OBS} reflects the
changing observing strategy over the course of the 10 VLBA sessions.  
For the first two sessions, four 8 MHz 
bands each at 24 and 43 GHz were recorded simultaneously for each 
scan using 2 bit sampling yielding a total bandwidth of 32~MHz.  
The remaining sessions used eight 8 MHz bands with 1 bit sampling for 
a total bandwidth of 64~MHz.  For these sessions, scans alternated
between K and Q bands (sessions 3 and 5), simultaneous S/X band and 
K band (sessions 6 and 8), or just K band alone (sessions 4, 9, and 10).
The first K-only session observed a large number of potential sources 
found in other VLBI surveys in order to expand the total number 
of available high-frequency astrometric targets.  In the final two K-only sessions, 
candidate sources near the ecliptic plane were added for potential future use in 
deep space navigation.  A detailed discussion of the evolving observing strategy 
and its relevance to the astrometric goals of the program is presented 
in Paper I. 

All of the observations were made in a bandwidth synthesis mode 
to facilitate group delay measurements for astrometry -- the multiplicity of 
channels allows for the determination of a precise group delay 
\citep{ROGERS:70}.  Observations in this mode also allow snapshot imaging.  
Source scans were typically 2 minutes in duration at K band and 3 
minutes for Q band.  Most sources were observed in three or more scans -- 
the one exception being the K-band survey session on 2003 May 22 in 
which many of the sources were observed only once or twice. 
The raw data bits were correlated with the VLBA correlator at the 
Array Operations Center in Socorro, New Mexico.

The correlated data were calibrated and corrected for residual delay
and delay rate using the NRAO's Astronomical Image Processing System
(AIPS). Initial amplitude calibration for each intermediate frequency (IF) was 
accomplished using system temperature measurements taken during the 
observations combined with NRAO supplied gain curves.  Fringe-fitting was done
in AIPS using solution intervals equal to the scan durations and a
point source model in all cases.  After correction for residual delay
and delay rate, the data were written to FITS disk files. All
subsequent image processing was carried out using the Caltech VLBI imaging
software, primarily DIFMAP.

The visibility data for each frequency band were self-calibrated,
Fourier inverted, and CLEANed using DIFMAP in an automatic mode
\citep{PSTM:94}. DIFMAP combines the visibilities
for each IF of an observation in the ({\it u,v})-plane during gridding, taking
into account frequency differences. However, DIFMAP makes no attempt
to correct for spectral index effects. The spanned bandwidth of the
IFs in each band is relatively small (0.5 GHz -- 2\% of the fractional
bandwidth), so we assume that source structure and flux density
variations across each of the two frequency bands are
negligible. After phase self-calibration with a point source model,
the 4 s correlator records were coherently averaged to 12 s
records.  The averaged data were subsequently edited to remove 
outliers due to poor data and times in which the antennas were 
not on source.

The amplitude calibration within each particular session at each frequency 
band was improved through observations of a strong, compact source. 
A single amplitude gain
correction factor was derived for each antenna for each IF, based on
fitting a simple Gaussian source model to the visibility data of these
compact sources after applying only the initial calibration based on
the measured system temperatures and gain curves. Gain correction
factors were calculated based on the differences between the observed
and model visibilities. The resulting amplitude gain correction
factors were typically less than 5\% and were applied to the visibility 
data of all sources.  

The data were self-calibrated following the hybrid-mapping technique
\citep{PR:84} to correct for residual amplitude and phase
errors. The data were initially phase self-calibrated and mapped using
uniform weighting in the {\it u,v}-plane before switching to natural
weighting after several iterations. A point source model was used as a
starting model for the iterative procedure in all cases. Convergence
was defined basically as the iteration when the peak in the residual
image became less than about six to eight times the root-mean-square
(rms) noise of the residual image from the previous iteration. The 1\%--2\% 
of sources with emission structure too complex or too extended for the automatic
imaging script to handle were imaged by hand, i.e. in an interactive
mode, following the same prescription as that for the automatic
mode. Convergence for these sources was subjective and was based on
the iteration at which it was judged that further self-calibration
would not significantly improve the resultant image.

Subsequent to the calibration and imaging process, 
an analysis of the variations in the flux density scaling 
from one session to the next was performed using 65 sources observed 
in at least 6 of 10 sessions at K band and 31 sources observed in at least 
3 of 4 sessions at Q band.  For this analysis, the mean and standard 
deviation of the flux density for the ensemble of sources at each band
were computed. The results showed that, on average, the relative flux 
density scaling from one session to the next varied by $\sim$5\% at K band 
and $\sim$9\% at Q band.

\section{IMAGING RESULTS AND ANALYSIS \label{SEC:RES_ANLY}}

Table~\ref{TAB:OBS} provides a summary of the number of sources 
successfully imaged (out of the 351 total) from the 10 sessions of K-band 
observations and four sessions of Q-band observations recorded as 
part of the high-frequency CRF program.  The results of the imaging 
part of the program are presented graphically in the form of contour 
plots available via the USNO Radio Reference Frame Image Database
(RRFID)\footnote{http://rorf.usno.navy.mil/rrfid.shtml}.  
Shown in Figures~\ref{FIG:KBAND_CNTR} and 
\ref{FIG:QBAND_CNTR} are example contour maps from the most 
recent session of each of the 274 sources imaged at K band and of 
each of the 132 sources imaged at Q band.  For convenience, the 
resulting K-band maps are identified by only a single fiducial frequency 
(either $\sim$23.9 or 24.4~GHz depending on session) even though they 
come from data using all frequency channels.  Similarly, the Q-band maps 
are also identified by only a single fiducial frequency ($\sim$43.1~GHz).  

Tables~\ref{TAB:KBAND_IMG_PARAM} and \ref{TAB:QBAND_IMG_PARAM} 
list the parameters relevant to the final images at each 
frequency including: the observation session, the elliptical beam size and 
position angle, the peak flux density ($S_{\rm peak}$), the total flux density
($S_{\rm total}$), the rms of the residuals of the flux 
density ($S_{\rm rms}$), and the contour levels in the maps shown in 
Figures~\ref{FIG:KBAND_CNTR} and \ref{FIG:QBAND_CNTR}.  
The total flux density is the sum of all CLEAN 
component model flux densities obtained from the source images 
\citep{CLARK:80}.  From the data in the table, we determined the 
minimum, maximum and average dynamic range ($S_{\rm peak}$/$S_{\rm rms}$) 
for all of the K-band images 
to be $\sim$40:1, $\sim$1720:1, and $\sim$340:1, respectively.  The average rms 
noise at K band is $\sim$2.8~mJy~beam$^{-1}$.  The minimum, maximum, 
and average dynamic range for the Q-band images are 
$\sim$60:1, $\sim$910:1, and $\sim$250:1, respectively.  The average rms 
noise at Q band is $\sim$5.0~mJy~beam$^{-1}$.  For 
comparison, the expected thermal noise at K band assuming 
6 minutes on source ($\sim$2 minutes per scan times 
three scans) is estimated to be $\sim$0.7~mJy~beam$^{-1}$.  At Q band,
assuming 9 minutes on source ($\sim$3 minutes per scan times 
three scans) the expected thermal noise is $\sim$1.6~mJy~beam$^{-1}$. 

The CLEAN component models were also used to determine two 
additional parameters, the source structure index ($SI$) and 
the source compactness ($C$).  The structure index is described 
in detail in $\S$~\ref{SEC:STRUC_IND}.  The source compactness is 
defined as the percentage of the total CLEANed flux density contained 
within a radius approximately equal to one synthesized beam width 
centered on the peak emission in the image.  

\subsection{Source Flux Density \label{FLUX_DENS}}

For a source to be an effective astrometric reference (e.g. for spacecraft 
tracking and astronomical phase referencing) it must be detectable within 
a coherence time at the observing frequency; typically less than 2 minutes 
at K and Q bands.   Sources should, therefore, have sufficient 
flux density for the given observational sensitivity, and sufficiently small 
variability so as not to fall below the detection threshold.  From the CLEAN 
components models described above, the total source flux densities ($S_{\rm total}$)
were determined for all of the sources in all sessions in both frequency bands.  
From $S_{\rm total}$ we computed the mean flux density ($\bar{S}$) per 
source averaged over all sessions in which the source was observed.
Plotted in Figure~\ref{FIG:XKQ_ALL_FLUX} are the distributions of $\bar{S}$ 
for all of the sources imaged in the program including 138 sources 
at X band, 274 sources at K band, and 132 sources at Q band.  
The mean (median) values of the distributions are approximately 
1.5 (0.8)~Jy at X band, 1.1 (0.7)~Jy at K band, and 1.4 (0.9)~Jy at Q band.  
Shown in Figure~\ref{FIG:XKQ_97COMMON_FLUX} are the same 
distributions of $\bar{S}$ for only those 97 sources that are common 
to all three bands.  Here the mean (median) values are 1.7 (0.9), 
1.5 (0.9) and 1.5 (1.0)~Jy at X, K and Q bands, respectively.

It is not surprising that for the case of all sources and for the 97 sources
in common that the mean flux densities at X band are the highest, 
especially since the strongest ($>300$~mJy) most compact sources 
in the S/X-band ICRF program were selected for the K/Q-band 
observations in order to optimize the probability of detection 
at the higher frequencies (Paper I).  This selection effect is also, 
very likely, the reason that, on average, the Q-band mean flux 
densities are larger than those at K band for all sources 
(Figure~\ref{FIG:XKQ_ALL_FLUX}).  The Q-band sources were drawn strictly 
from this pool of strong, compact, ICRF sources.  However, as 
Table~\ref{TAB:OBS} shows, a significant fraction 
of the K-band sources were observed in a single-session survey that 
occurred on 2003 May 22.  For this survey, the X-band flux density criterion 
was relaxed from a nominal value of 300~mJy to 200~mJy in order to 
include more sources.  For the 97 sources in common 
to all three bands, the selection effect is reduced since many of
the K-band survey sources are excluded.  
Figure~\ref{FIG:XKQ_97COMMON_FLUX} shows the K- and Q-band
mean flux densities to be consistent with each other to within the 
$\sim$10\% session-to-session flux density scale variations mentioned 
previously. 

The extent to which the total flux density varies over time 
can be characterized by the flux density variability index ($\sigma_S/\bar{S}$), 
which is the standard deviation of the measured flux densities divided by 
the mean.  A minimum value, $\sigma_S/\bar{S}=0.0$, indicates no 
variation in the flux density with time.  Since a number of the sources 
at K and Q bands were observed in multiple sessions, it is possible to 
determine $\sigma_S/\bar{S}$ for these sources.   This was not the 
case for X band where only 25 sources were observed in both of the two
X-band sessions.  Distributions for the flux density variability index for 
the 235 sources at K band and 82 sources at Q band observed in 
two or more sessions are shown in Figure~\ref{FIG:KQ_FLUX_VAR}.   
The mean and median values of $\sigma_S/\bar{S}$ at K band are 
0.18 and 0.16, respectively.  The Q-band mean and median values of 
the flux density variability for sources observed in two or more sessions 
are 0.19 and 0.15, respectively.  

Given of the relatively few number of sessions thus far observed, the 
values above represent lower limits to the flux density variability.  
Increasing the cutoff for the minimum number of sessions per 
source to five, reduces the number of available K-band sources to 87 and 
yields a mean value of $\sigma_S/\bar{S}$ of 0.23.  Further increasing 
the minimum cutoff to 10 sessions, reduces the number of sources to 
18 and results in a mean flux density variability of 0.24.  At Q band 
there were 22 sources with data available from all four observing sessions.
The mean flux density variability for these 22 sources was 0.28.  Our
preliminary results therefore indicate that on average the flux density
varied by 20\%--25\% at K band, and 20\%--30\% at Q band.  These results are 
limited to the $\sim$5~year period over which our data were recorded and 
by the $\sim$10\% variations in the flux density scaling mentioned 
previously.

In addition to the flux density variability index, 
we also determined the ratio of the maximum to minimum flux density
($S_{\rm max}/S_{\rm min}$) for each source observed in more than 
one session.  Distributions for $S_{\rm max}/S_{\rm min}$ at K band 
and Q band are plotted in Figure~\ref{FIG:KQ_SMAX_SMIN}, with 
the mean and median ratio for each distribution indicated.  For the 
two distributions, the mean (median) values are 1.7 (1.5) at K band,
and 1.7 (1.4) at Q band.  The figure shows that over the 
five-year period in which the sources were monitored, the maximum to
minimum flux density ratio was greater than a factor of 2 for roughly
23\% of the sources at K band and 24\% of the sources at Q band.  
The ratio is greater than a factor of 3 for only 6\% of the sources
at K band and 4\% of the sources at Q.

Our flux density results can be compared to other extragalactic source 
monitoring programs.  The 14-m telescope at the Metsahovi Radio 
Observatory has monitored the flux densities of 
active galactic nuclei at 22, 37, and 87 GHz over the course
of 25$+$ years \citep[][and reference therein]{TERASRANTA:05}.  
Statistical analysis of this data using several different methods 
\citep{HOVATTA:08, HOVATTA:07} has shown that smaller flux density variations 
occur on shorter timescales of 1--2 years, while larger outbursts are
rare occurring on average every 6 years.  Their sample consisted of 
80 sources with a minimum flux density of 1~Jy and a minimum 
monitoring time of ten years.  Although the time span of our data is 
shorter, the sampling much sparser, and the minimum flux density 
lower at 200--300~mJy, our K- and Q-band results seem to be consistent 
with the Metsahovi data.  The variability index results show that the 
variations are relatively small over the five-year monitoring period, while the 
$S_{\rm max}/S_{\rm min}$ results indicate very few large-scale 
outbursts over the same time period. 

\subsection{Source Structure Index  \label{SEC:STRUC_IND}}

As previously demonstrated by \cite{CHARLOT:90}, the contribution of
intrinsic source structure to a VLBI bandwidth synthesis delay
measurement can be significant.   It depends on the exact form of the
spatial brightness distribution of the extended radio source relative
to the geometry of the VLBI baseline vector projected onto the plane
of the sky.  The overall magnitude of the source structure effect for a given
source is most easily estimated by calculating corrections to the
bandwidth synthesis delay based upon the observed source structure, for
a range of ({\it u,v}) coordinates ($u$ and $v$ are
coordinates of the baseline vector projected onto the plane of the sky
and are expressed in units of the observing wavelength).  Following
such a scheme, \cite{FC:97} defined a source ``structure
index'' according to the median value of the structure delay
corrections, $\tau_{\rm median}$, calculated for all projected VLBI
baselines that could be possibly observed with Earth-based VLBI (i.e.,
for all baselines with $\sqrt{u^2+v^2}$ less than the diameter of the
Earth), separating the sources into four classes as follows:

\[ \begin{array}{c} {\rm Structure \,\, Index} 
   \end{array} 
 = \left\{ 
   \begin{array}{ll}
  1, & {\rm if\  0\  ps} \le \tau_{\rm median} < 3\  {\rm ps,}\\
  2, & {\rm if\  3\  ps} \le \tau_{\rm median} < 10\  {\rm ps,}\\
  3, & {\rm if\  10\  ps} \le \tau_{\rm median} < 30\  {\rm ps,}\\
  4, & {\rm if\  30\  ps} \le \tau_{\rm median} < \infty.
   \end{array} 
\right. \]

Based on this definition, the structure index at each of the three bands 
(X, K, and Q) was obtained for each source/session, thus providing an 
indication of the magnitude of the effects of intrinsic source structure in 
the session of observation.  Detailed information regarding the computation 
of the structure delay corrections and the source structure index may
be found in \cite{CHARLOT:90} and \cite{FC:97}. False color images of 
the structure corrections as a function of 
({\it u,v}) distance are available for each source via the Bordeaux VLBI Image 
Database\footnote{http://www.obs.u-bordeaux1.fr/BVID/}.

From the source CLEAN components models, 
structure indices for 165 images of 138 sources imaged at X band, 
1072 images of 274 sources at K band, and 267 images of
132 sources at Q band have been determined in a manner similar to 
that presented in \cite{FC:97, FC:00}.  For each source that 
was successfully observed, Table~\ref{TAB:STRUC_INDX} lists the 
number of sessions imaged along with the minimum and maximum 
values of the source structure index 
over the course of those sessions at each of the three frequency bands. 
In most of the cases, the difference between the maximum and minimum 
$SI$ is either 0 or 1 for X, K and Q bands.  For a few sources at K and Q,
the difference is 2, while there are no sources with a difference of 3.  Although 
the data are limited in number of sessions, this seems to indicate that 
the $SI$ is relatively stable over time. 

As reported in \S\ref{SEC:RES_ANLY}, average image dynamic ranges 
were $\sim$340:1 and $\sim$250:1 for K and Q band, respectively.
The effects of dynamic range on $SI$ were investigated by varying the 
number of low-level image CLEAN components used to determine the median 
structure corrections and the structure index classification.  Variation of the 
CLEAN components resulted in changes in the median 
structure correction on the order of a few tenths of a picosecond (0.7~ps 
rms), with changes larger than 1~ps occurring for only about 5\% 
of the 1073 images at K band.  These changes in the median structure
correction resulted in a change in $SI$ by 1 unit for roughly 4\% of the
1073 structure indices computed.   Results for Q band were similar to 
those at K band.  Thus, the structure indices determined in this study 
are not significantly affected by the dynamic range of the images.

Figures \ref{FIG:SI_HIST_ALL} and \ref{FIG:SI_HIST_COMMON} compare
the distributions of the source structure indices derived from our VLBA 
observations at the three bands, X, K, and Q.
In Figure \ref{FIG:SI_HIST_ALL} we plot histograms of the maximum 
$SI$ measured for all sources in each of the three bands.  The maximum
is drawn from all available sessions in which the structure index for a 
particular source was determined.  There were a total of 138 sources 
at X band, 274 sources at K band and 132 sources at Q band for 
which the $SI$ was measured.  The figure shows a greater number 
of $SI=1,2$ (the most compact) sources at progressively higher 
frequencies.  The percentage of $SI=1,2$ sources is 71, 85, and 92\%
for X-, K-, and Q-band, respectively. 
Figure \ref{FIG:SI_HIST_COMMON} plots histograms of the maximum 
structure index measured for only the 97 sources common to all
three frequencies in our VLBA observations.   Again, the obvious trend 
toward lower structure index at higher frequencies is seen.  For 
the 97 common sources, the percentage having $SI=1,2$ is 71\% 
at X, 87\% at K, and 93\% at Q band.  These results indicate that the 
impact of source structure should be reduced at higher frequencies.

\subsection{Source Compactness \label{SEC:COMPACT}}

The source structure index and its underlying quantity, the 
median structure delay correction, have proven to be highly 
useful measures of source structure and its effects on astrometric 
observations \citep{CHARLOT:05, FEY:04, SOVERS:02}.  
In addition to these quantities, we introduce here 
a second indicator of structure called the source compactness, 
$C$, described earlier in \S~\ref{SEC:RES_ANLY}.
The compactness, which is proportional to $S_{\rm beam}/S_{\rm total}$, 
provides an indication of how point-like or extended the radio emission 
from a particular source is on a continuous scale 
from 0.0 to 1.0, where 1.0 represents a source with no extended emission.   

The compactness was determined for all sources in all sessions for 
each of the three observing frequencies.  Session-based compactness
values were used to determine the average compactness over all 
sessions ($\bar{C}$) for each source.  Shown in 
Figure~\ref{FIG:XKQ_ALL_COMPACT} are the 
distributions for $\bar{C}$ for all sources imaged in the program including 
138 sources at X band,  274 sources at K band, and 132 sources at 
Q band.  The mean and median values of the distributions at X band are 
0.83 and 0.86, or greater than 80\% of the maximum possible compactness 
$\bar{C}=1.0$.  The mean and median values at K band are
slightly lower with values of 0.81 and 0.83, respectively.   For Q band, 
the results are nearly identical to those at K band, with a value of 0.81 
for both the mean and median of the distribution of $\bar C$.   In 
Figure~\ref{FIG:XKQ_97COMMON_COMPACT} 
the distributions of $\bar{C}$ are plotted for the 97 sources common 
to all three observing bands.  The results here are very similar to those
shown for all sources with mean (median) values of 0.84 (0.87), 
0.79 (0.81), and 0.80 (0.80) for the X-, K-, and Q-band distributions, 
respectively.  While these results suggest only minor differences between 
the distributions of the mean source compactness at the three frequencies, 
one should recall that the compactness is dependent upon the 
flux density contained within one beam-width, $S_{\rm beam}$, 
which, in turn, is dependent upon the frequency of the observations.  
Therefore, a constant value for the compactness with increasing
frequency still implies a reduced intrinsic source size.

The source compactness also provides a convenient way to track 
changes in the source structure from one observing session to the next.  
To measure these changes, we computed the source compactness variability 
index ($\sigma_C/\bar{C}$), which is simply the standard deviation of the 
compactness divided by its mean.  A
minimum value, $\sigma_C/\bar{C}=0.0$, corresponds to no variation in  
the compactness over time.  Shown in Figure~\ref{FIG:KQ_COMPACT_VAR} 
are the distributions of source compactness variability index for sources
imaged in more than one session at K and Q bands, respectively.  
The mean (median) compactness variability for the 235 sources  
at K band is 0.08 (0.06).  For the 82 sources at Q band mean (median)
values of $\sigma_C/\bar{C}$ are slightly smaller 0.07 (0.05).  
These variations are small (no more than about 8\% on average), 
however the probability that the compactness variability is significantly 
different from zero has not been determined.  As in the case of the flux density 
variability, the compactness variability results are limited to the $\sim$5~yr 
period over which our VLBA data were recorded.

\section{COMPARISONS AND DISCUSSION}

In the previous section, we discussed results for several quantities
that emerged from our analysis of the VLBA images, namely:  the 
source flux density, the source structure index, the source compactness, 
and the time variability of each of these quantities.   In this section, 
we compare the source structure index to the compactness 
in order to verify that the two quantities are related.
In addition, we compare both the structure index and the compactness
to two astrometric quantities derived in Paper I, namely: the
formal position uncertainties and position variability of the sources.  
These comparisons were made to determine whether the reduced 
structure effects seen at high frequency correspond to more precise 
astrometric positions.

\subsection{Structure Index and Source Compactness \label{SEC:SI_SC}}

As an initial test of the source compactness, we compared $\bar C$ to the 
source structure index, $SI$, at each of the two frequencies.  Shown in 
Figure~\ref{FIG:K_COMPACT_SI} are the distributions of the mean 
source compactness for the 274 sources imaged at K band separated 
in terms of maximum source structure index.  Figure~\ref{FIG:Q_COMPACT_SI} 
shows similar distributions for the 132 sources imaged at Q band.  The three 
panels in each figure show $SI$ = 1, 2 and 3 sources, respectively.  
There are two sources at K band and one source at Q band 
with $SI=4$ that are not shown in the figures.  In both figures, it is evident
that the distributions in compactness among the three structure indices
are quite different with the distribution for $SI = 1$ being strongly peaked
at both frequencies, and the distributions broadening with increasing 
values of $SI$.  Within each panel of Figures~\ref{FIG:K_COMPACT_SI} and 
\ref{FIG:Q_COMPACT_SI} the mean and median of the distribution 
are given.  These values are also summarized in 
Table~\ref{TAB:COMPACT_SI}.  From the table we see that for both 
frequencies, the mean and median source compactness 
is directly related to structure index.  The correspondence between $SI$ and 
$\bar C$ is not unexpected, since both quantities provide an indication 
of the source structure as derived from the VLBI imaging. 

We also compared the compactness variability index to the source structure 
index and the results are shown in Figure~\ref{FIG:K_COMPACT_VAR_SI}.  
Plotted in the figure is $\sigma_C/\bar{C}$ at K band 
as a function of each of the three structure indices, $SI$ = 1, 2 and 3.
Recall that the compactness variability index was determined for only those
sources observed in more than one session (235 sources at K band and 
82 sources at Q band).  The equivalent plot for the Q band data was not
produced because of the greatly reduced number of sources.
As in the case of the source compactness, the 
variability in the compactness shows a clear trend with $SI$, with the 
$\sigma_C/\bar{C}$ distribution being the most narrow for $SI = 1$ and the 
broadest for $SI=3$.   The mean (median) values for each distribution
are 0.04 (0.04), 0.08 (0.07) and 0.15 (0.14) for the $SI$ = 1, 2 and 3
sources, respectively.  A variability index $\sigma_C/\bar{C} = 0$ indicates no 
variability in the source compactness from one session to the next.
This relationship between the compactness variability index and the 
structure index suggests that the sources with the most structure as 
measured by $SI$ exhibit increased variability in their structure as 
measured by $\sigma_C/\bar{C}$. 

\subsection{Structure Index, Compactness and Source Position Uncertainty \label{SEC:SI_ACCUR}}

The impact of source structure on the high-frequency 
CRF can be further studied by comparing structure index with the 
formal precision of the source positions comprising a potential CRF.
The formal position uncertainties were taken from the K-band astrometric
solution detailed in Paper I.  For this solution, we used the 
CALC/SOLVE software package maintained by the NASA Goddard Space Flight 
Center (GSFC) to perform a least-squares astrometric solution for the K-band
data.  The 10 diurnal K-band experiments encompassed 82,334 measurements 
of bandwidth synthesis (group) delay and phase delay rate.  Geodetic parameters
estimated for each session include:  station positions, 20-minute 
piecewise linear continuous troposphere zenith parameters, tropospheric gradients 
in the east--west and north--south directions, linear in time, estimated once 
per 24 hr session, quadratic clock polynomials to model the gross clock 
behavior, and 60-minute piecewise linear continuous clock parameters.
Corrections for ionospheric refraction drawn from Global Positioning System 
(GPS) total electron content (TEC) maps were applied to K-band data as
discussed in Paper I.   Positions for sources having three or more measurements 
of group delay were the only global parameters that were estimated.

The K-band catalog derived from the astrometric solution is comprised 
of positions and associated formal uncertainties for 268 sources.  
Because there were too few sources at Q-band (131) to separate 
into the four structure index categories, we chose only to compare the
K-band uncertainties with structure index and compactness.  An 
initial comparison of all 268 sources showed the
position uncertainties plotted as a function of $SI$ category to be 
sensitive to a few outliers with relatively few observations.  We, 
therefore, decided to restrict the comparison to sources with 100
or more group delay measurements.  This same restriction was 
used by \cite{FC:00} in a similar study performed at X band.

Shown in Figure~\ref{FIG:POS_UNCER_HIST_100} 
are the distributions of the formal position uncertainties in a) $\alpha\cos\delta$ 
and b) $\delta$ for 193 sources with 100 or more group delay measurements 
at K band.  The distributions are separated into the three $SI$ categories 
1, 2, and 3.  There were two $SI = 4$ sources with greater than 100 delay 
measurements that are not shown.  There is good agreement between the 
mean and median values suggesting little or no dependence on outliers.  
The results show no significant difference in the mean and median position 
uncertainties from one structure index category to the next.  In addition, 
we find the mean and median position uncertainties in $\delta$ are roughly twice 
the uncertainties in $\alpha\cos\delta$ for all three $SI$ categories.  

In \cite{FC:00}, it was shown that the mean and median X-band source
position uncertainties in both $\alpha\cos\delta$ and $\delta$  increased 
regularly as a function of structure index from $SI$=1 to 4.  In addition
to the 100 group delay measurement restriction, \cite{FC:00} {used 
the formal position uncertainties from \cite{MA:98}, which were adjusted
by the standard ICRF inflation factor described therein.
In order to accurately compare our K-band 
values, we applied the same inflation factor (the root sum square (rss) of 
1.5 times the formal uncertainty and 0.25 mas) to the K-band formal 
uncertainties.  Table~\ref{TAB:POS_UNCER} lists the mean and median position 
uncertainties at both X and K bands as a function of $SI$ category.   
Again we see that there is no increase in position uncertainty from 
structure index 1 to 3 at K-band unlike the earlier findings at X-band.  
More importantly, the table shows that for a similar number of sources, 
the inflated mean/median position uncertainties are smaller at K band 
than at X band for all structure index categories.  It should be noted, 
that there have been significant improvements in both the number 
and quality of the VLBI observations since the construction of 
the ICRF, and the \cite{FC:00} values reported in Table~\ref{TAB:POS_UNCER}
do not reflect the accuracy of the current X-band CRF.

In addition to the comparison of the $SI$ and the formal uncertainties, 
we compared the mean source compactness ($\bar C$) to the formal
uncertainties.   Recall that the maximum possible value of the source 
compactness $C = 1.0$ indicates that all of the source flux density is 
contained within one synthesized beam.  The mean compactness is 
just an average over all of the sessions in which the source was imaged.
Figure~\ref{FIG:POS_COMPACT_SCATTER} shows the formal uncertainties
in the position versus the source compactness.  The plotted formal position 
uncertainty is the rss of the uncertainties in 
$\alpha\cos\delta$ and $\delta$.  The figure also shows the structure 
index for each source by point color and type, 
with green circles, blue triangles, and red squares indicating 
maximum $SI$ values of 1, 2, and 3 respectively. 
Figure \ref{FIG:POS_COMPACT_SCATTER} does not show a clear trend in 
the formal position uncertainty versus compactness.  This is to be expected 
since it was previously shown that the $\bar C$ and $SI$ are correlated, 
and no significant relationship was found for the position uncertainty as a 
function of $SI$ (e.g. Figure~\ref{FIG:POS_UNCER_HIST_100}).   

\subsection{Structure Index, Compactness, and Source Position Stability \label{SEC:SI_STAB}}

Another useful comparison can be made between the stability of the 
source positions over time and the source structure 
as traced by either the structure index or the source compactness.  
Time variation of the astrometric coordinates of
CRF sources has previously been attributed to variability in the intrinsic
source structure \citep[c.f.][]{CHARLOT:94, CHARLOT:02, FEK:97, FC:00}. 
Thus, it is natural to search for any relationship between such variations
and the structure index at higher frequencies.  A measure of source 
position stability is the weighted root-mean-square (wrms) position 
variation.  These variations were obtained from a series of K-band 
astrometric solutions described in Paper I.  In these solutions, a fraction 
of the sources were treated as local parameters (i.e. an estimate 
of the position was derived for each session in which the source was 
observed) with the remaining sources treated as global.  The estimation 
of source position stability was limited to 88 sources that were 
observed in five or more of the 10 VLBA sessions.  Five was 
considered a sufficient number of sessions (position estimates) 
per source to derive reliable statistics.   

From these solutions, a time series of source positions was 
generated for each source, and from these time series the wrms
position variations were computed.  Shown in Figure~\ref{FIG:TS_HIST} 
are the distributions of the wrms position variations in $\alpha\cos\delta$ 
(a) and $\delta$ (b), respectively.  The figure also lists the mean and median 
position variations for each distribution.  In $\alpha\cos\delta$, the 
mean and median of the distribution are 0.14 and 0.12 mas, respectively.
For $\delta$ the mean and median are roughly twice as large at
0.30 and 0.25 mas, respectively.  As stated in Paper I, the larger 
wrms position variations in $\delta$ are likely due to the combination of 
ionospheric/tropospheric effects and network geometry, specifically the 
lack of long north--south baselines for the VLBA.

It would have been desirable to separate the wrms position variations 
into structure index categories, and to plot the distributions as a function
of $SI$ as was done in $\S$\ref{SEC:SI_ACCUR}.  Because the structure 
index is quantized into four categories, there were simply too few sources
observed in five or more sessions to determine reliable statistics for the 
separate $SI$ categories.  Instead, in Figure~\ref{FIG:TS_COMPACT_SCATTER},
we plot the wrms variation against the mean source compactness previously 
described in $\S$\ref{SEC:COMPACT}.  The wrms variation plotted is the
rss of the variations in $\alpha\cos\delta$ and $\delta$.  In addition, the 
structure index for each source is shown by point type with green circles, blue triangles, 
and red squares indicating maximum $SI$ of 1, 2, and 3 respectively. 
Figure \ref{FIG:TS_COMPACT_SCATTER} does not suggest a clear trend in 
the compactness and wrms position stability. While it does appear that 
there are a number of sources clustered in a region of high compactness 
and low wrms variations, there are also sources that are less compact 
with equally low wrms variations.  A more quantitative comparison between the
$SI$ and the position stability will have to wait for additional high-frequency
observations.  

\section{SUMMARY}

We have produced a combined total of 1339 K- and Q-band images 
of 274 extragalactic sources from 10 sessions of VLBA data 
observed as part of a program to select high-frequency sources for 
use in spacecraft navigation and future CRFs.  
A detailed analysis of the images has allowed us to determine 
several quantities that can serve as useful indicators of both detectability 
and astrometric quality including: source flux density, flux density variability, 
structure index, source compactness, and compactness variability.
The results of our imaging and analysis can be summarized as 
follows:

\begin{enumerate}

\item{The sources described herein were selected from 
previous catalogs at lower (e.g. 8.4 GHz) frequencies.   The distributions 
of source flux density at X, K, and Q bands show mean values greater 
than 1~Jy for each of the three bands.  These results suggest a sufficient 
number of suitable sources for astrometric use at high frequencies. }

\item{For sources imaged at 24 GHz, we find that, on average, 
1$\sigma$ variations in the session-to-session flux density
were 20\%--25\% over the five-year period of our observations.  
Comparatively, variations of the flux density scale from one 
session to the next were less than 10\%.  Variations 
in the source flux density are, thus, relatively small and it should 
be expected that sources will be observable from one session to the next. }

\item{There is a clear dependence of source structure index on 
observational frequency.  The percentage of sources with $SI=1, 2$ 
(the most compact) increases from $\sim$70\% at X band, to 
$\sim$85\% at K band, and to $\sim$92\% at Q band.   This suggests 
that the higher resolution afforded by the higher observing frequencies
yields less structure at a fixed angular scale.  This result is corroborated
by the mean source compactness which shows that 80\% of the 
total flux density is contained within the core at all three frequencies.} 

\item{On average, the 1$\sigma$ variations in the source compactness 
at K and Q-band were no more than about 8\% at either frequency over 
the five-year period of our VLBA observations.}  

\item{For the K-band data, we find a trend between the source
compactness variability index (i.e., the variation of the compactness 
over time) and the source structure index indicating that sources with 
larger values of $SI$ exhibit increased variation in the structure as 
measured by $\sigma_C/\bar{C}$.}

\item{No clear trend was found between the formal uncertainties 
in the astrometric source positions and the source structure index at 
K band in contrast to the trend found at X band by \cite{FC:00}.  }

\end{enumerate}

The implications of the above results with regards to spacecraft 
navigational needs are twofold.  First, there are numerous sources available
at high frequencies, and these sources persist over relatively long (five years)
periods of time.  Second, the sources are more compact at these 
higher frequencies and should provide high-quality astrometric 
reference points for spacecraft navigation.  Such 
sources also provide a foundation upon which to build a future CRF 
and are of benefit to radio astronomy, in general, as fiducial references.  
Future imaging observations should continue to improve our 
understanding of source structure at K and Q band and allow us to better 
characterize the variability of the structure over time and its effect on astrometric 
positional stability. 

\acknowledgements

This research was partially supported through NASA contracts with 
the California Institute of Technology and the U.S. Naval Observatory.     
This research made use of the USNO Radio Reference
Frame Image Database (RRFID).

\clearpage

\begin{sidewaysfigure}[phbt]
\centering
\includegraphics[height=8.0in,angle=90]{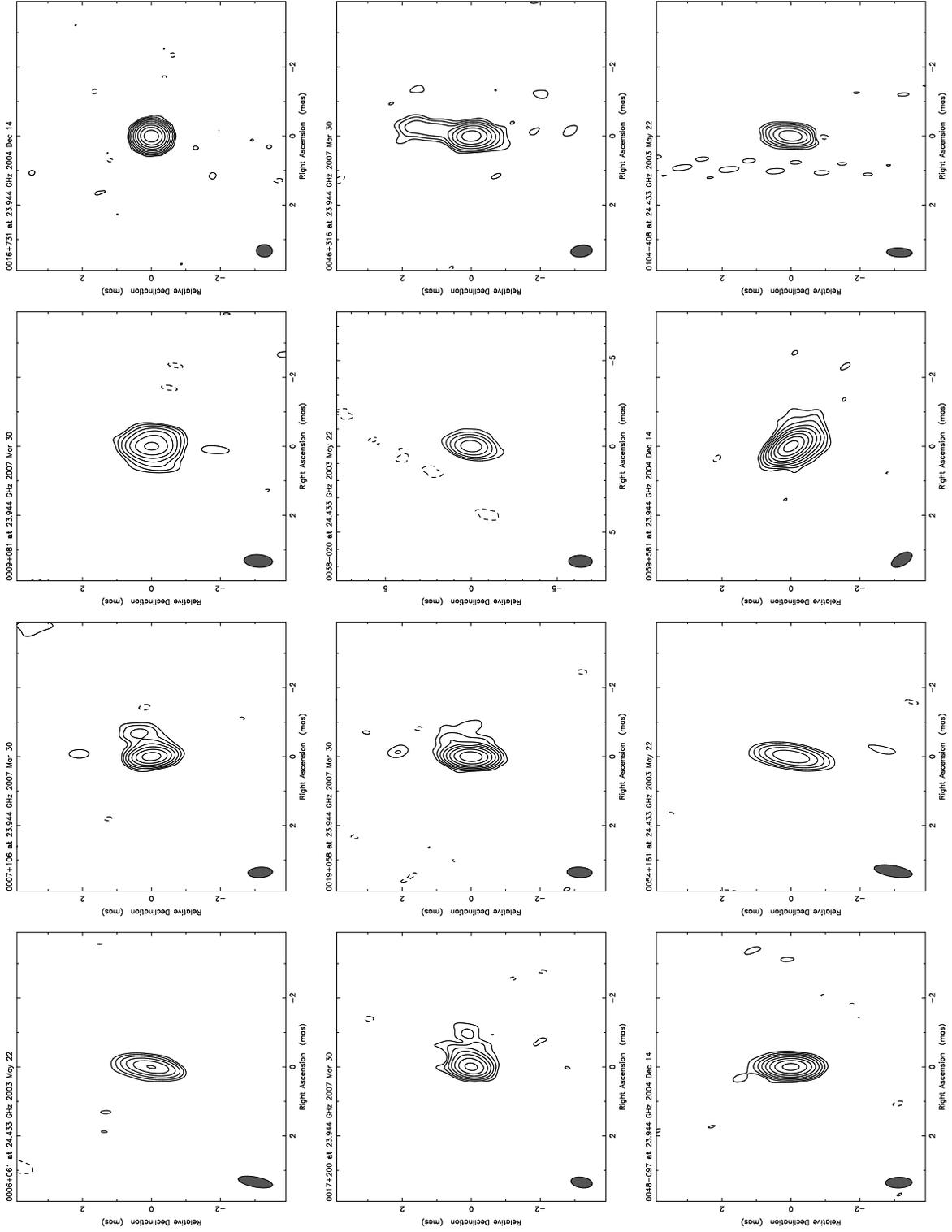}
\figcaption{Contour plots of 274 extragalactic radio sources
at K band.  Image parameters are listed in
Table~\ref{TAB:KBAND_IMG_PARAM}.  The
scale of each image is in milliarcseconds. The FWHM Gaussian restoring
beam applied to the images is shown as a hatched ellipse in the lower
left of each panel. For convenience, the images for each band are
labeled only by a single fiducial frequency ($\sim$23.9 or 24.4~GHz,
depending on session) even though they were made using the data from all
frequency channels (see \S\,\ref{SEC:OBS}). 
\label{FIG:KBAND_CNTR}}
\end{sidewaysfigure}
\clearpage
\begin{sidewaysfigure}[phbt]
\centering
\includegraphics[height=8.0in,angle=90]{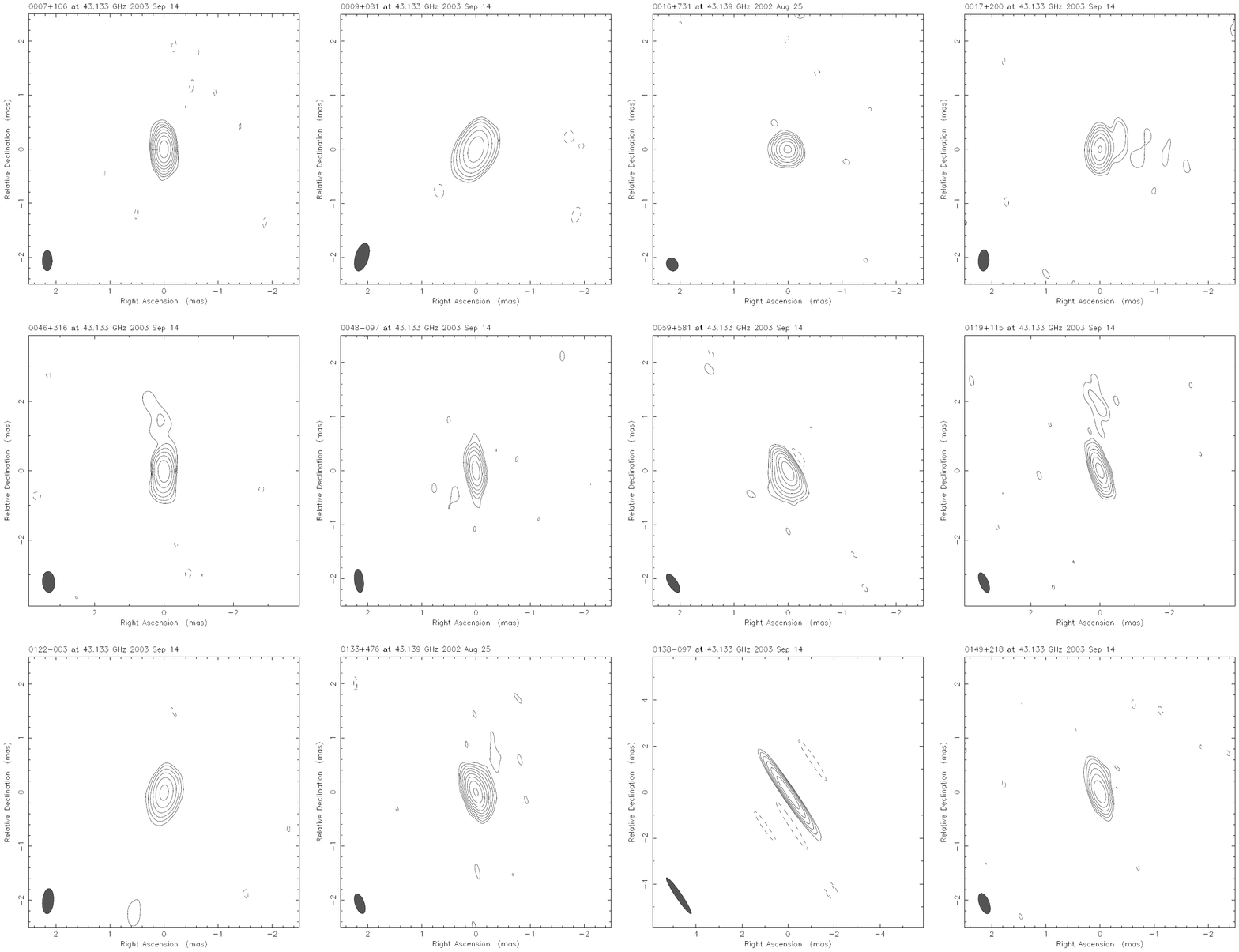}
\newline\newline Fig. 1.--- {\it Continued}
\end{sidewaysfigure}
\clearpage
\begin{sidewaysfigure}[phbt]
\centering
\includegraphics[height=8.0in,angle=90]{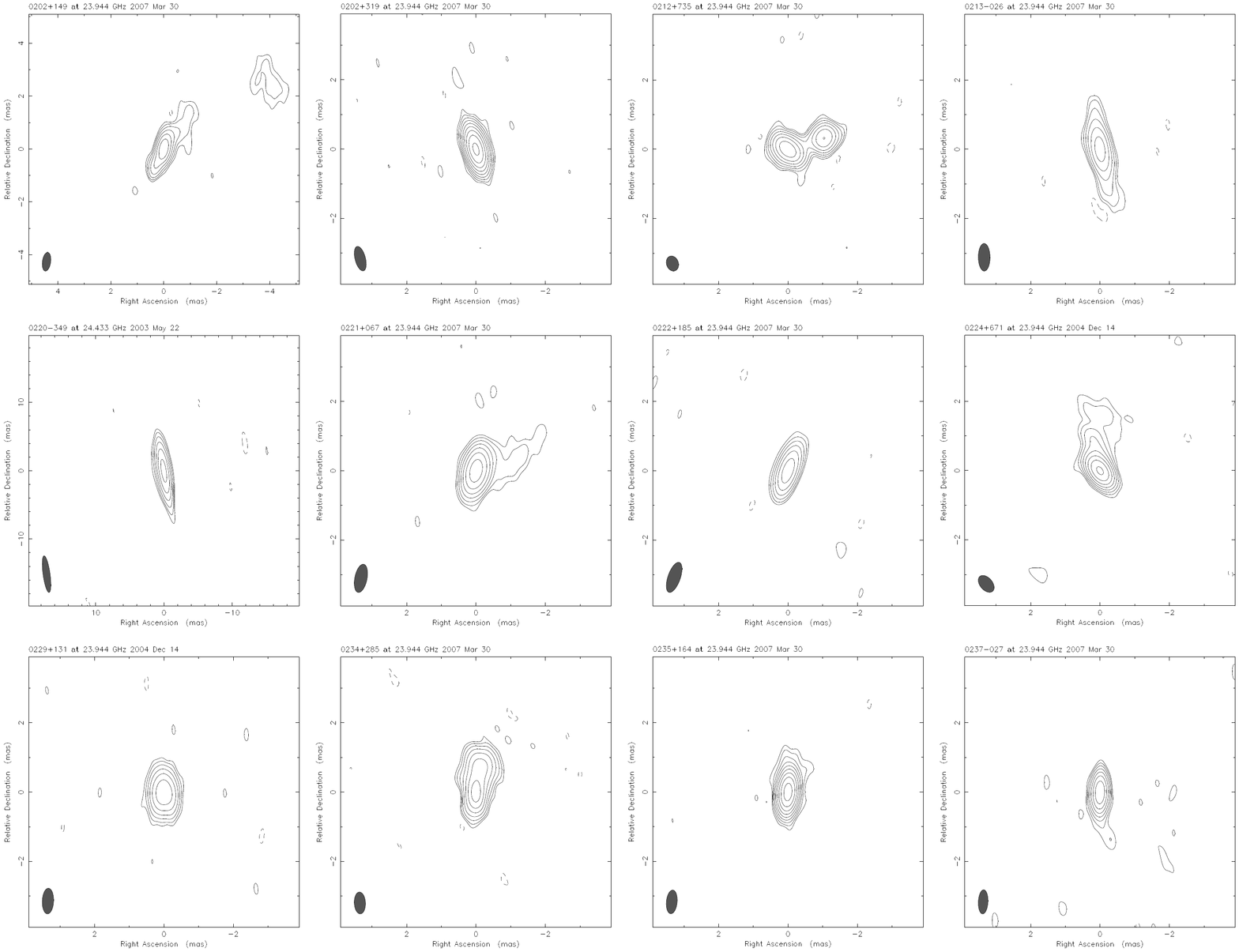}
\newline\newline Fig. 1.--- {\it Continued}
\end{sidewaysfigure}
\clearpage
\begin{sidewaysfigure}[phbt]
\centering
\includegraphics[height=8.0in,angle=90]{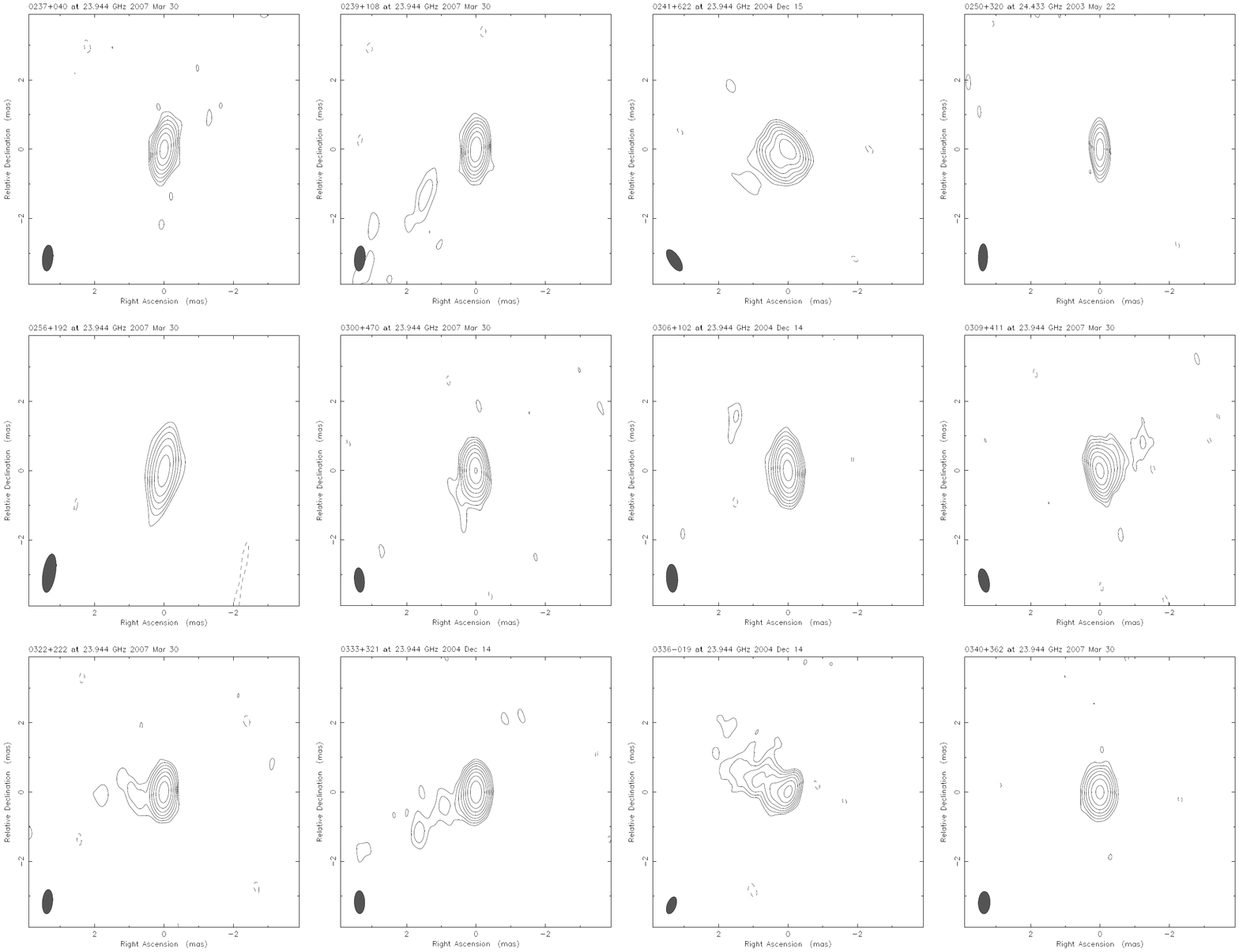}
\newline\newline Fig. 1.--- {\it Continued}
\end{sidewaysfigure}
\clearpage
\begin{sidewaysfigure}[phbt]
\centering
\includegraphics[height=8.0in,angle=90]{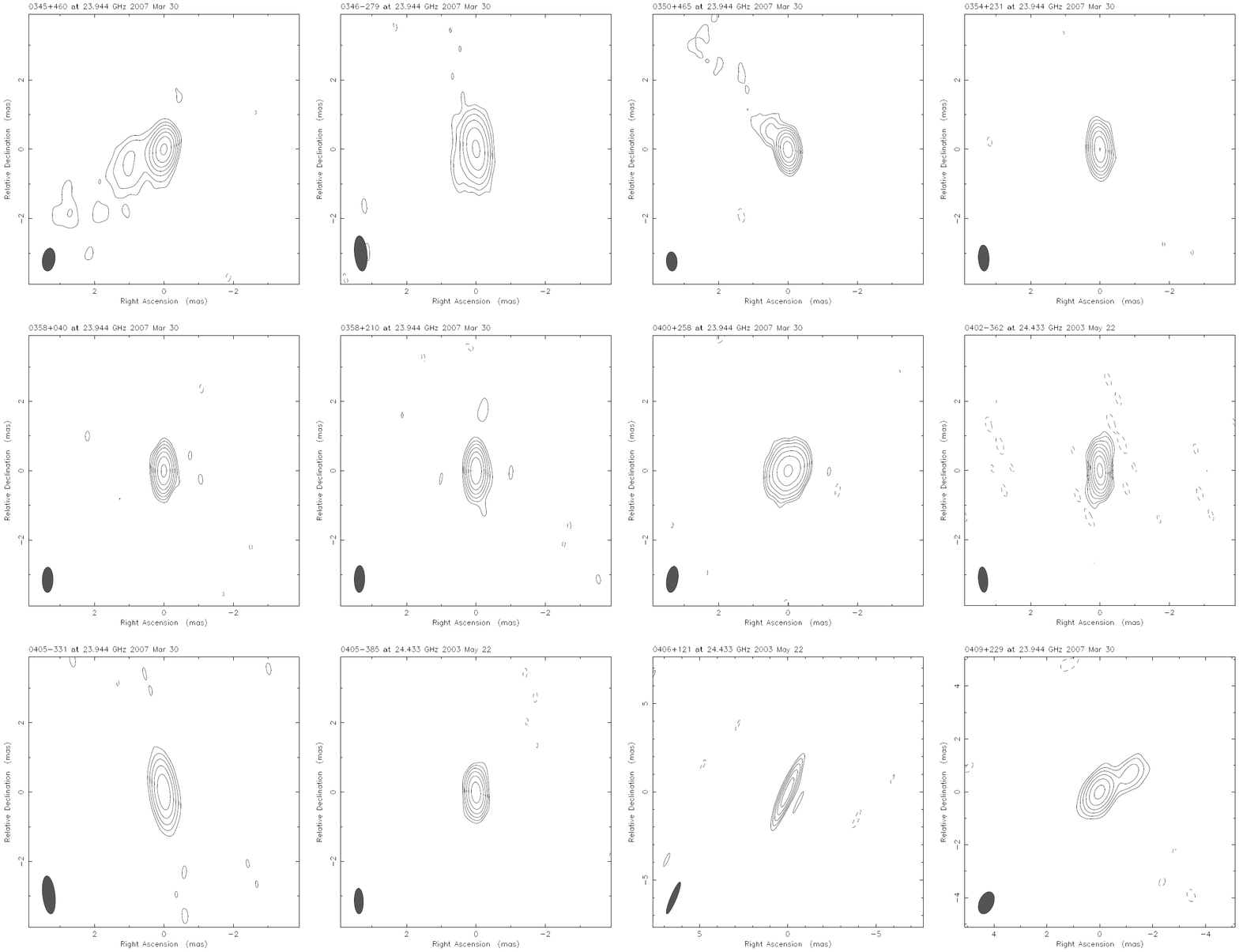}
\newline\newline Fig. 1.--- {\it Continued}
\end{sidewaysfigure}
\clearpage
\begin{sidewaysfigure}[phbt]
\centering
\includegraphics[height=8.0in,angle=90]{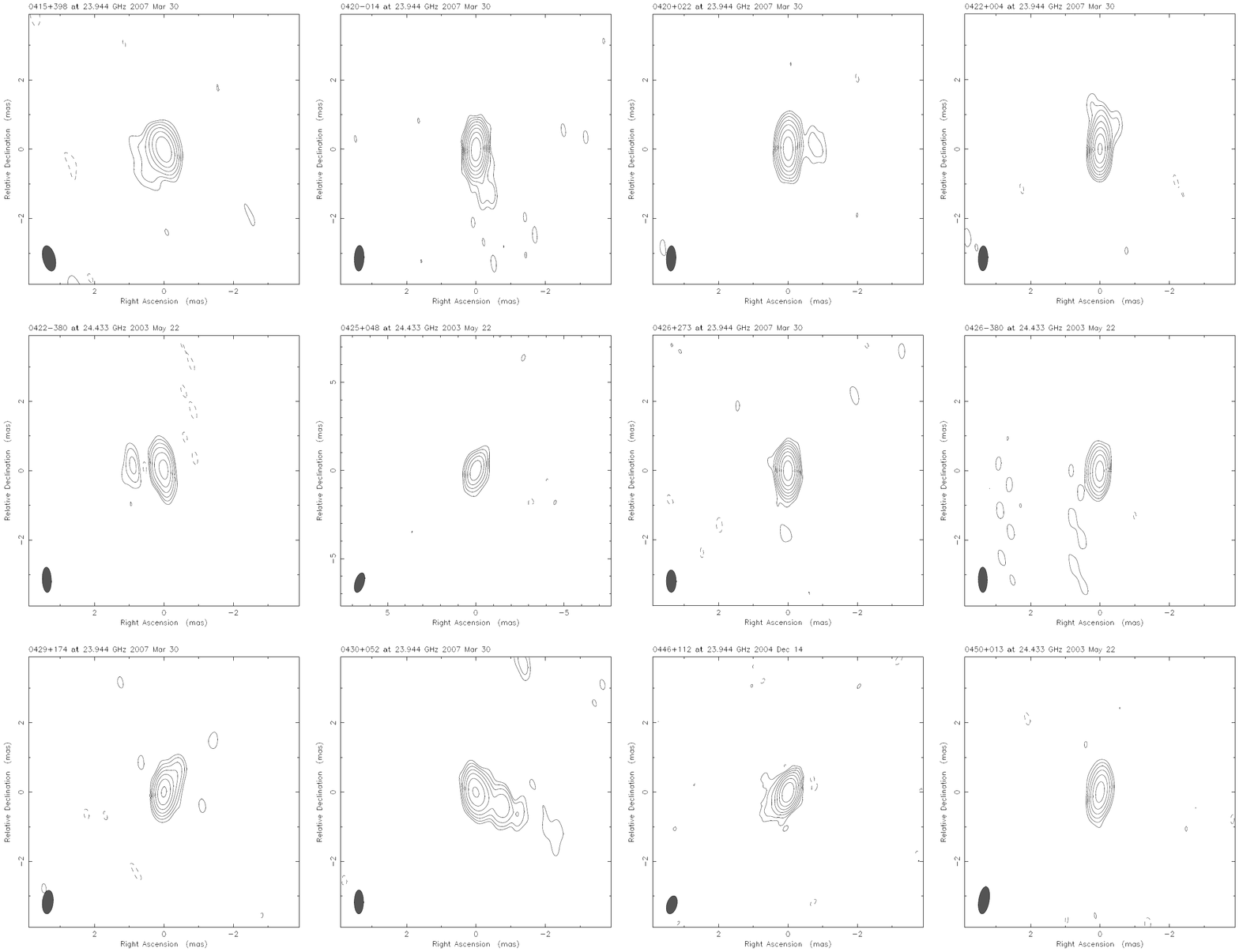}
\newline\newline Fig. 1.--- {\it Continued}
\end{sidewaysfigure}
\clearpage
\begin{sidewaysfigure}[phbt]
\centering
\includegraphics[height=8.0in,angle=90]{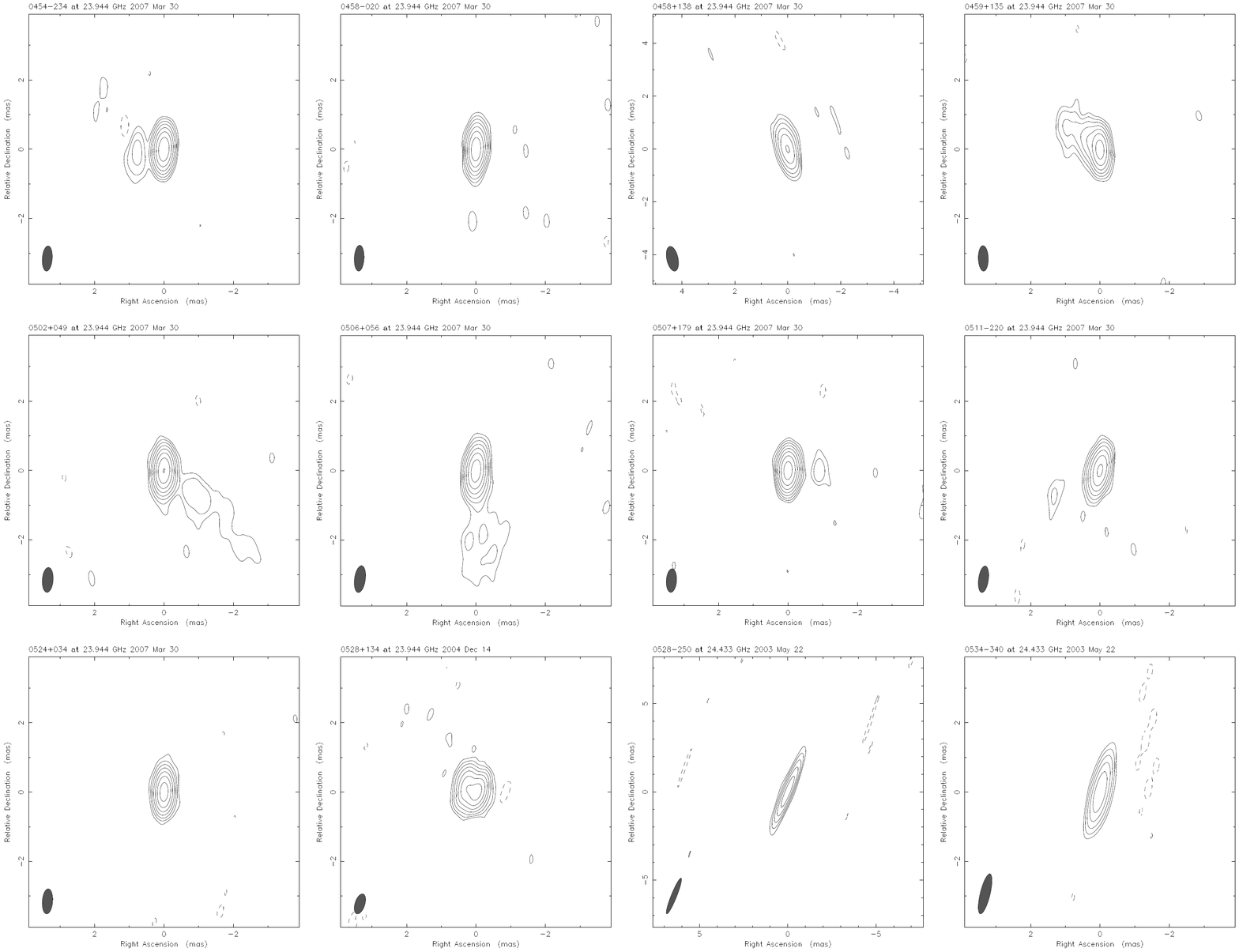}
\newline\newline Fig. 1.--- {\it Continued}
\end{sidewaysfigure}
\clearpage
\begin{sidewaysfigure}[phbt]
\centering
\includegraphics[height=8.0in,angle=90]{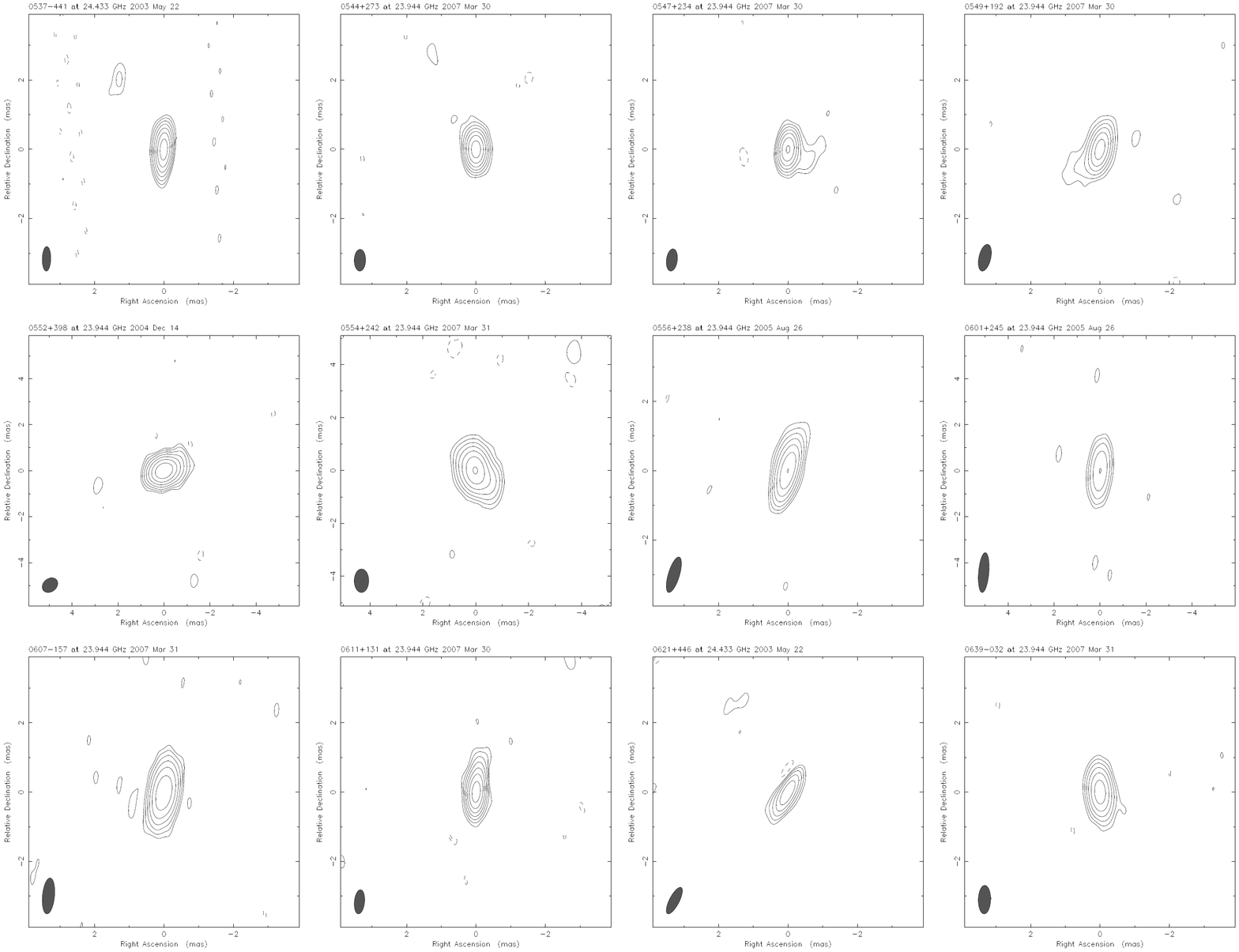}
\newline\newline Fig. 1.--- {\it Continued}
\end{sidewaysfigure}
\clearpage
\begin{sidewaysfigure}[phbt]
\centering
\includegraphics[height=8.0in,angle=90]{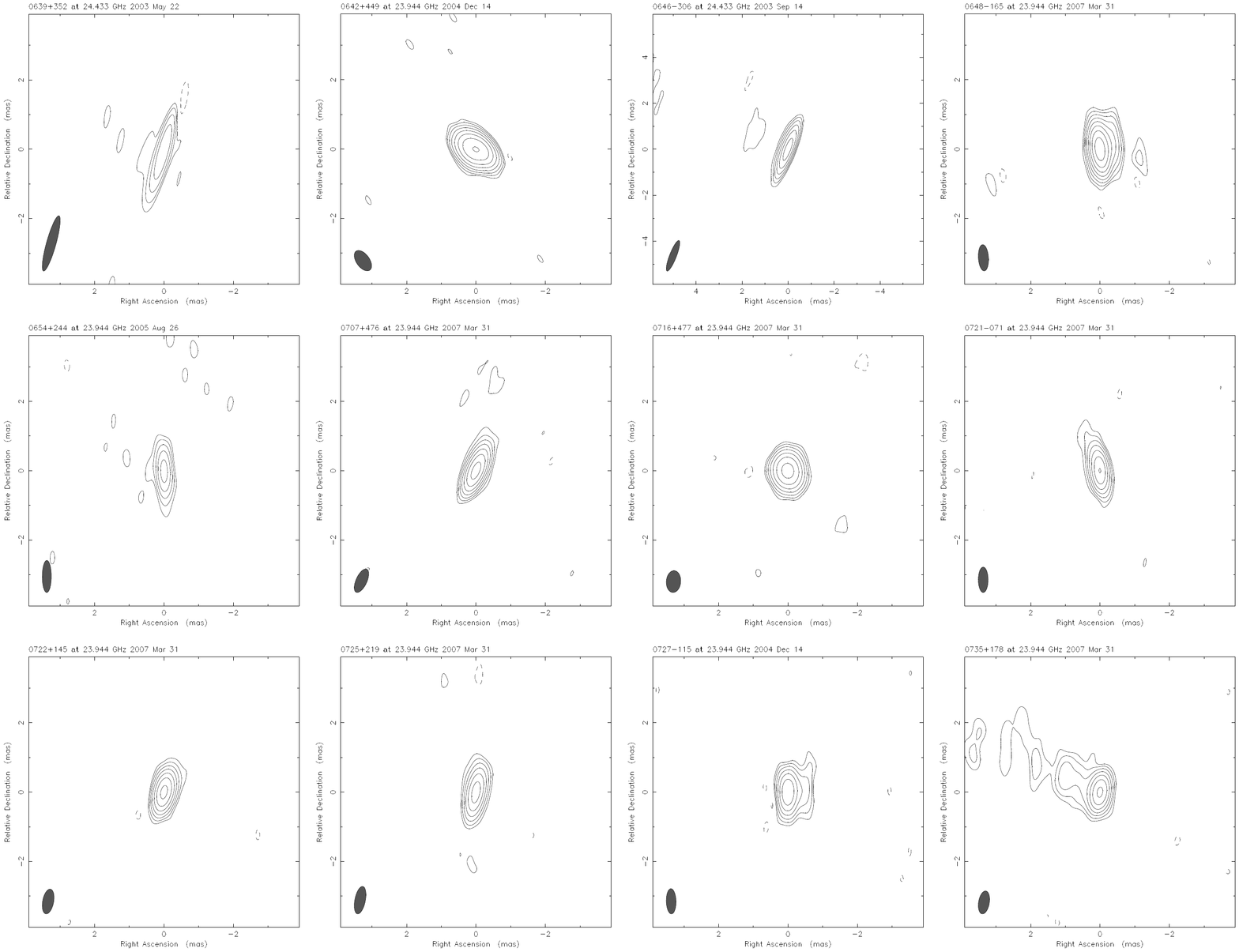}
\newline\newline Fig. 1.--- {\it Continued}
\end{sidewaysfigure}
\clearpage
\begin{sidewaysfigure}[phbt]
\centering
\includegraphics[height=8.0in,angle=90]{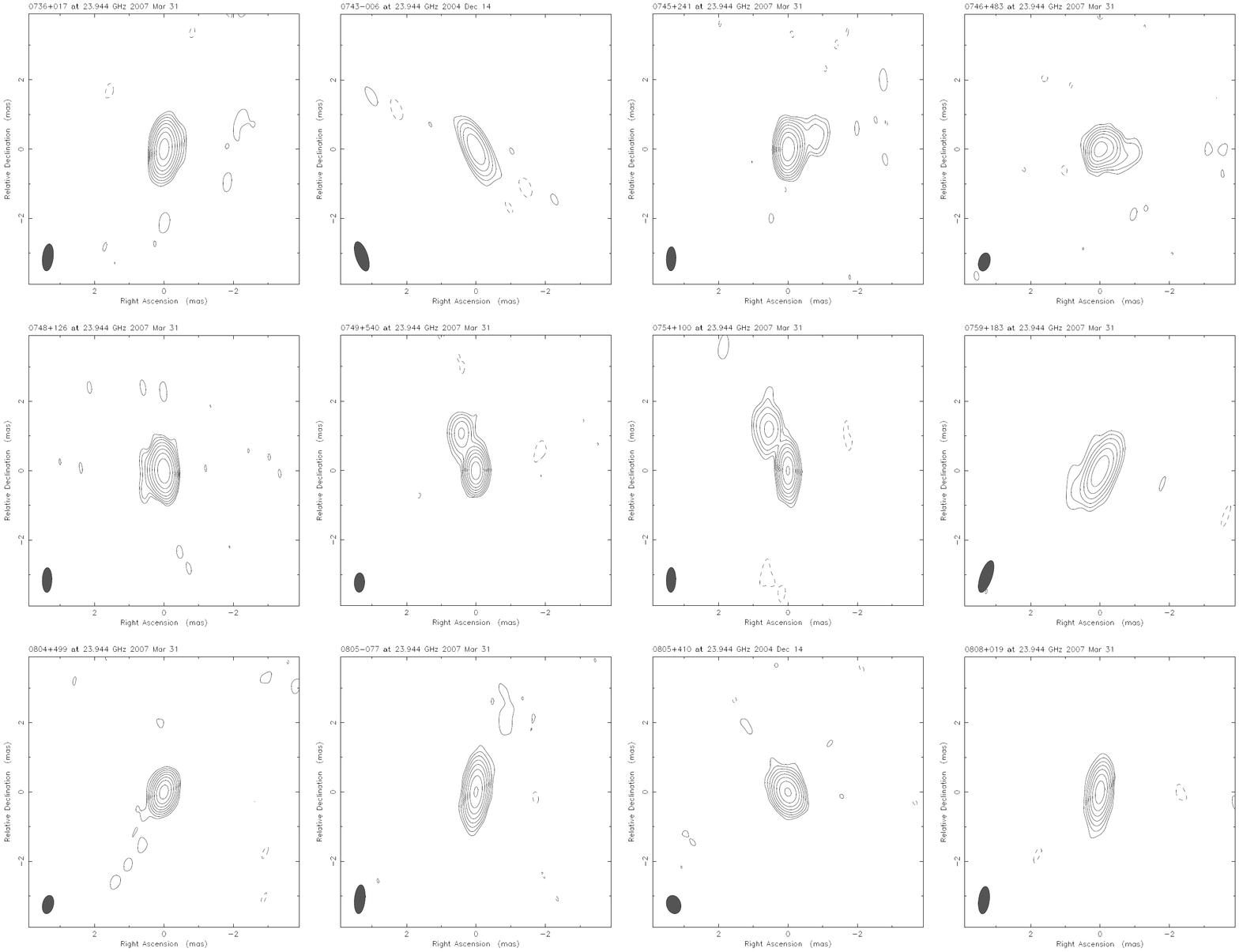}
\newline\newline Fig. 1.--- {\it Continued}
\end{sidewaysfigure}
\clearpage
\begin{sidewaysfigure}[phbt]
\centering
\includegraphics[height=8.0in,angle=90]{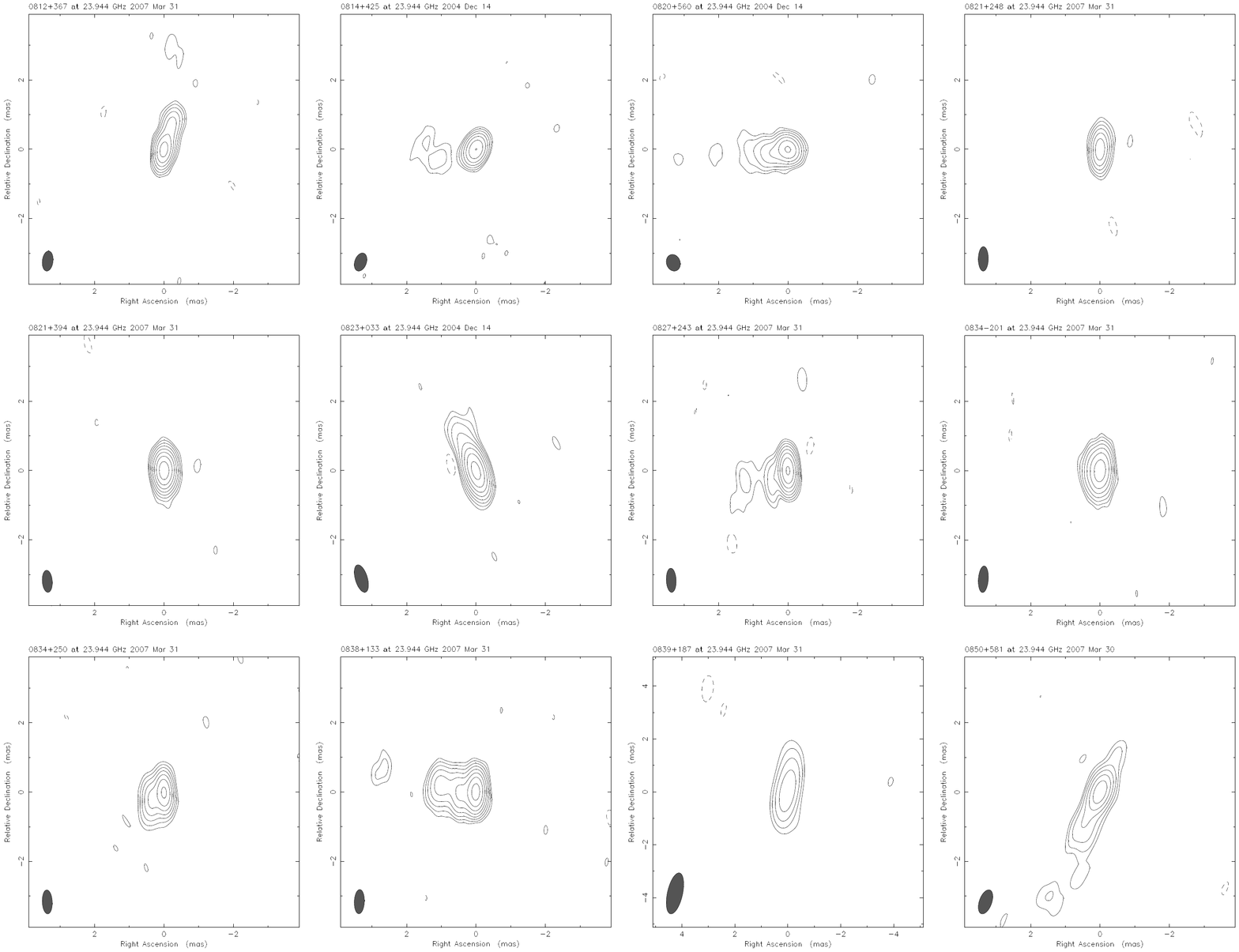}
\newline\newline Fig. 1.--- {\it Continued}
\end{sidewaysfigure}
\clearpage
\begin{sidewaysfigure}[phbt]
\centering
\includegraphics[height=8.0in,angle=90]{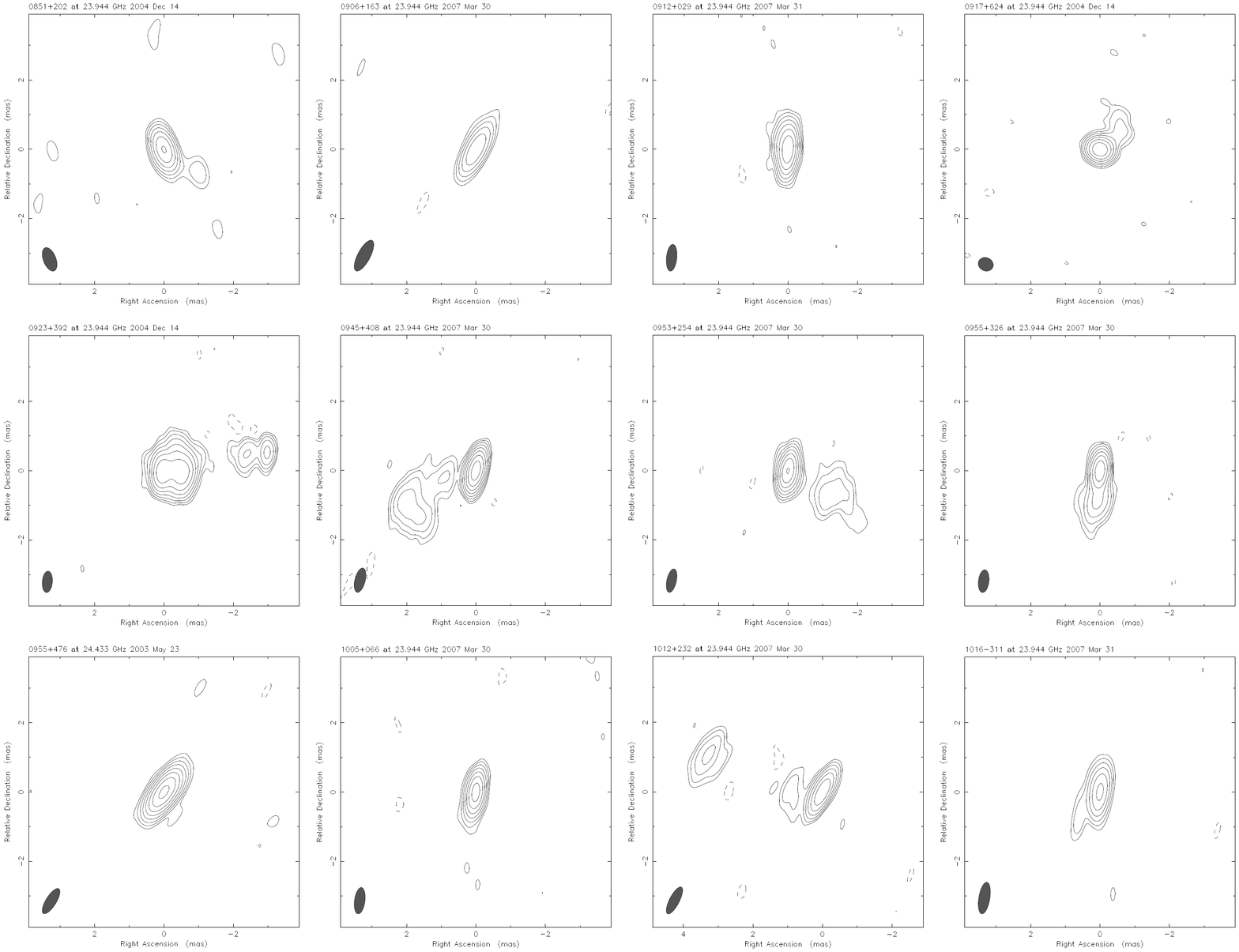}
\newline\newline Fig. 1.--- {\it Continued}
\end{sidewaysfigure}
\clearpage
\begin{sidewaysfigure}[phbt]
\centering
\includegraphics[height=8.0in,angle=90]{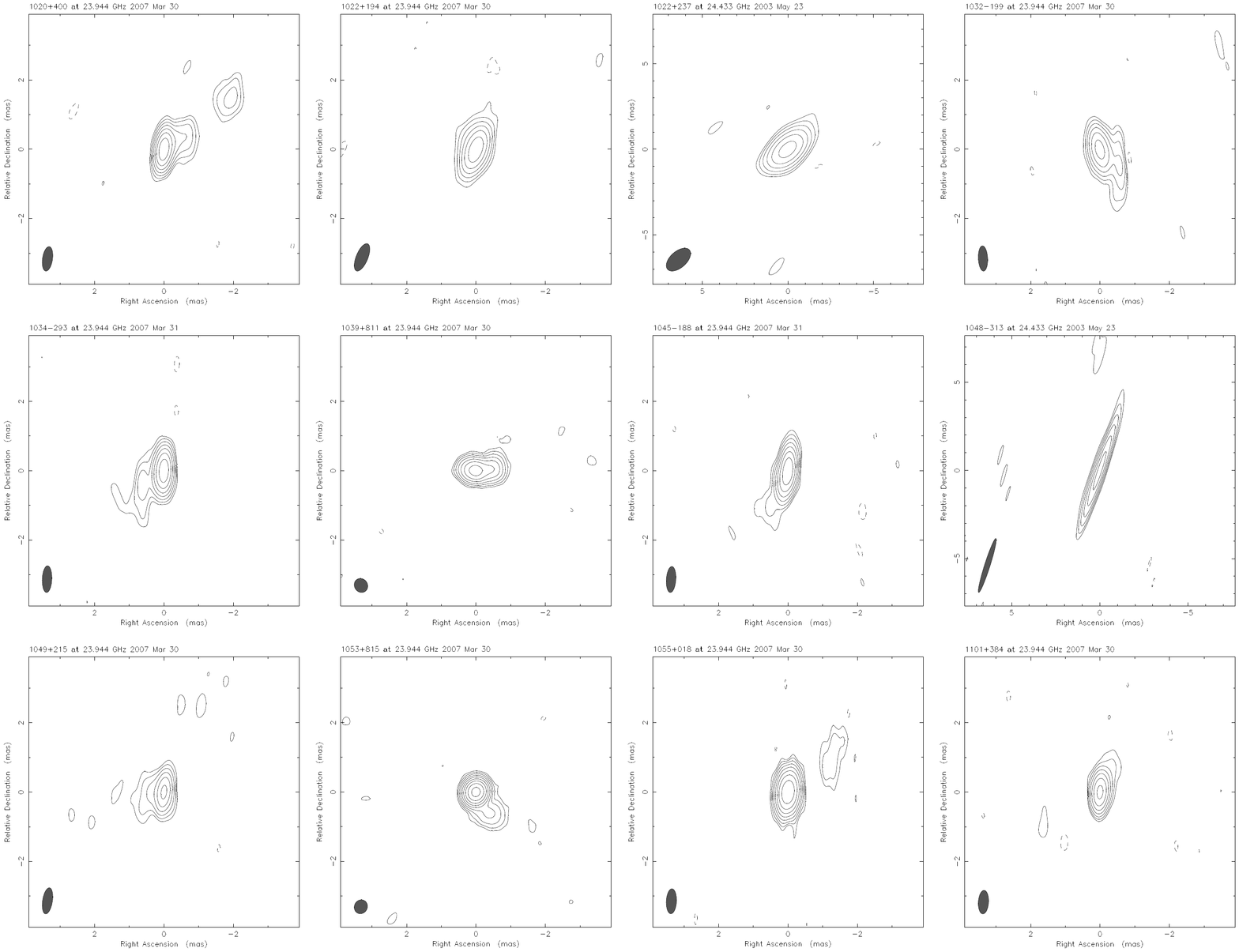}
\newline\newline Fig. 1.--- {\it Continued}
\end{sidewaysfigure}
\clearpage
\begin{sidewaysfigure}[phbt]
\centering
\includegraphics[height=8.0in,angle=90]{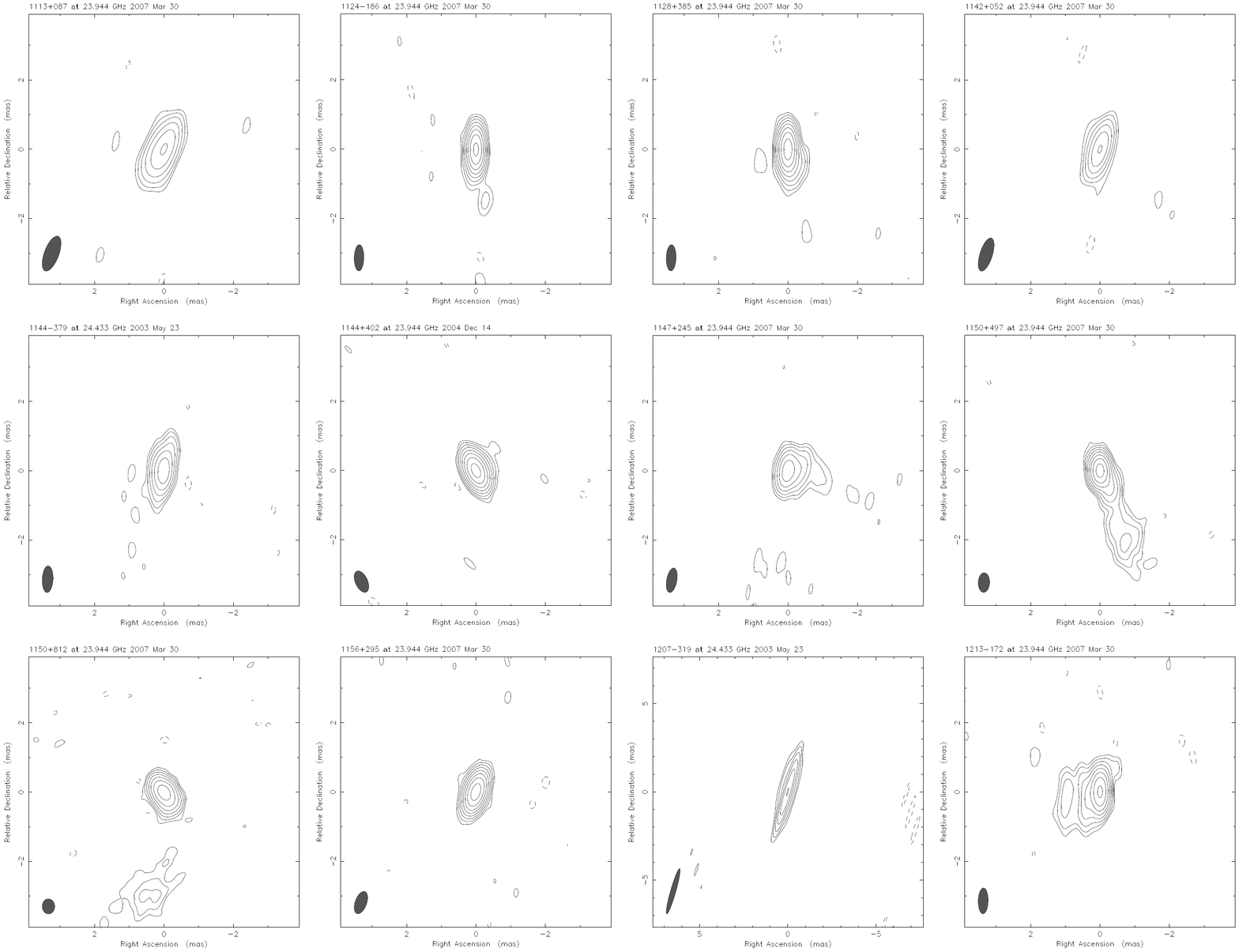}
\newline\newline Fig. 1.--- {\it Continued}
\end{sidewaysfigure}
\clearpage
\begin{sidewaysfigure}[phbt]
\centering
\includegraphics[height=8.0in,angle=90]{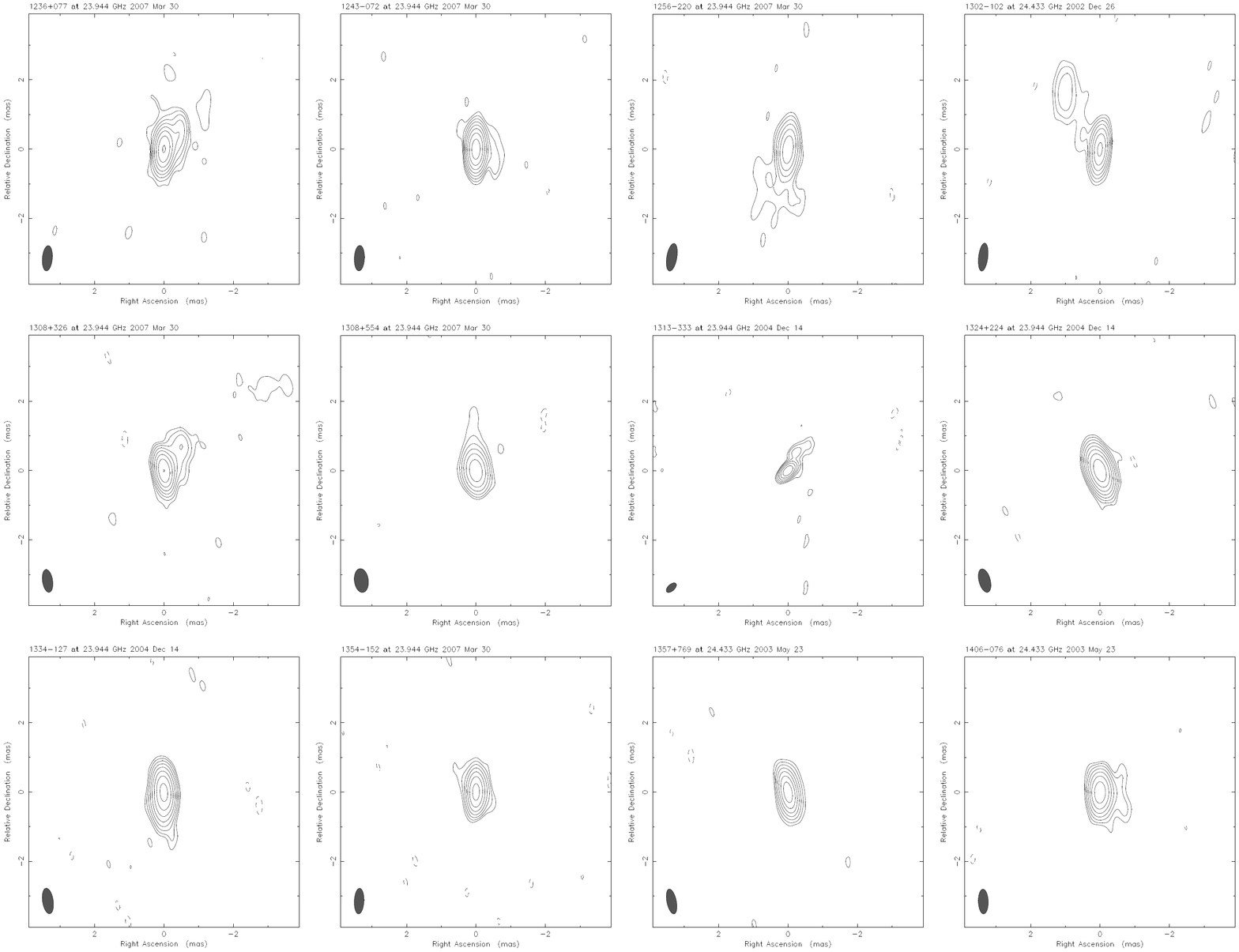}
\newline\newline Fig. 1.--- {\it Continued}
\end{sidewaysfigure}
\clearpage
\begin{sidewaysfigure}[phbt]
\centering
\includegraphics[height=8.0in,angle=90]{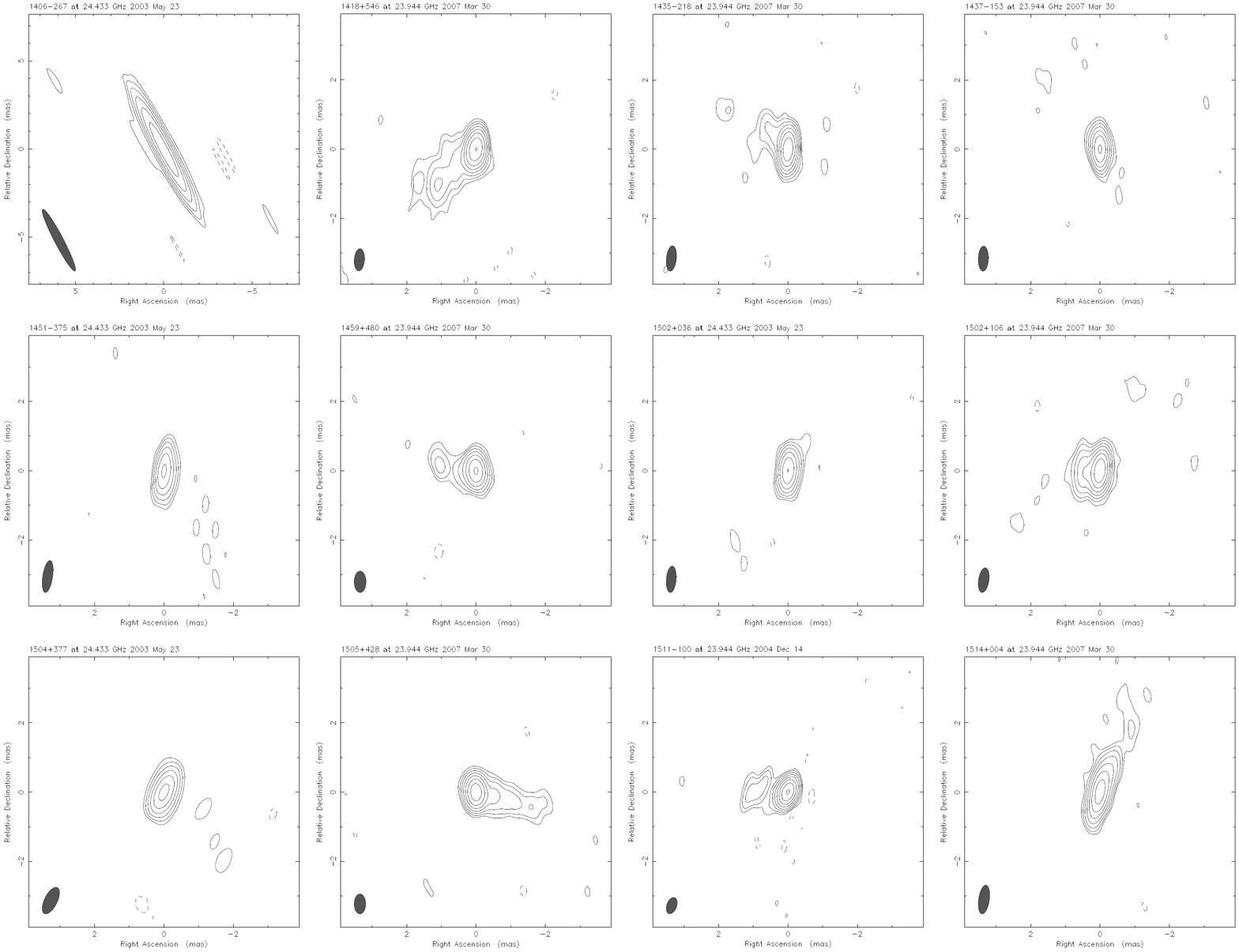}
\newline\newline Fig. 1.--- {\it Continued}
\end{sidewaysfigure}
\clearpage
\begin{sidewaysfigure}[phbt]
\centering
\includegraphics[height=8.0in,angle=90]{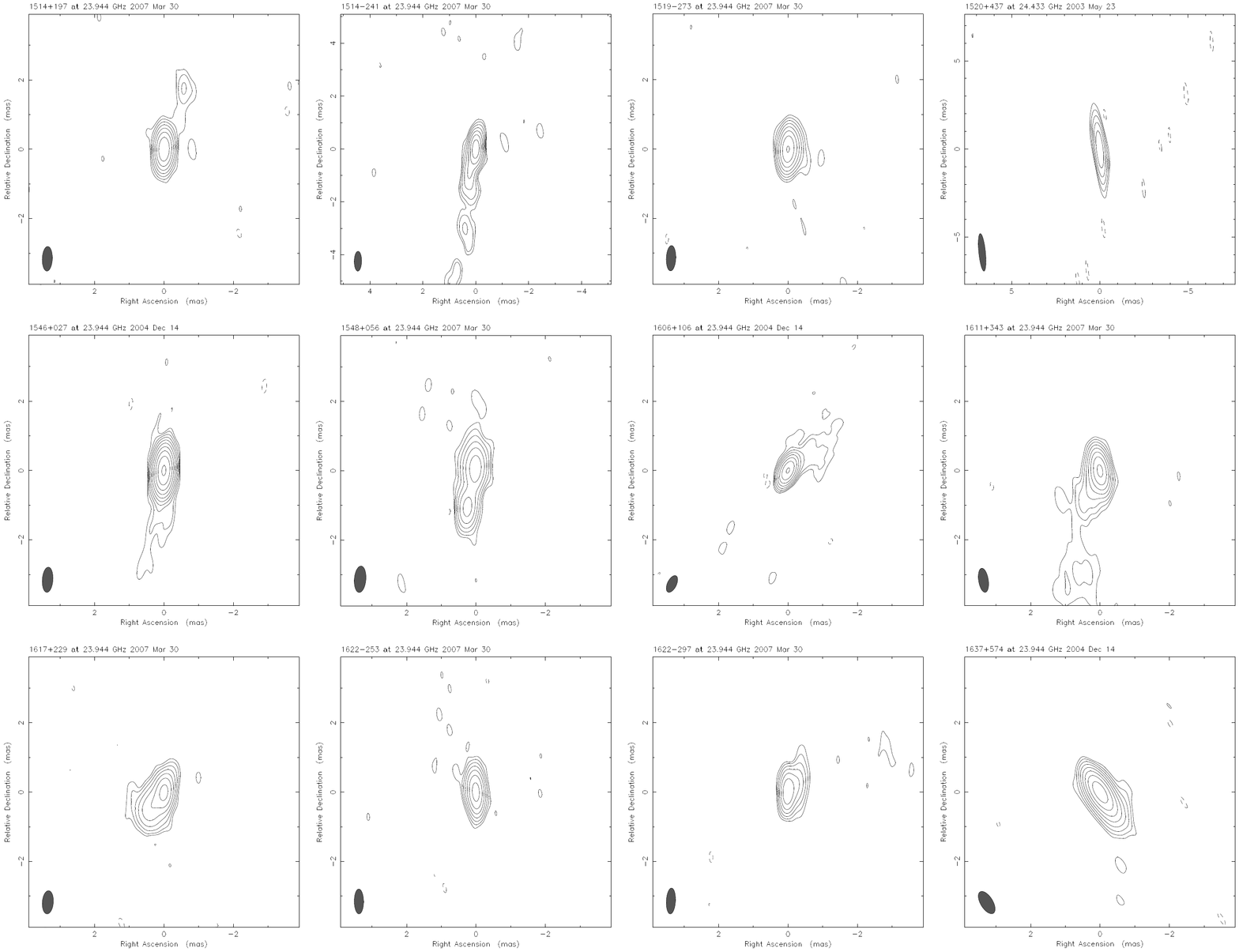}
\newline\newline Fig. 1.--- {\it Continued}
\end{sidewaysfigure}
\clearpage
\begin{sidewaysfigure}[phbt]
\centering
\includegraphics[height=8.0in,angle=90]{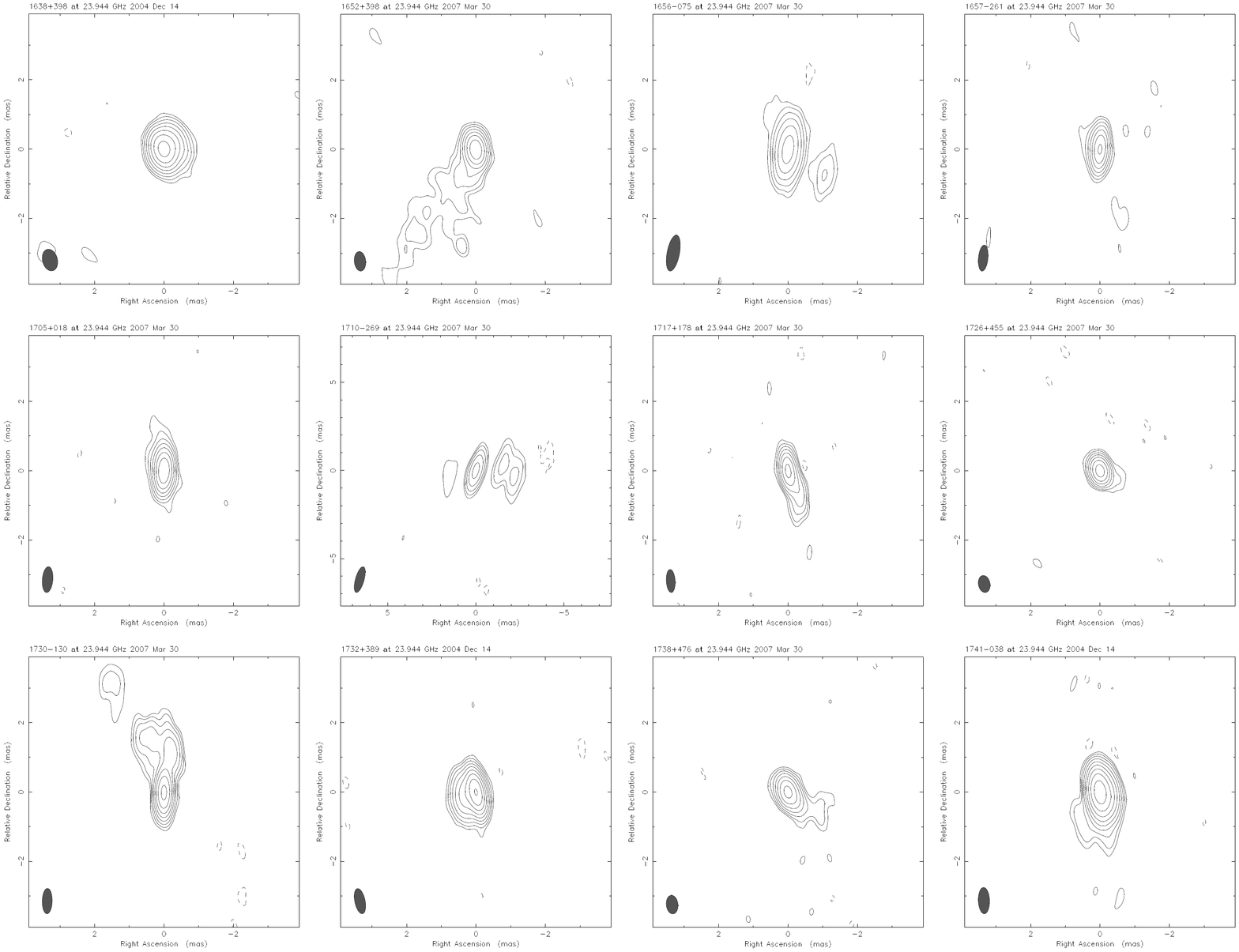}
\newline\newline Fig. 1.--- {\it Continued}
\end{sidewaysfigure}
\clearpage
\begin{sidewaysfigure}[phbt]
\centering
\includegraphics[height=8.0in,angle=90]{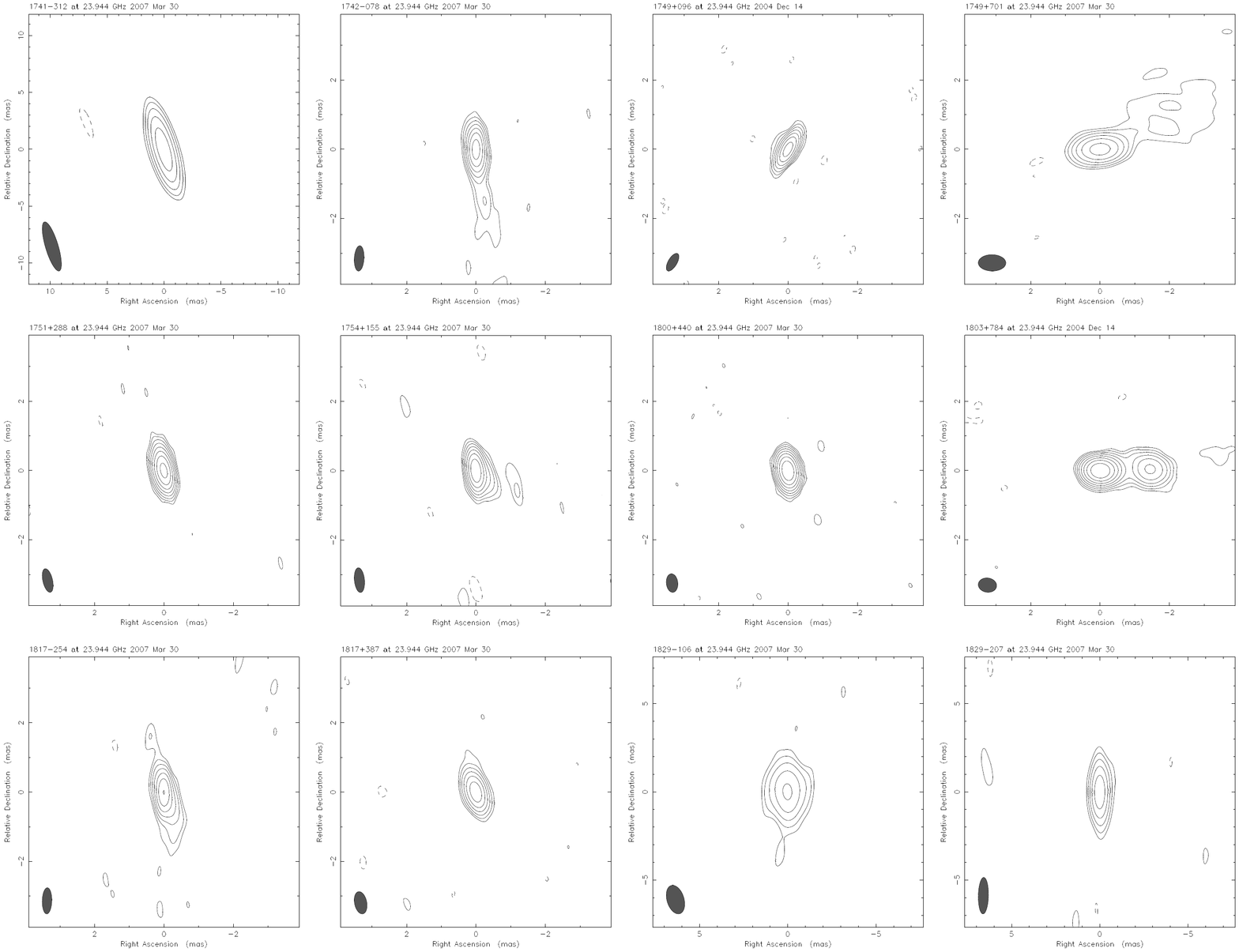}
\newline\newline Fig. 1.--- {\it Continued}
\end{sidewaysfigure}
\clearpage
\begin{sidewaysfigure}[phbt]
\centering
\includegraphics[height=8.0in,angle=90]{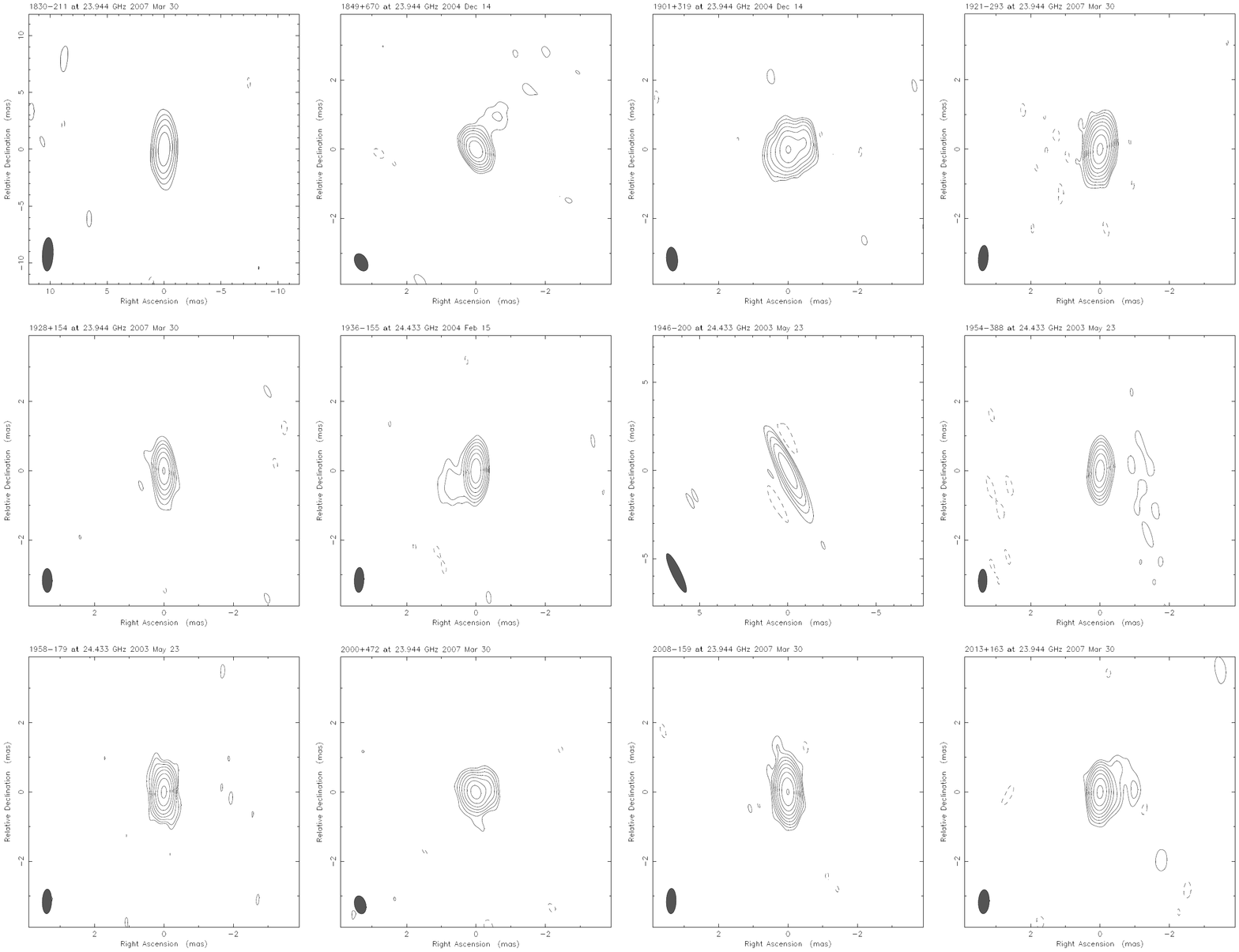}
\newline\newline Fig. 1.--- {\it Continued}
\end{sidewaysfigure}
\clearpage
\begin{sidewaysfigure}[phbt]
\centering
\includegraphics[height=8.0in,angle=90]{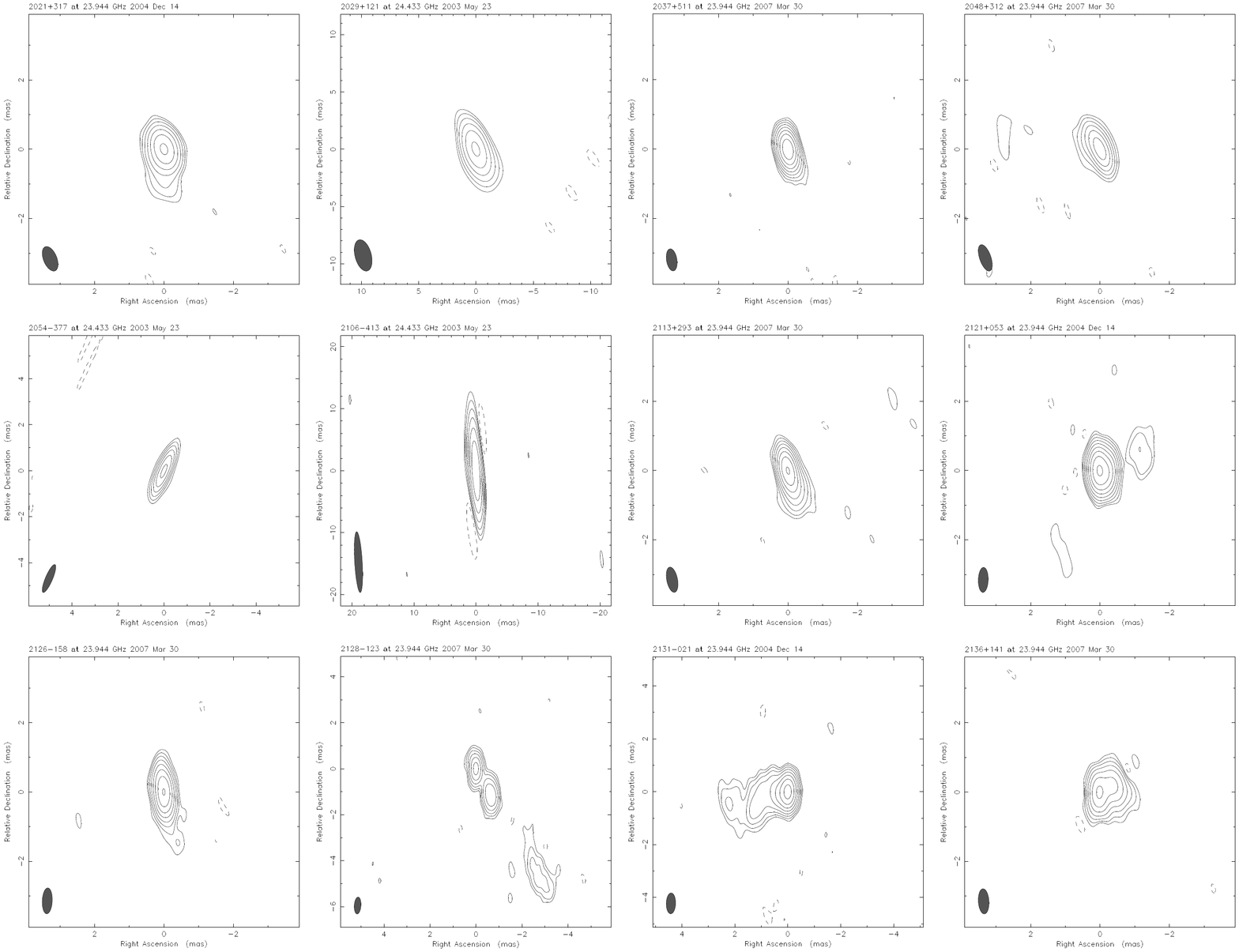}
\newline\newline Fig. 1.--- {\it Continued}
\end{sidewaysfigure}
\clearpage
\begin{sidewaysfigure}[phbt]
\centering
\includegraphics[height=8.0in,angle=90]{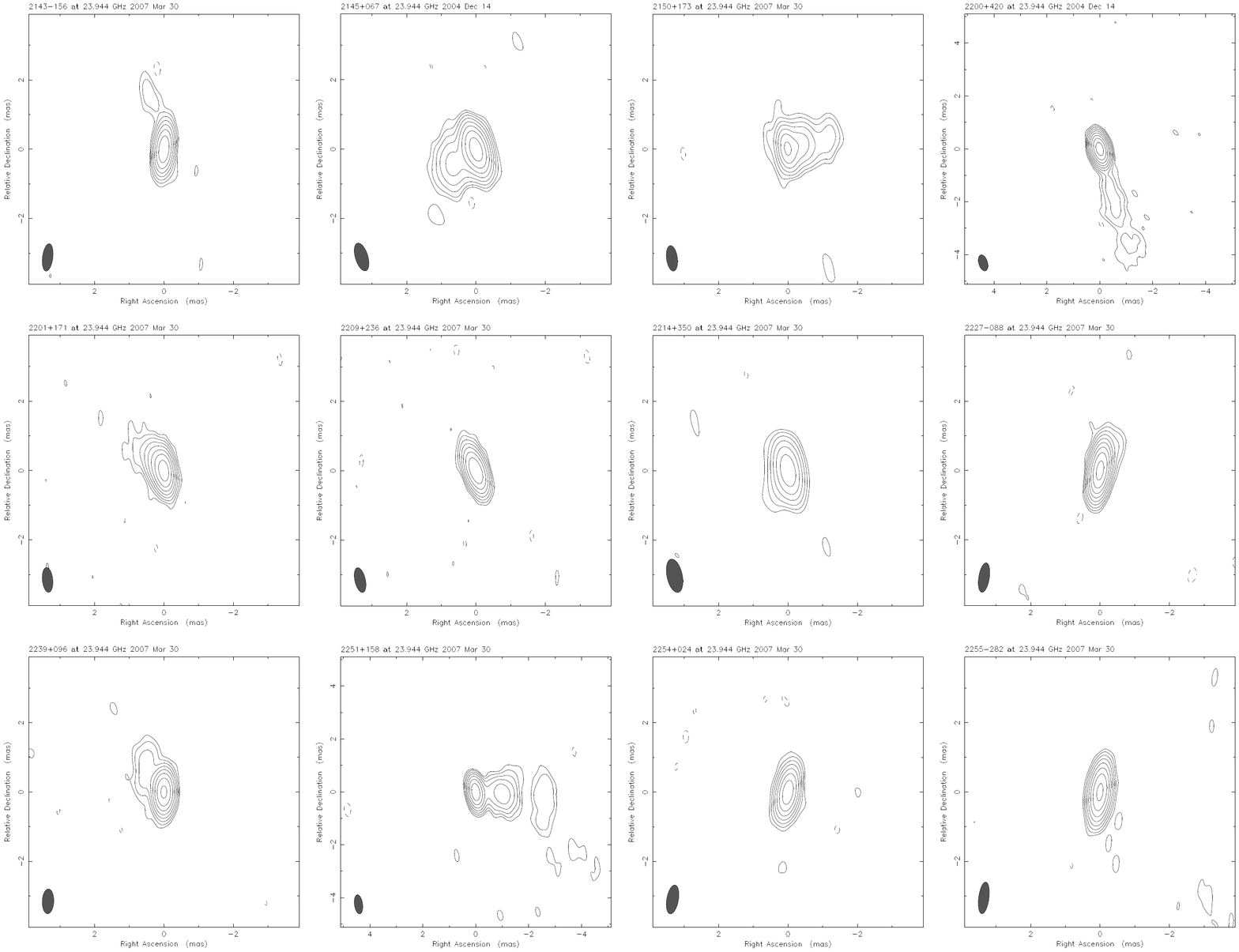}
\newline\newline Fig. 1.--- {\it Continued}
\end{sidewaysfigure}
\clearpage
\begin{sidewaysfigure}[phbt]
\centering
\includegraphics[height=8.0in,angle=90]{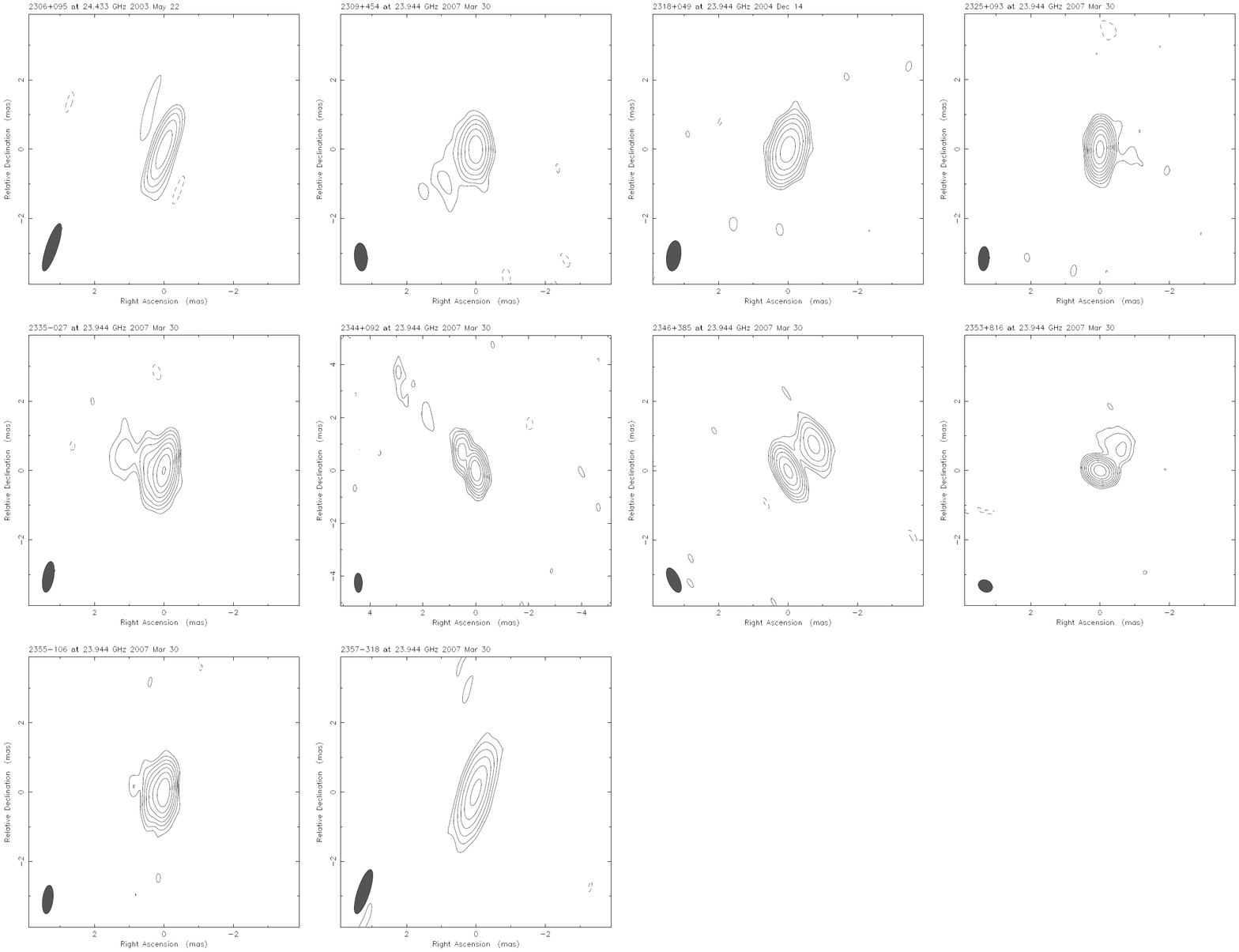}
\newline\newline Fig. 1.--- {\it Continued}
\end{sidewaysfigure}

\begin{sidewaysfigure}[phbt]
\centering
\includegraphics[height=8.0in,angle=90]{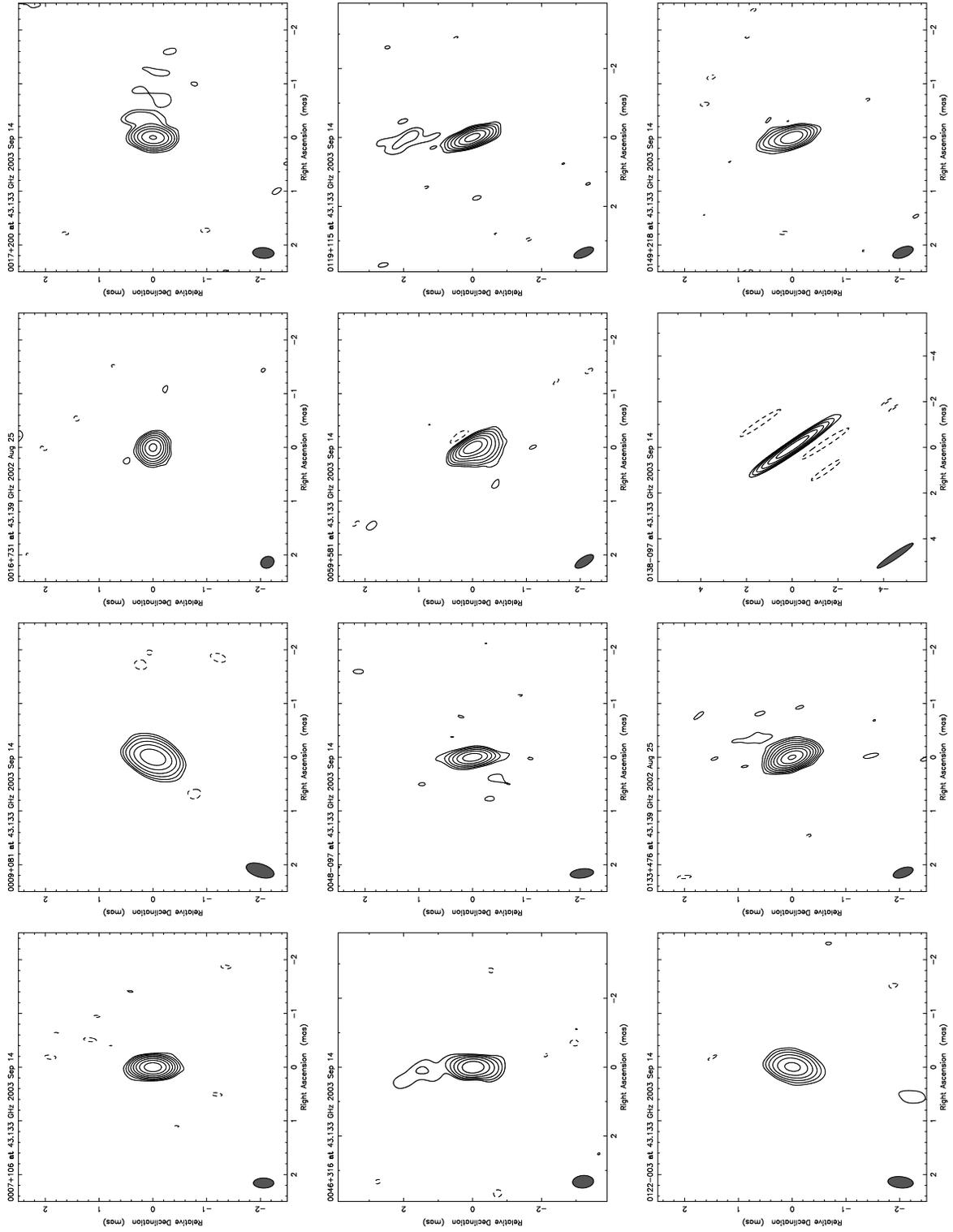}
\figcaption{Contour plots of 132 extragalactic radio sources
at Q band.  Image parameters are listed in
Table~\ref{TAB:QBAND_IMG_PARAM}.  The
scale of each image is in milliarcseconds. The FWHM Gaussian restoring
beam applied to the images is shown as a hatched ellipse in the lower
left of each panel. For convenience, the images for each band are
labeled only by a single fiducial frequency ($\sim$43.1~GHz) even though 
they were made using the data from all frequency channels 
(see \S\,\ref{SEC:OBS}).  \label{FIG:QBAND_CNTR}}
\end{sidewaysfigure}
\clearpage
\begin{sidewaysfigure}[phbt]
\centering
\includegraphics[height=8.0in,angle=90]{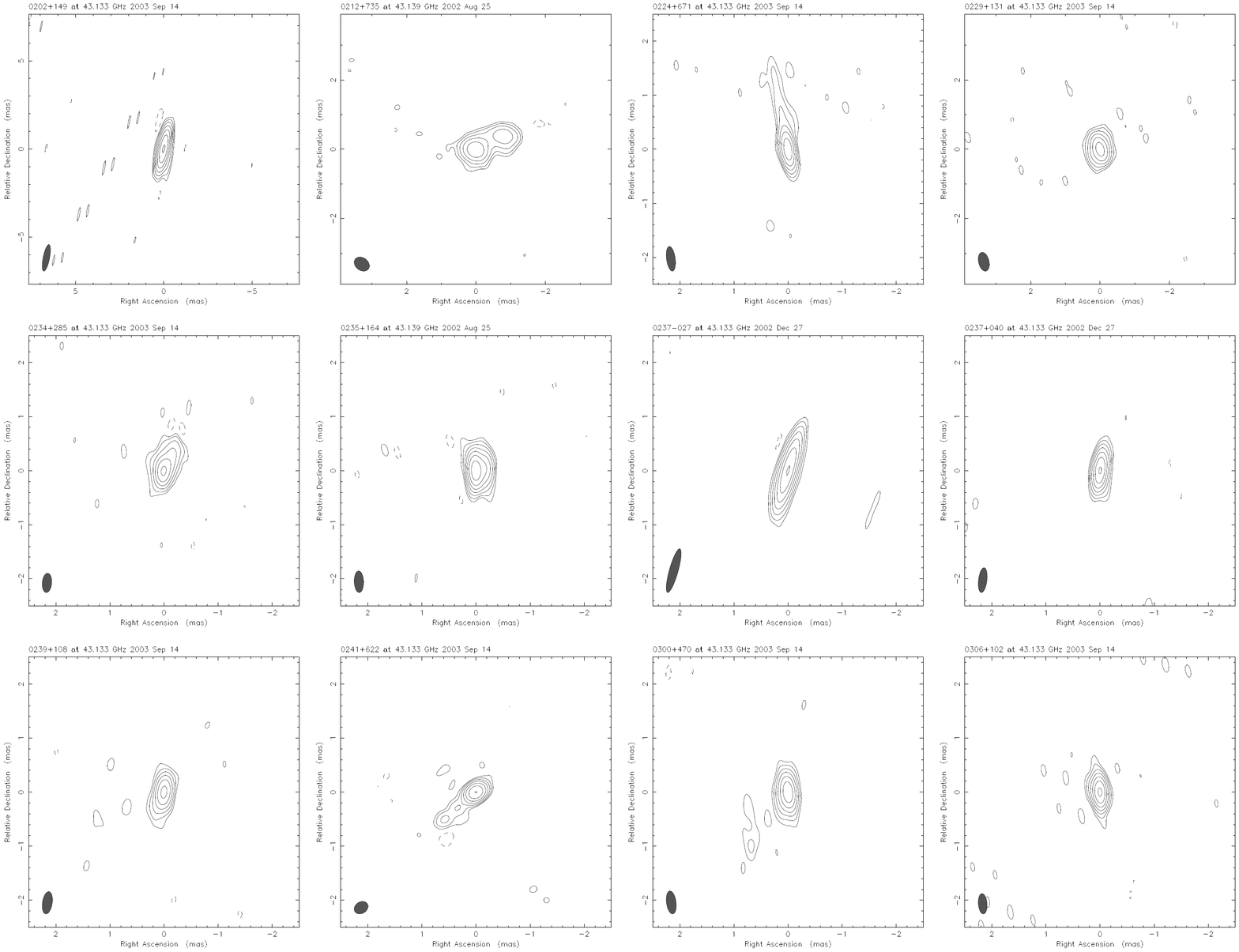}
\newline\newline Fig. 2.--- {\it Continued}
\end{sidewaysfigure}
\clearpage
\begin{sidewaysfigure}[phbt]
\centering
\includegraphics[height=8.0in,angle=90]{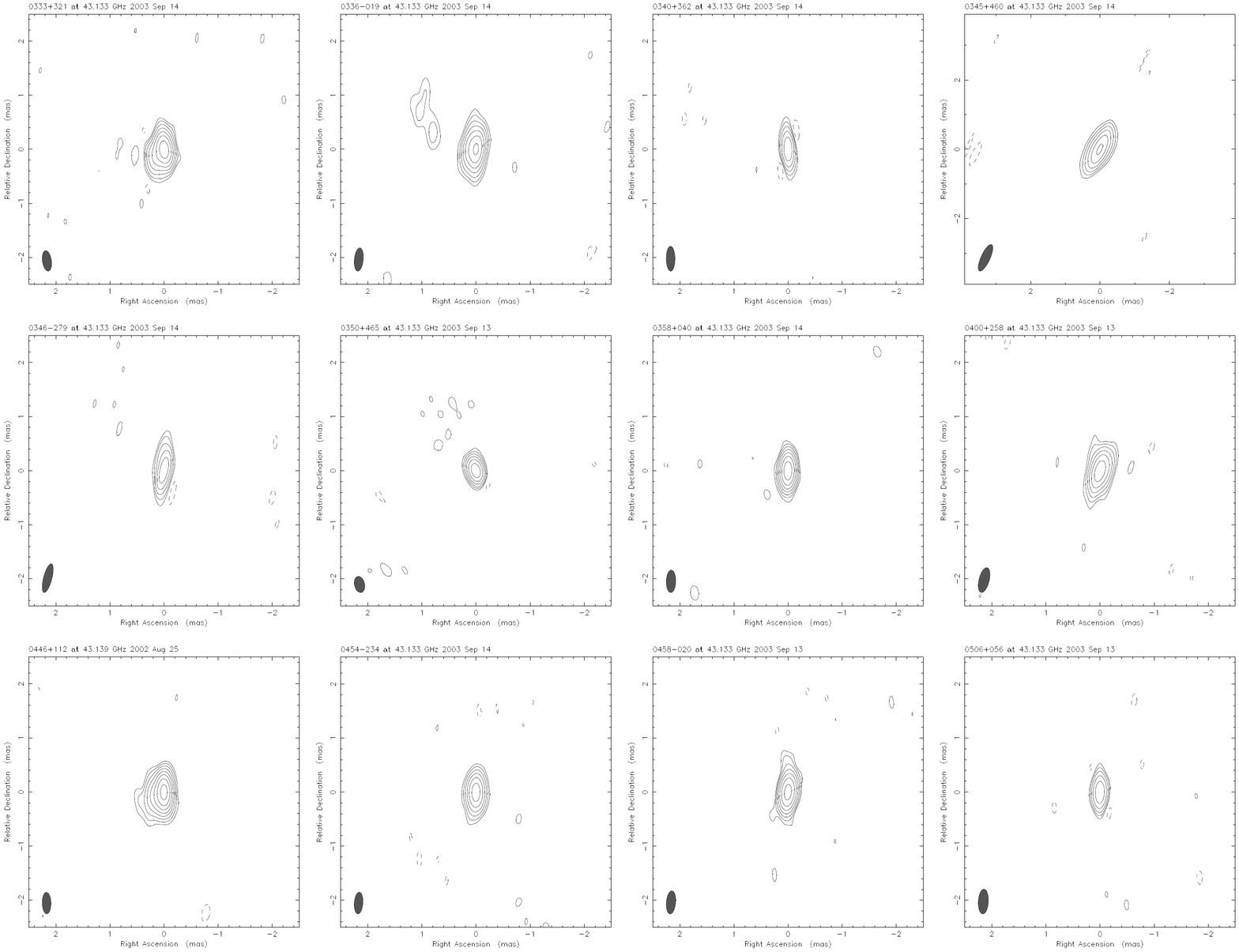}
\newline\newline Fig. 2.--- {\it Continued}
\end{sidewaysfigure}
\clearpage
\begin{sidewaysfigure}[phbt]
\centering
\includegraphics[height=8.0in,angle=90]{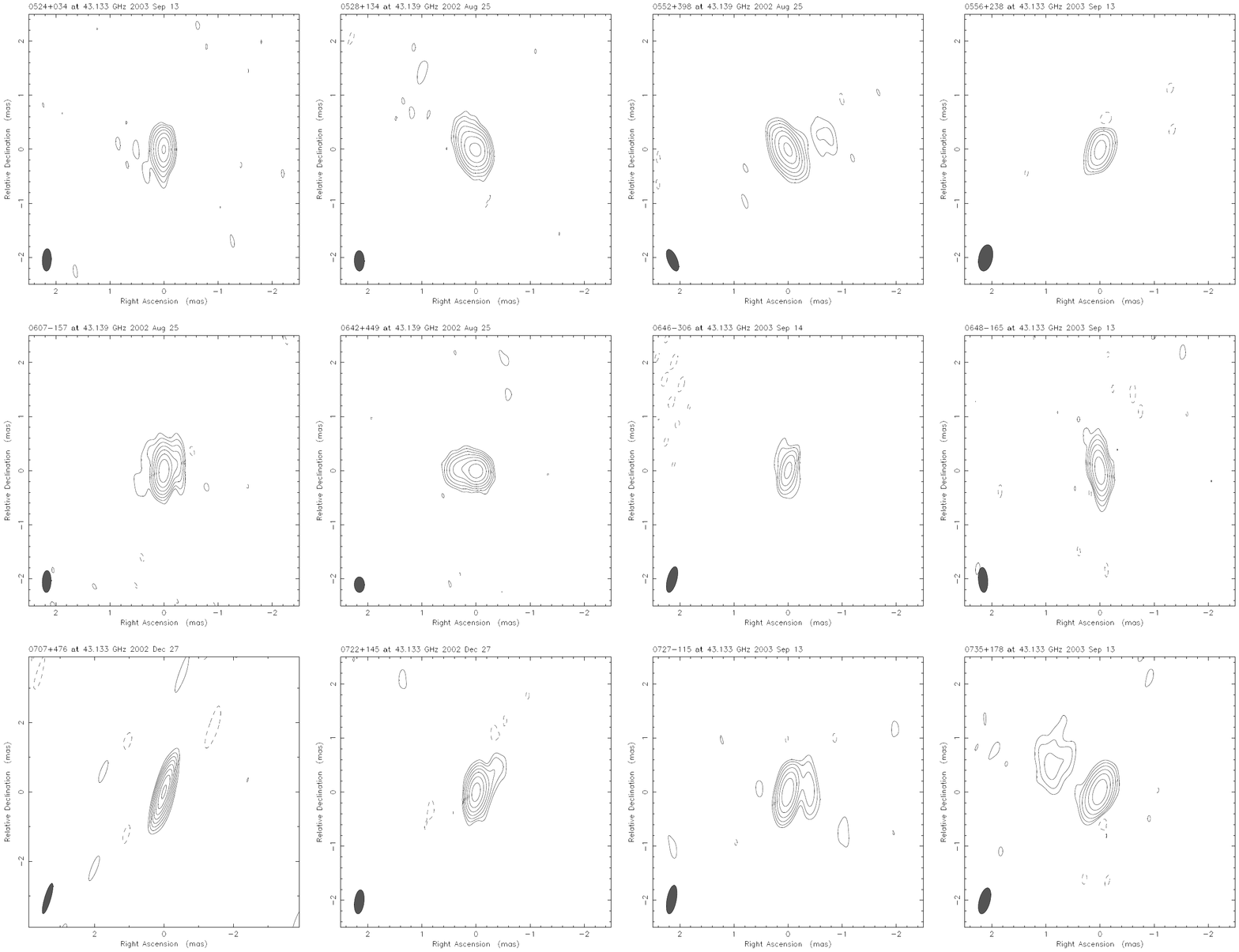}
\newline\newline Fig. 2.--- {\it Continued}
\end{sidewaysfigure}
\clearpage
\begin{sidewaysfigure}[phbt]
\centering
\includegraphics[height=8.0in,angle=90]{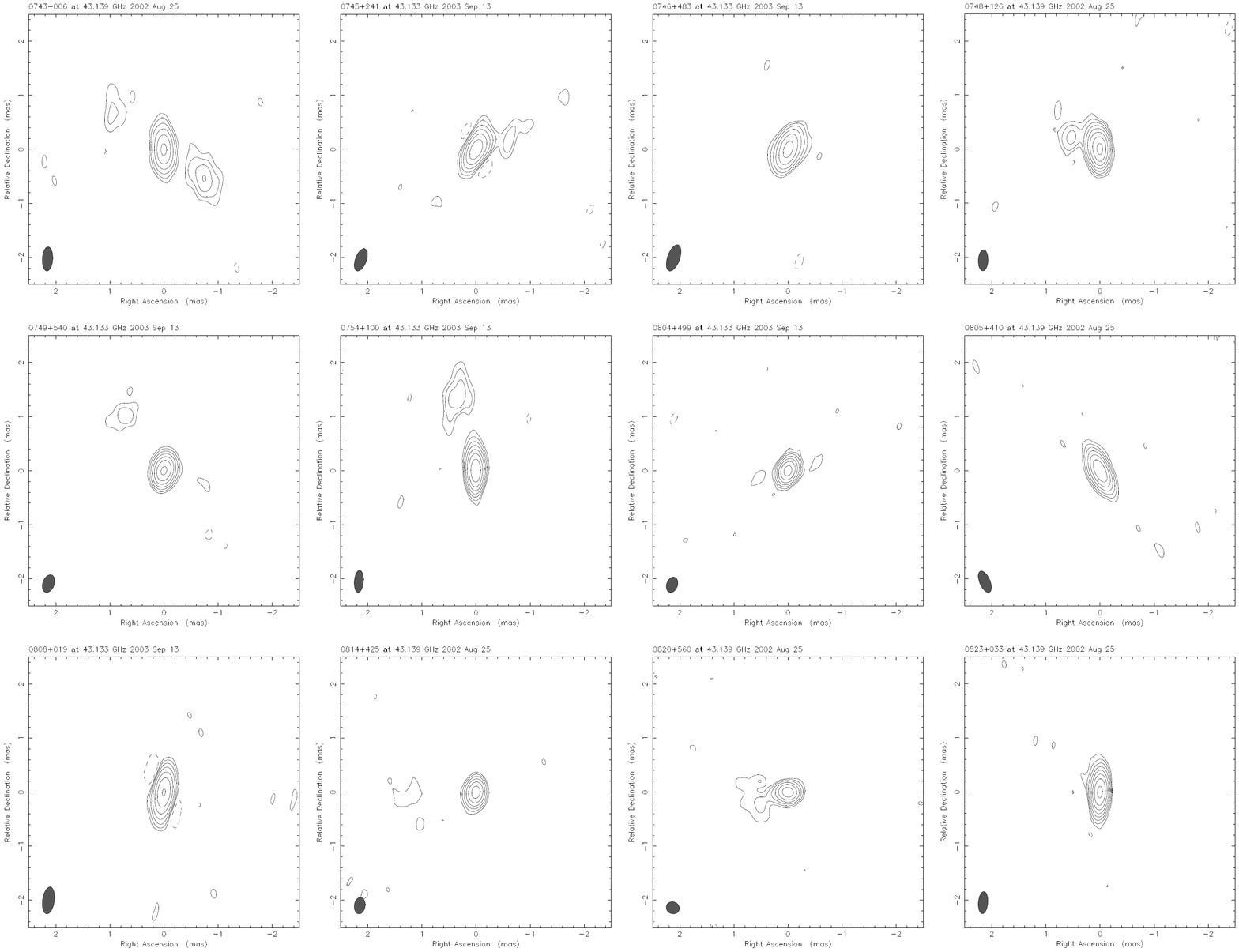}
\newline\newline Fig. 2.--- {\it Continued}
\end{sidewaysfigure}
\clearpage
\begin{sidewaysfigure}[phbt]
\centering
\includegraphics[height=8.0in,angle=90]{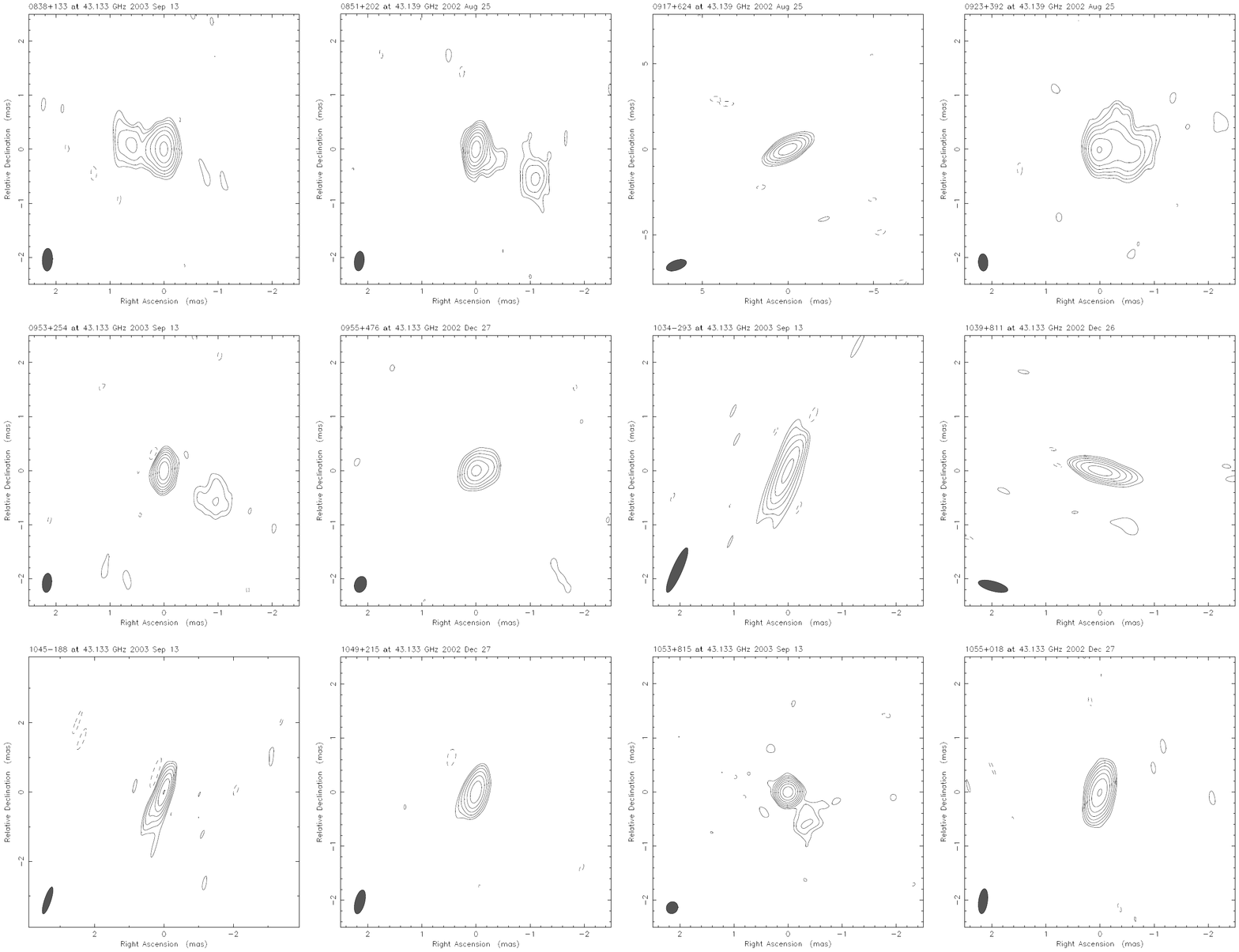}
\newline\newline Fig. 2.--- {\it Continued}
\end{sidewaysfigure}
\clearpage
\begin{sidewaysfigure}[phbt]
\centering
\includegraphics[height=8.0in,angle=90]{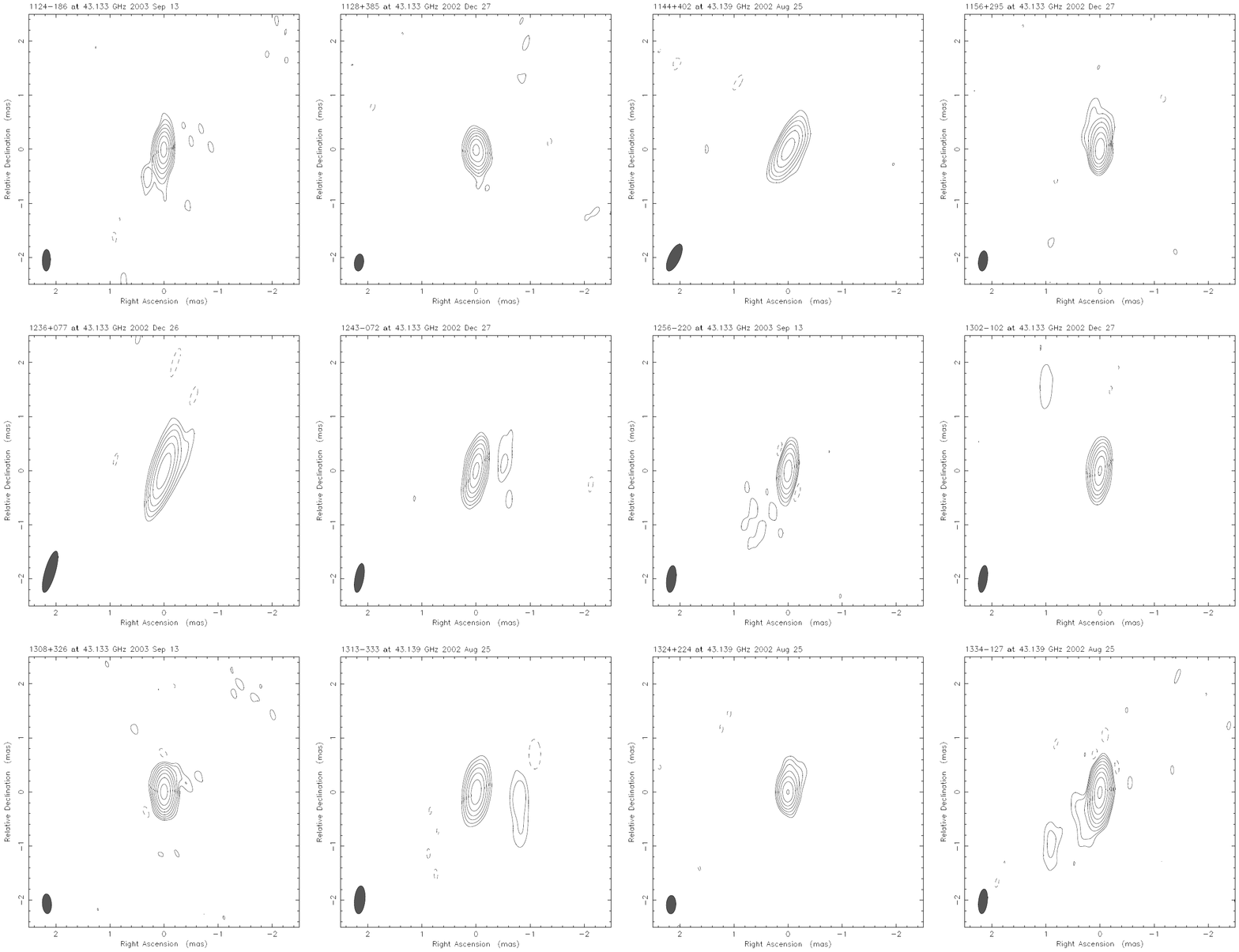}
\newline\newline Fig. 2.--- {\it Continued}
\end{sidewaysfigure}
\clearpage
\begin{sidewaysfigure}[phbt]
\centering
\includegraphics[height=8.0in,angle=90]{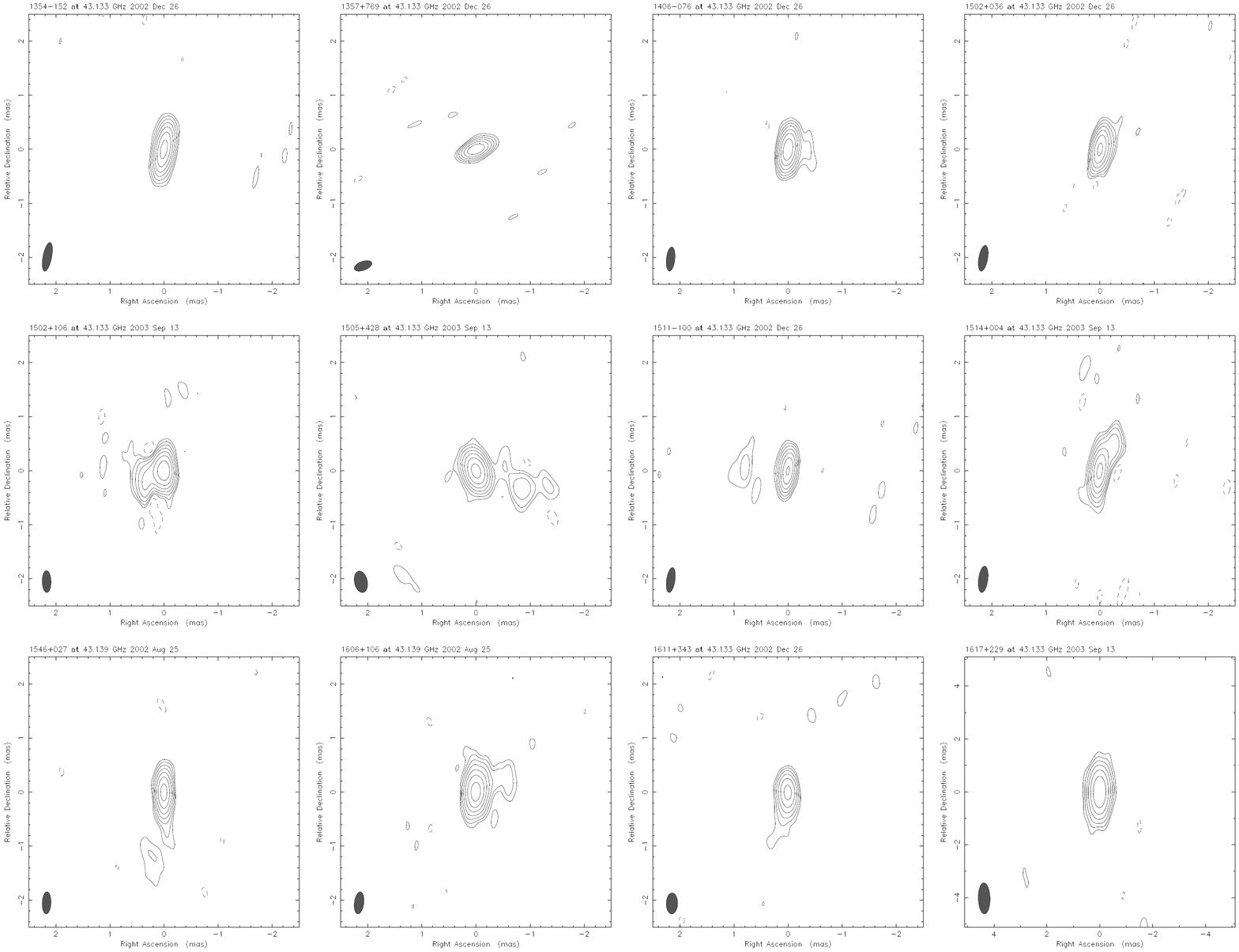}
\newline\newline Fig. 2.--- {\it Continued}
\end{sidewaysfigure}
\clearpage
\begin{sidewaysfigure}[phbt]
\centering
\includegraphics[height=8.0in,angle=90]{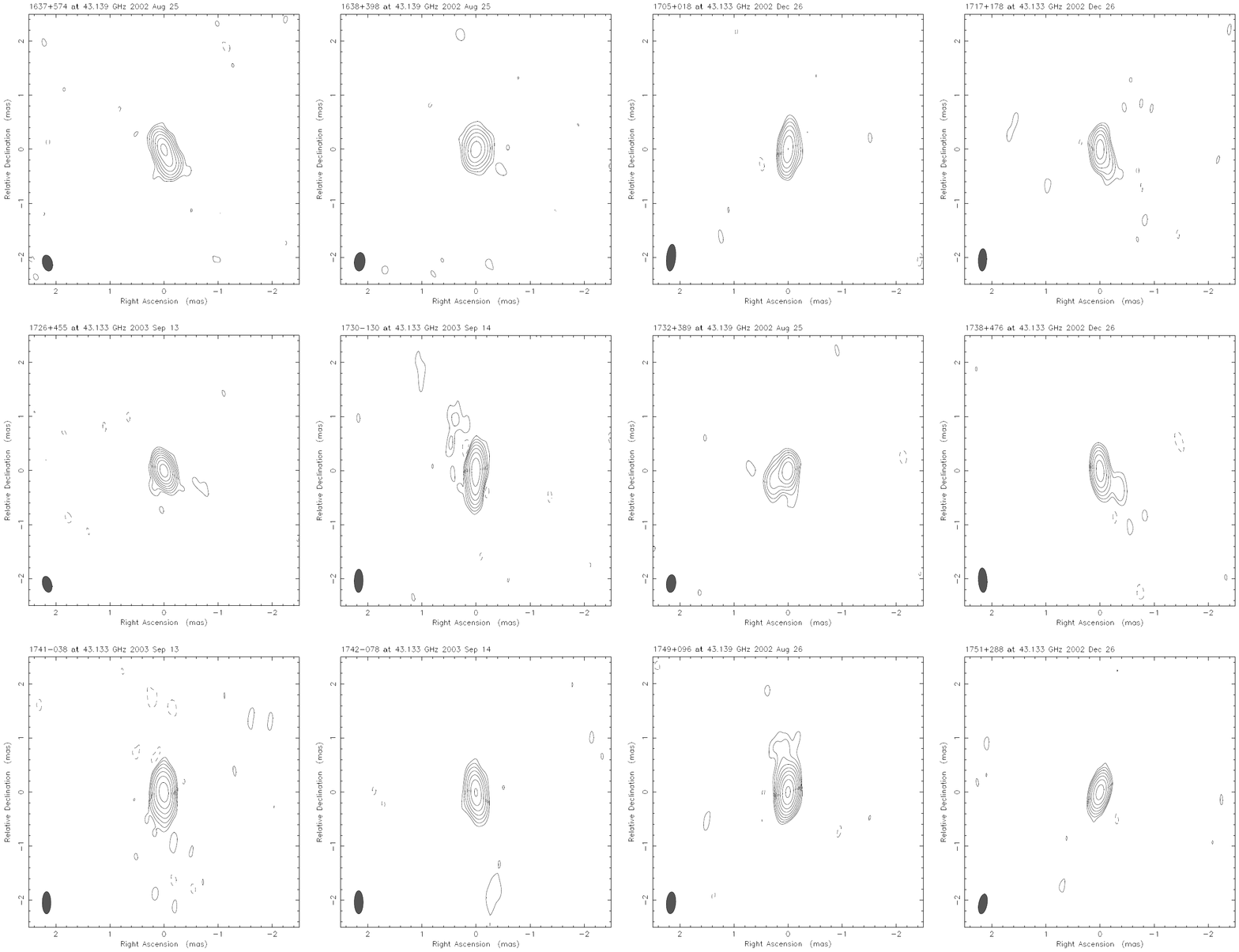}
\newline\newline Fig. 2.--- {\it Continued}
\end{sidewaysfigure}
\clearpage
\begin{sidewaysfigure}[phbt]
\centering
\includegraphics[height=8.0in,angle=90]{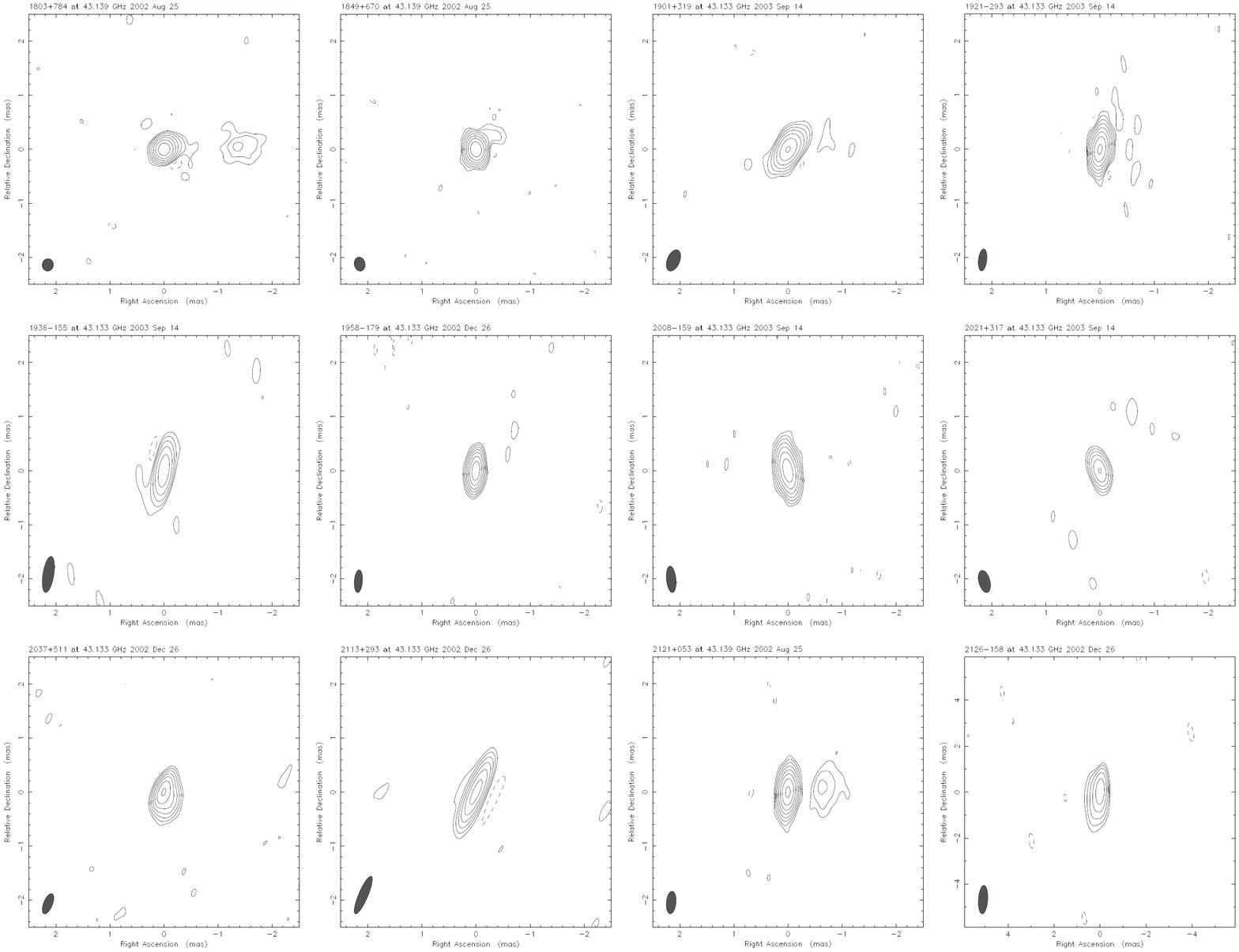}
\newline\newline Fig. 2.--- {\it Continued}
\end{sidewaysfigure}
\clearpage
\begin{sidewaysfigure}[phbt]
\centering
\includegraphics[height=8.0in,angle=90]{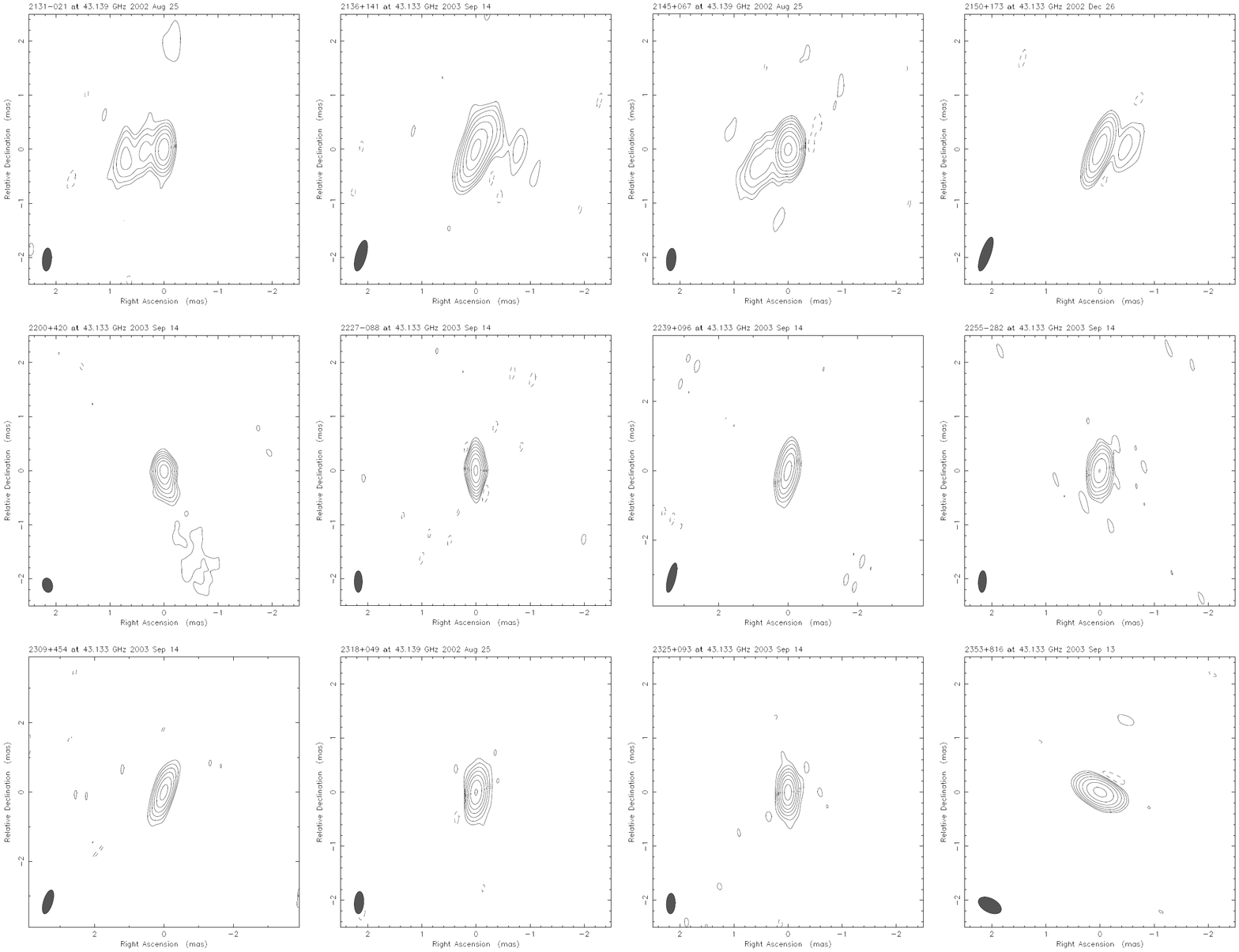}
\newline\newline Fig. 2.--- {\it Continued}
\end{sidewaysfigure}

\begin{figure}[htb!]
\plotone{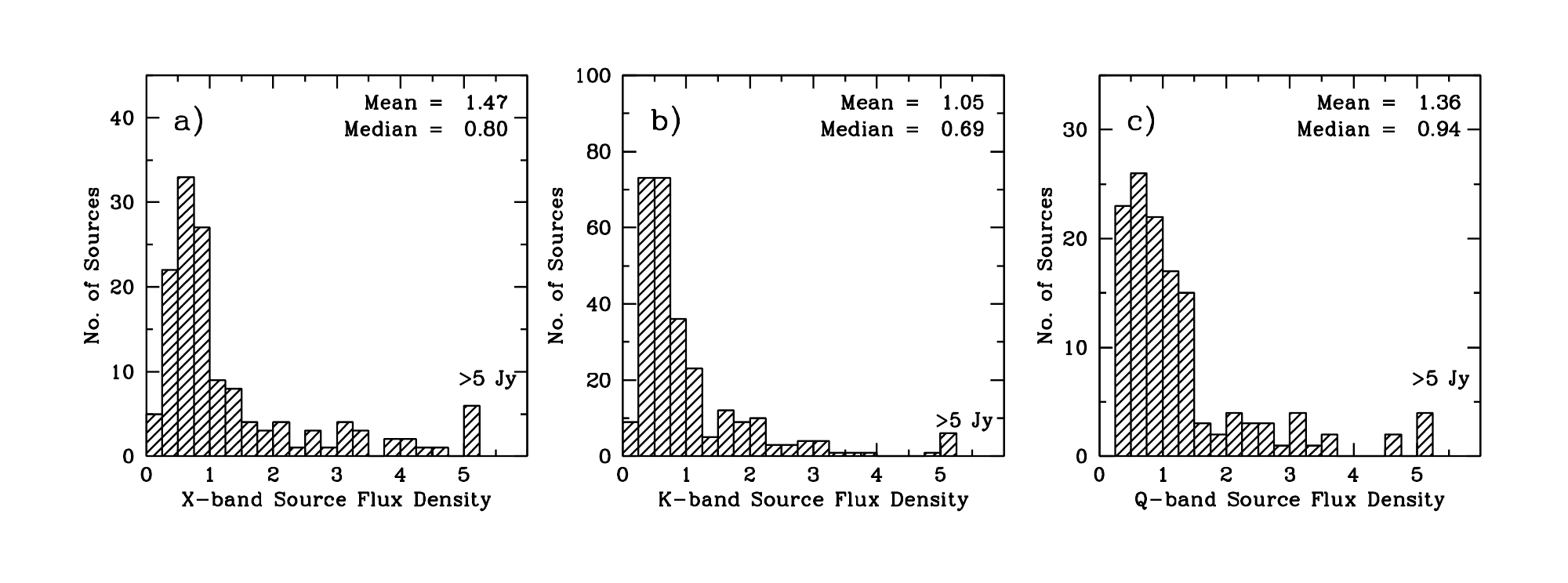}
\figcaption{Distributions of the mean (averaged over all sessions) source flux
density ($\bar S$) at a) X band, b) K band, and b) Q band. 
There were a total of 138, 274, and 132 sources, respectively, for which 
the flux density was measured at each of the three bands in our
VLBA data set.  
\label{FIG:XKQ_ALL_FLUX}}
\end{figure}

\begin{figure}[htb!]
\plotone{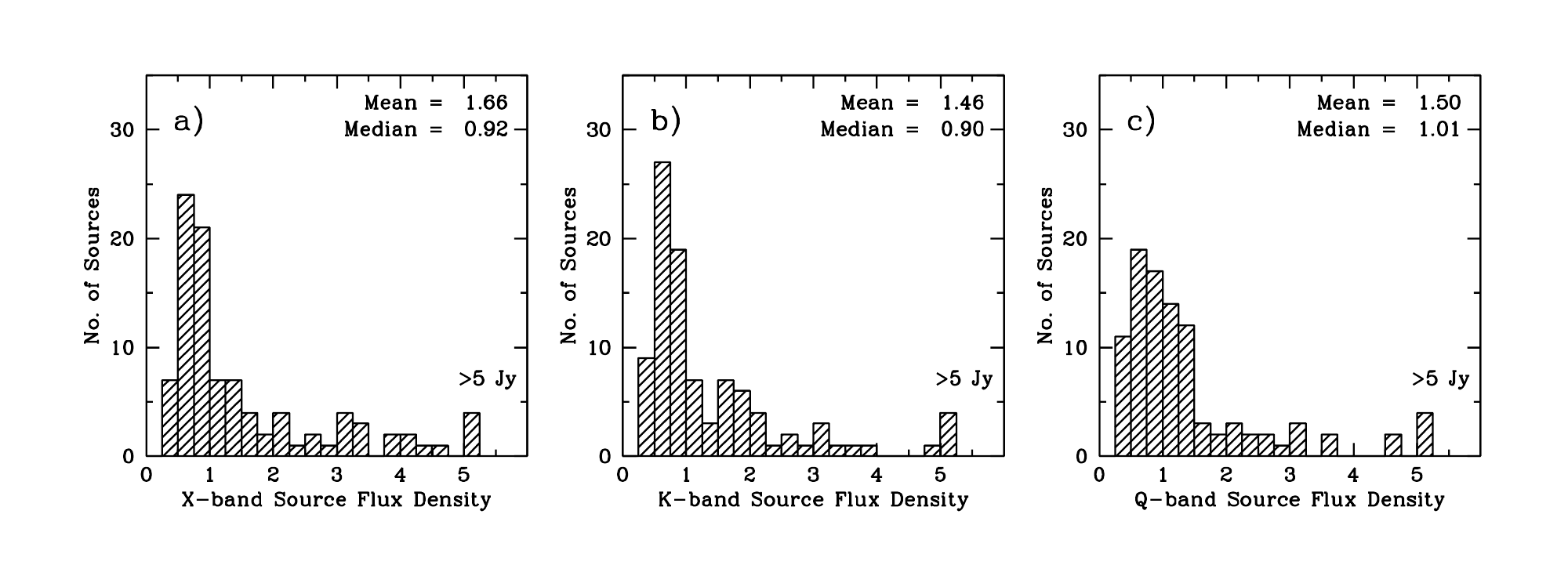}
\figcaption{Distributions of the mean (averaged over all sessions) source flux
density ($\bar S$) at a) X band, b) K band, and b) Q band for the 97 
sources common to all three frequencies within our VLBA data set. 
\label{FIG:XKQ_97COMMON_FLUX}}
\end{figure}

\begin{figure}[htb!]
\epsscale{0.9}
\plotone{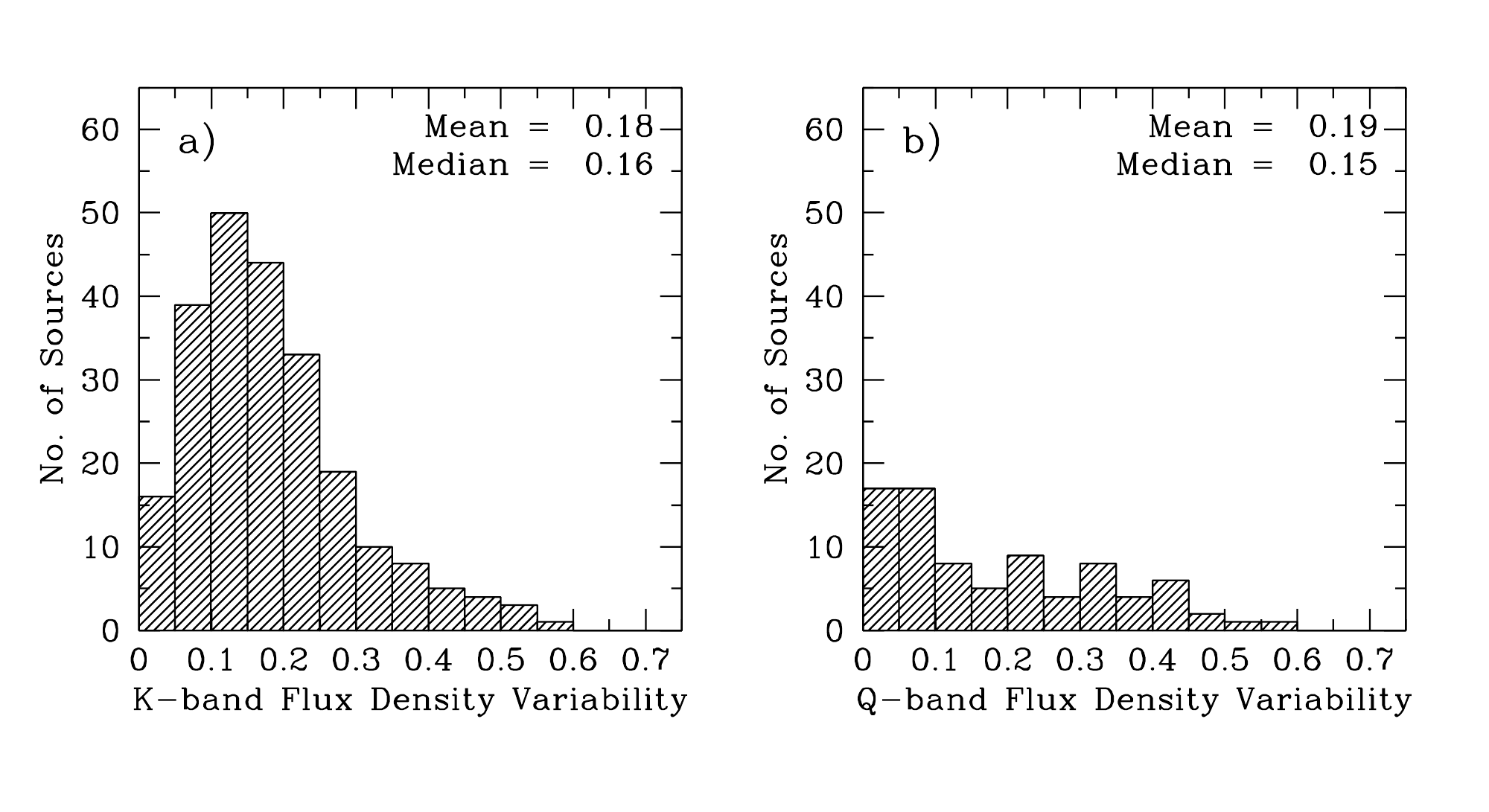}
\figcaption{Distributions of the source flux density variability index
($\sigma_S/\bar{S}$) at a) K band and b) Q band. There are a total of 
235 sources observed at K band in more than one session and a total
of 82 sources observed at Q band in more than one session. 
\label{FIG:KQ_FLUX_VAR}}
\end{figure}

\begin{figure}[htb!]
\epsscale{0.9}
\plotone{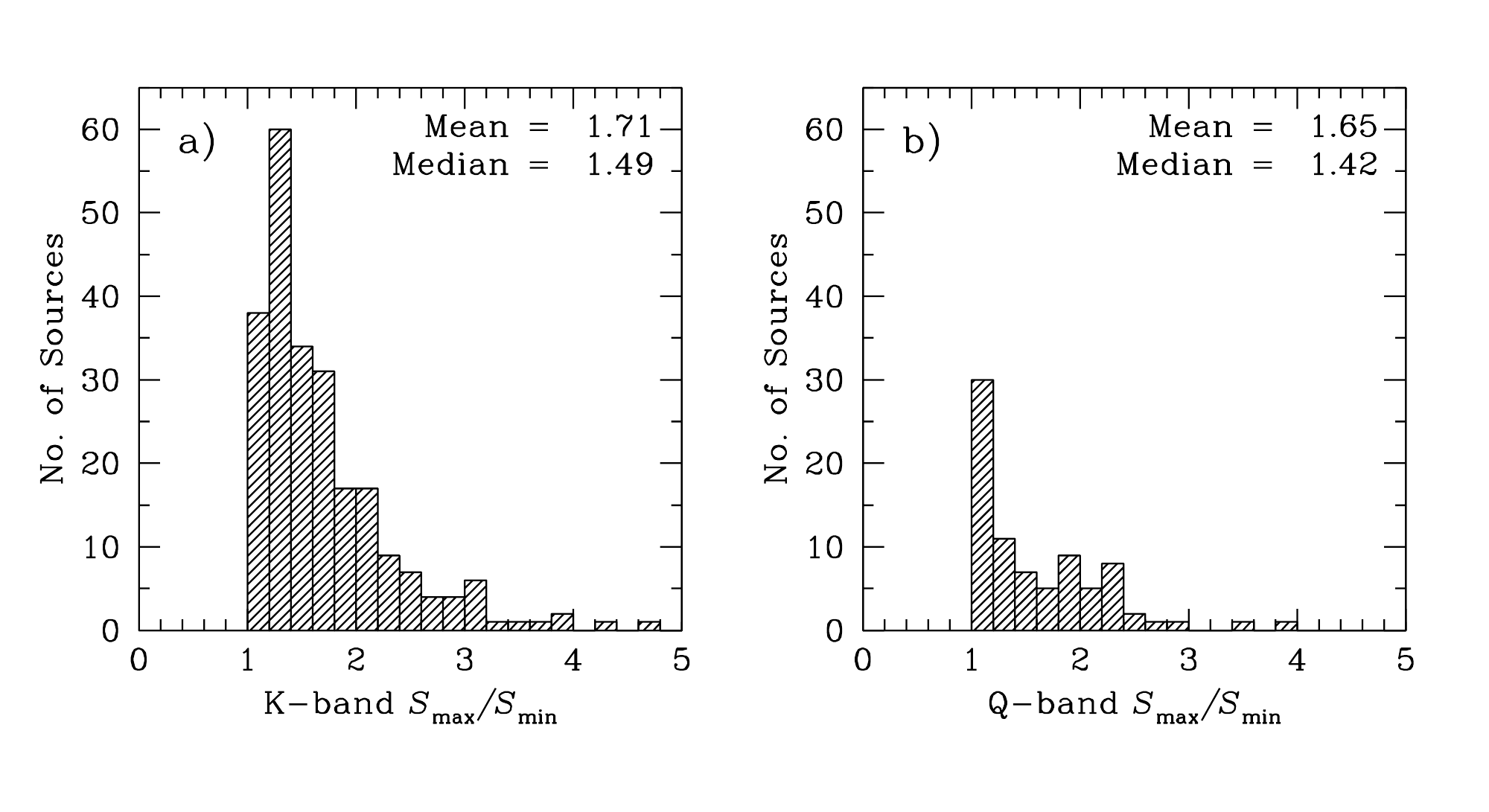}
\figcaption{Distributions of the ratio of the maximum to minimum source flux 
density ($S_{\rm max}/S_{\rm min}$) at a) K band and b) Q band. There are a 
total of 235 sources observed at K band in more than one session and a total
of 82 sources observed at Q band in more than one session. 
\label{FIG:KQ_SMAX_SMIN}}
\end{figure}

\begin{figure}[hbt]
\epsscale{0.9}
\plotone{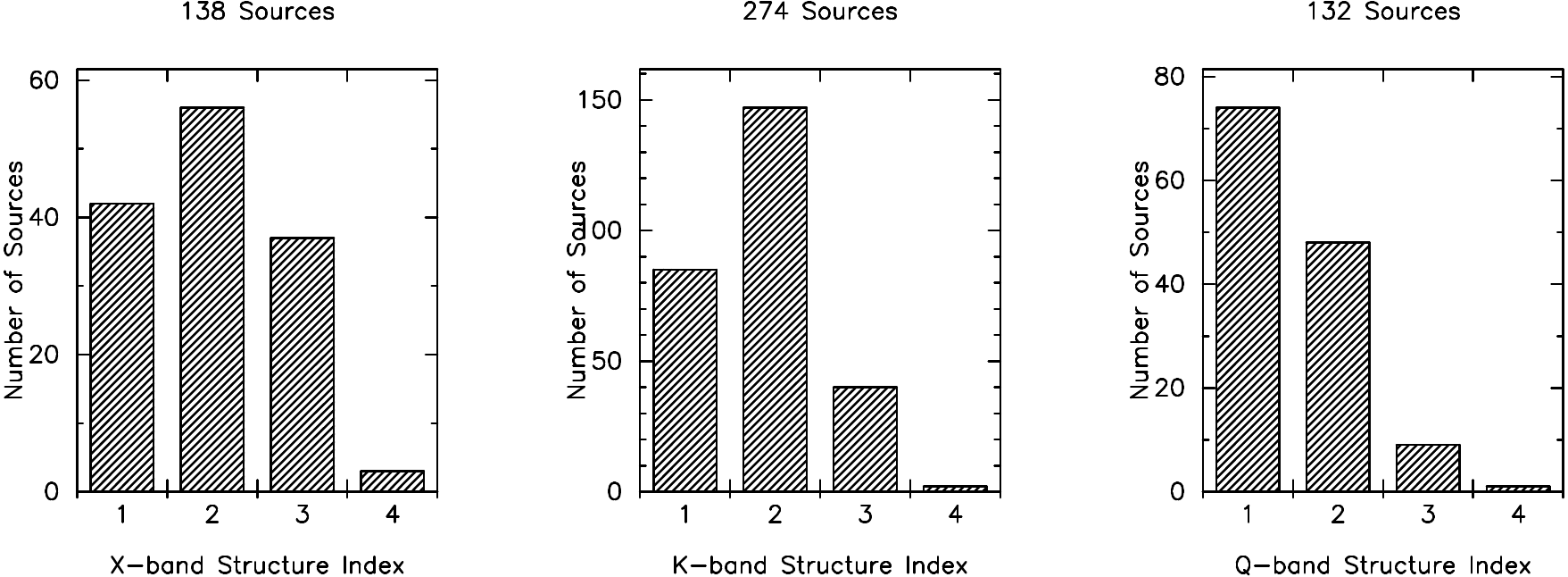}
\figcaption{Distributions of values for the maximum source structure index 
 ($SI$) at (a) X band, (b) K band and (c) Q band.  There were a total of 138, 274, and 
132 sources, respectively, for which the flux density was measured at each 
of the three bands in our VLBA data set.  
\label{FIG:SI_HIST_ALL}}
\end{figure}

\begin{figure}[hbt]
\epsscale{0.9}
\plotone{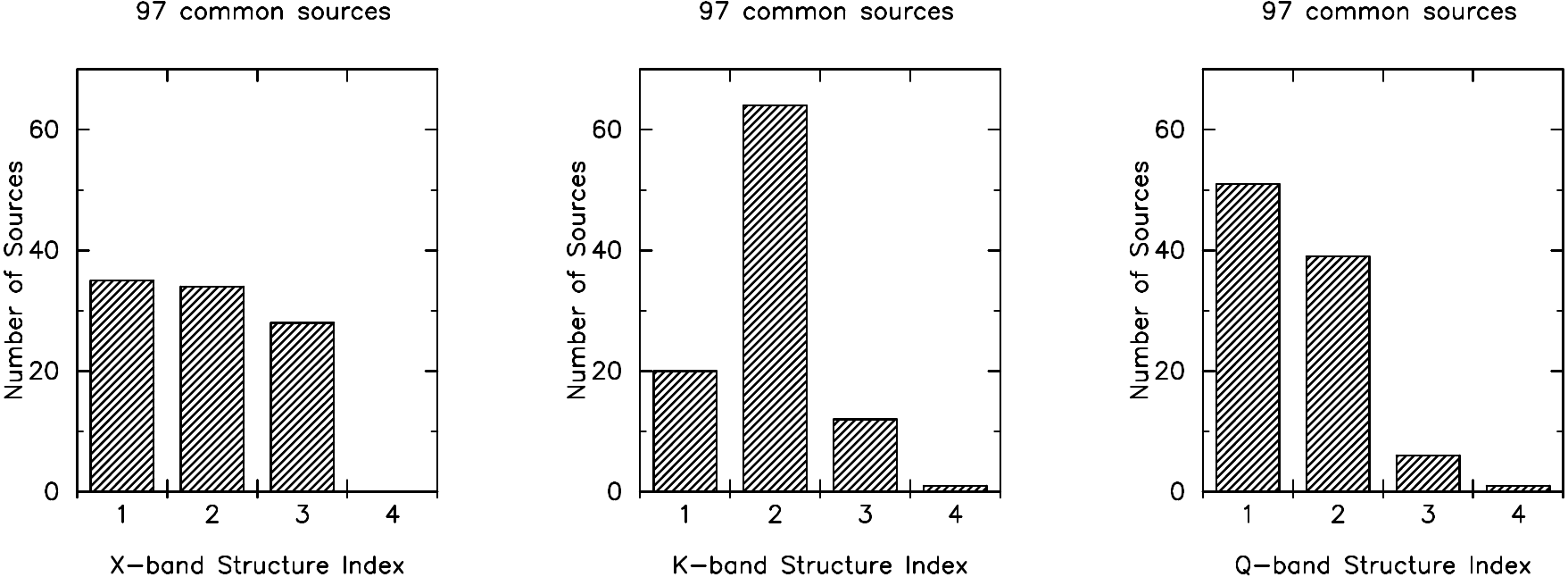}
\figcaption{Distributions of values for the maximum source structure 
index ($SI$) at (a) X band, (b) K band and (c) Q band for the 97 sources common 
to all three frequencies within our VLBA data set. \label{FIG:SI_HIST_COMMON}}
\end{figure}

\begin{figure}[htb!]
\epsscale{1.0}
\plotone{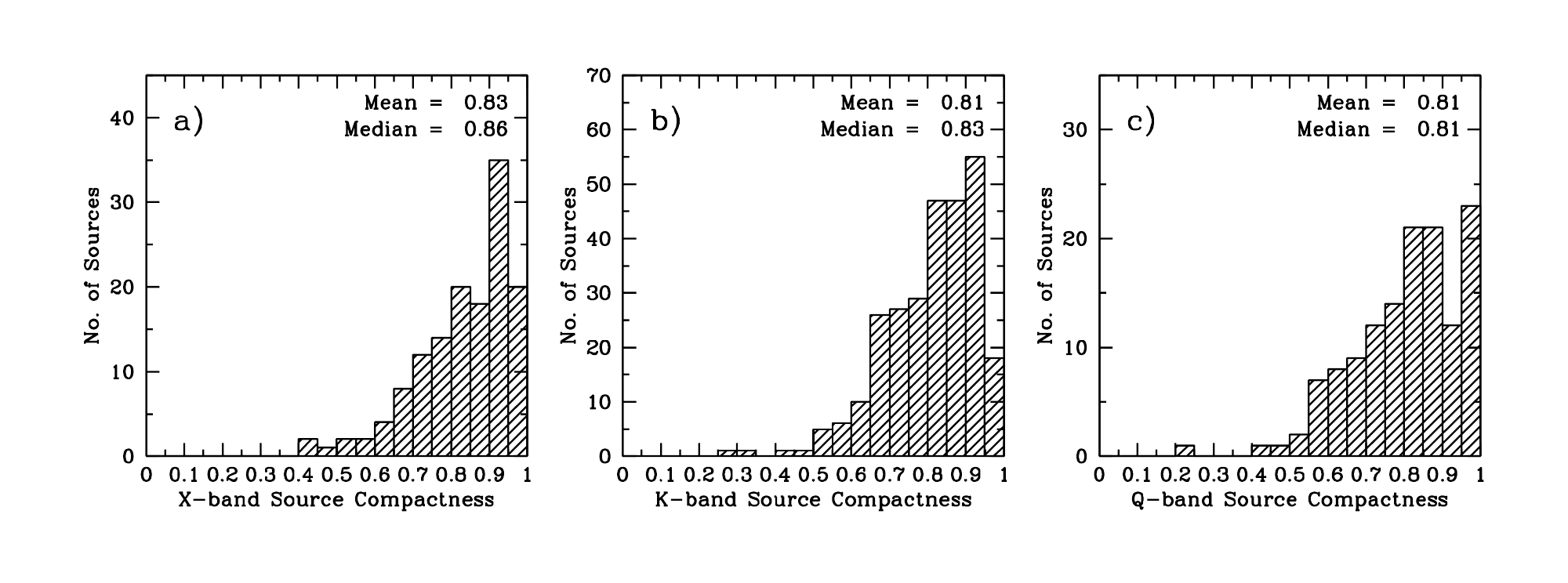}
\figcaption{Distributions of the mean (averaged over all sessions)
source compactness ($\bar C$) at (a) X band, (b) K band, and 
(c) Q band.  There were a total of 138, 274, and 132 sources for 
which the compactness was determined at each of the three bands, 
respectively.  A maximum compactness, $C=1.0$, indicates that all 
of the flux is contained within one synthesized beam.
\label{FIG:XKQ_ALL_COMPACT}}
\end{figure}

\begin{figure}[htb!]
\epsscale{1.0}
\plotone{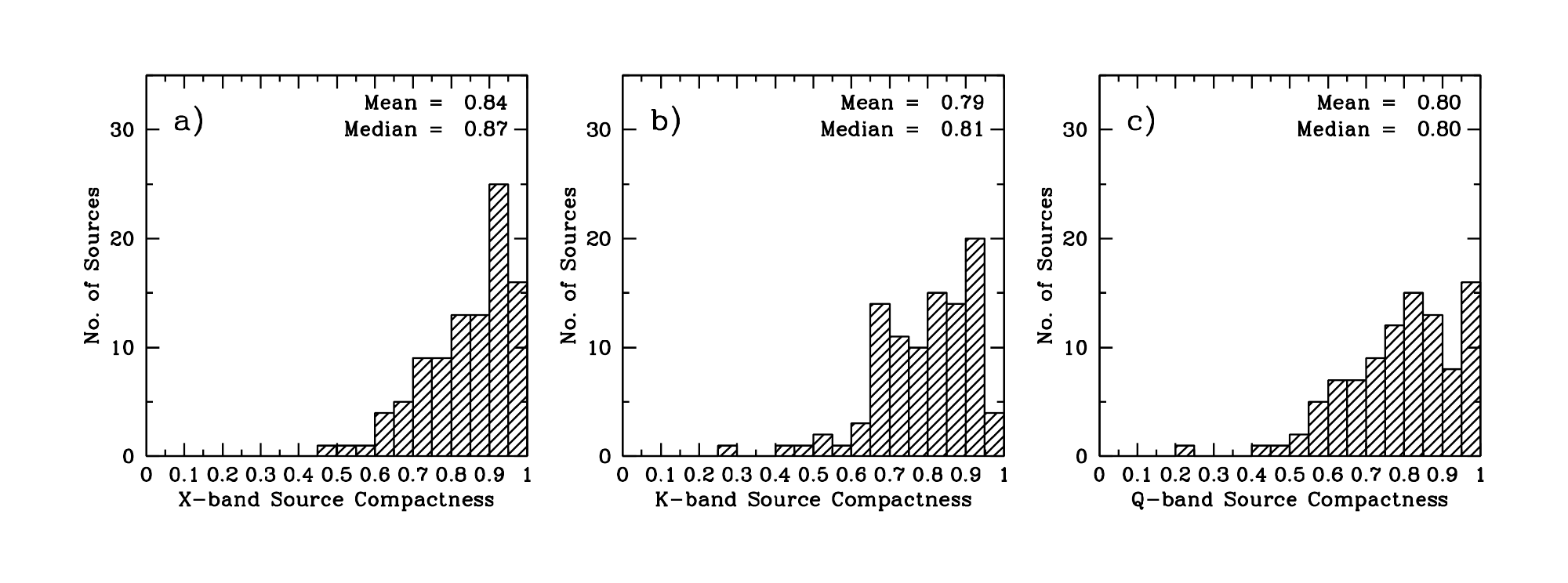}
\figcaption{Distributions of the mean (averaged over all sessions)
source compactness ($\bar C$) at (a) X band, (b) K band, and 
(c) Q band for the 97 sources common to all three frequencies within 
our VLBA data set.  A maximum compactness, $C=1.0$, indicates that all 
of the flux is contained within one synthesized beam.
\label{FIG:XKQ_97COMMON_COMPACT}}
\end{figure}

\begin{figure}[htb!]
\epsscale{1.0}
\plotone{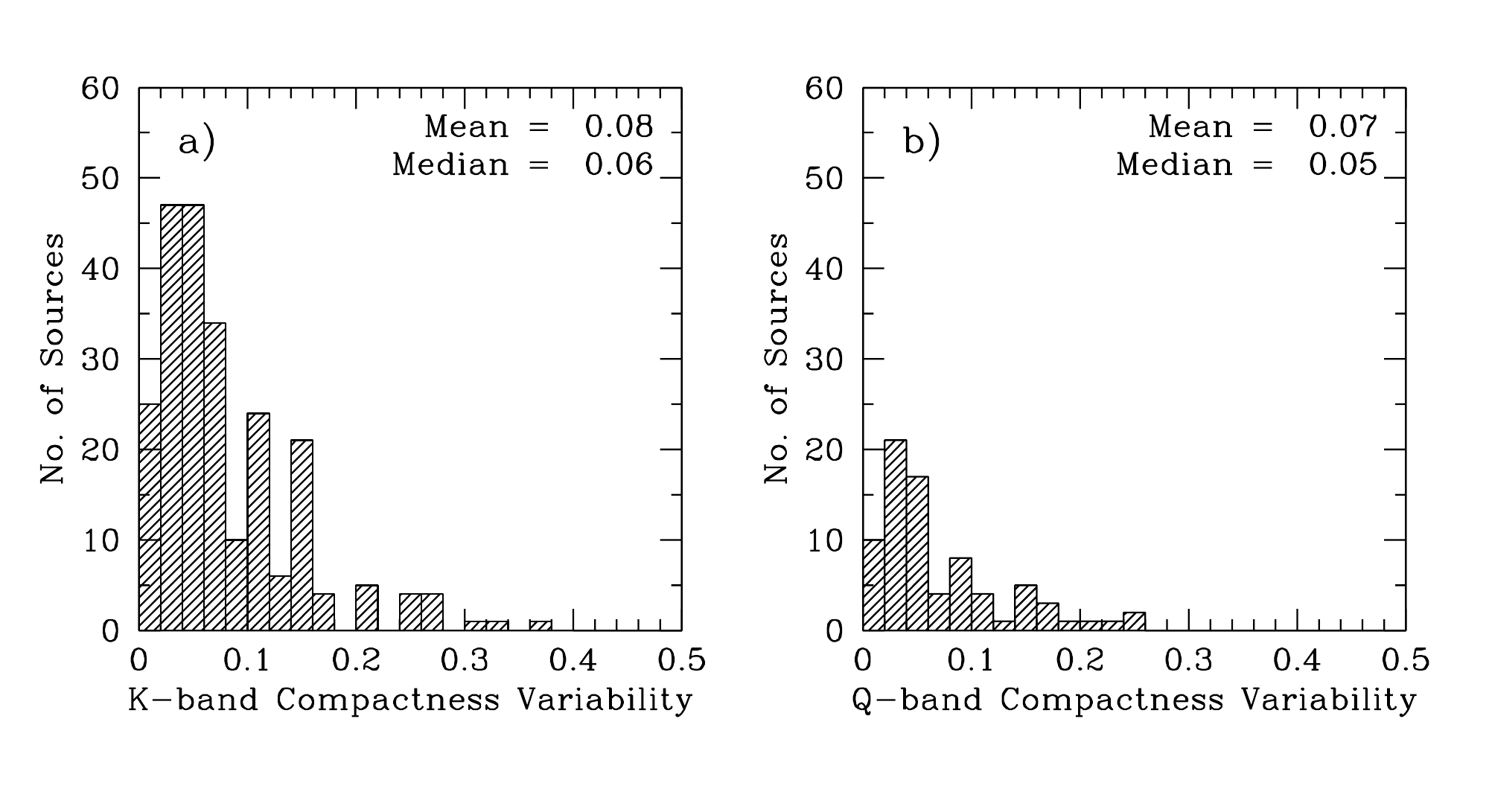}
\figcaption{Distributions of the source
compactness variability index ($\sigma_C/\bar{C}$) for sources 
at (a) K band and (b) Q band.  There are a total of 235 sources observed in
more than one session at K band, and a total of 82 sources observed in 
more than one session at Q band. A minimum variability, 
$\sigma_C/\bar{C}=0.0$, indicates no variation in the compactness over time.
\label{FIG:KQ_COMPACT_VAR}}
\end{figure}

\begin{figure}[hbt]
\epsscale{1.0}
\plotone{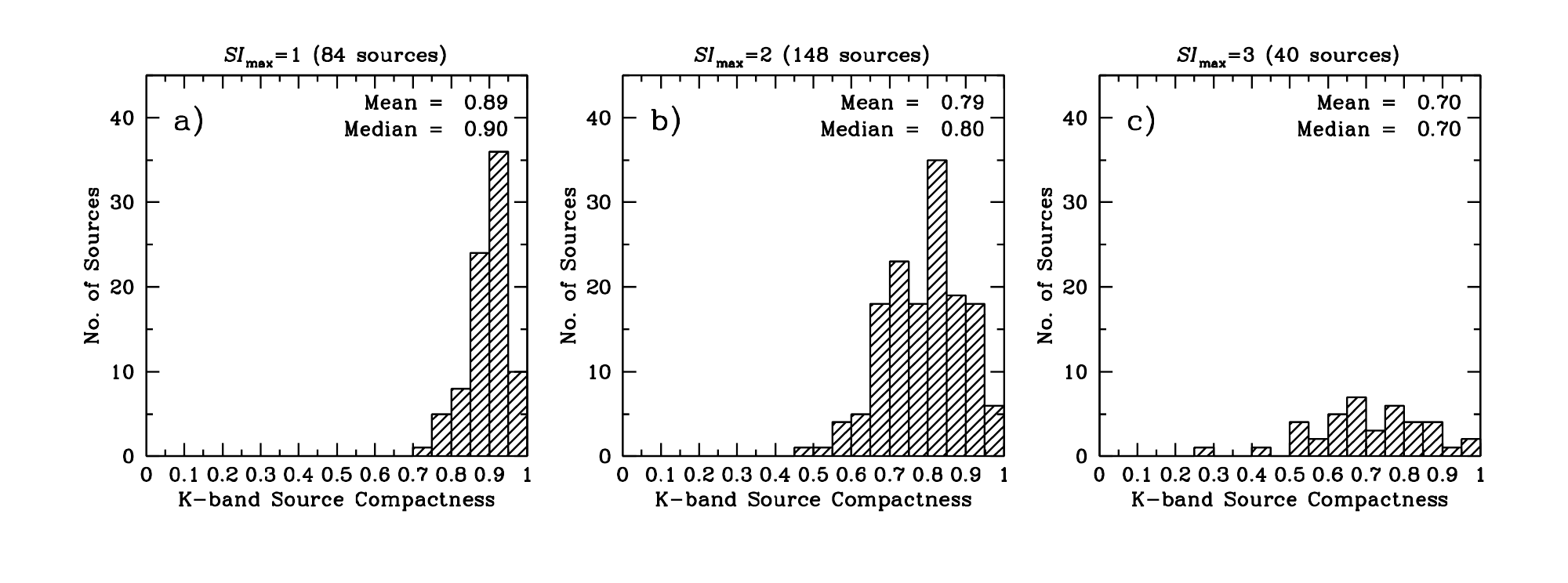}
\figcaption{Distributions of the mean (averaged over all sessions) 
source compactness ($\bar C$) separated by maximum structure 
index ($SI_{\rm max}$) for the 274 sources imaged at K band (24 GHz).  
The three panels represent sources with (a) $SI_{\rm max}=1$, 
(b) $SI_{\rm max}=2$, and (c) $SI_{\rm max}=3$, respectively.  
There are two sources with a $SI_{\rm max}=4$ that are not shown.
\label{FIG:K_COMPACT_SI}}
\end{figure}

\begin{figure}[hbt]
\epsscale{1.0}
\plotone{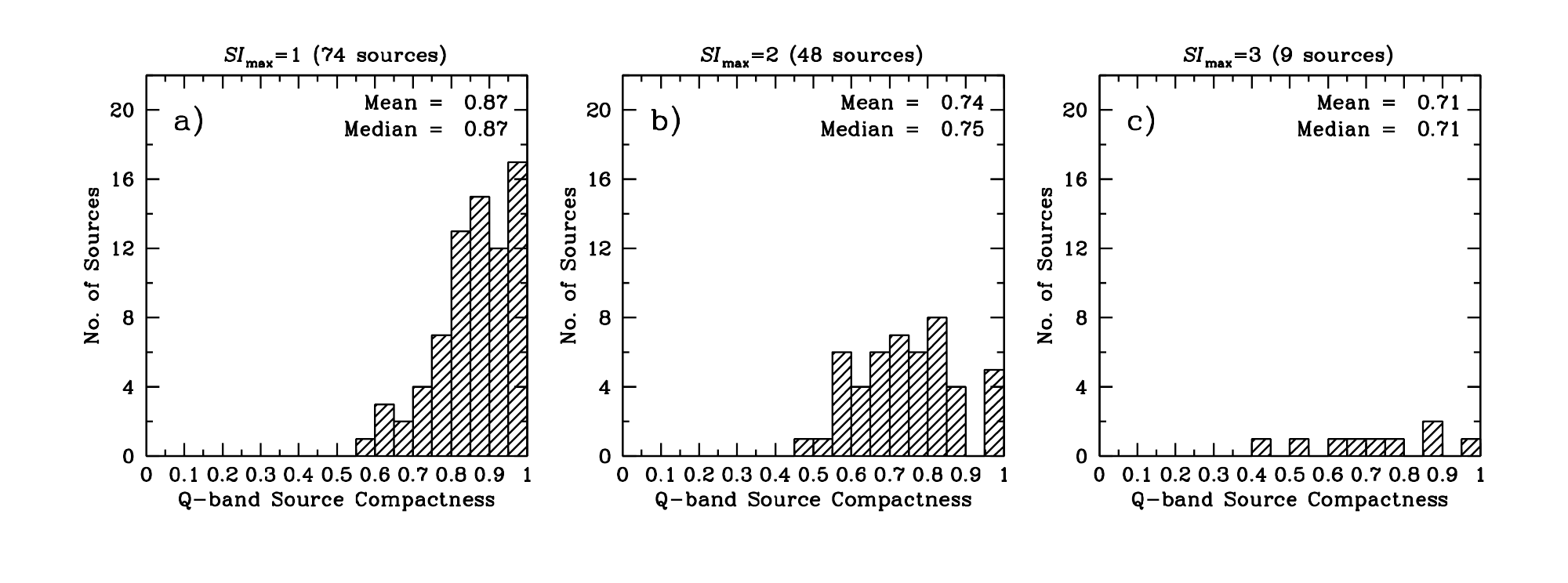}
\figcaption{Distributions of the mean (averaged over all sessions) 
source compactness ($\bar C$) separated by maximum structure 
index ($SI_{\rm max}$) for the 132 sources imaged at Q band (43 GHz).  
The three panels represent sources with (a) $SI_{\rm max}=1$, 
(b) $SI_{\rm max}=2$, and (c) $SI_{\rm max}=3$, respectively.  
There is one source with a $SI_{\rm max}=4$ that is not shown.
\label{FIG:Q_COMPACT_SI}}
\end{figure}

\begin{figure}[hbt]
\epsscale{1.0}
\plotone{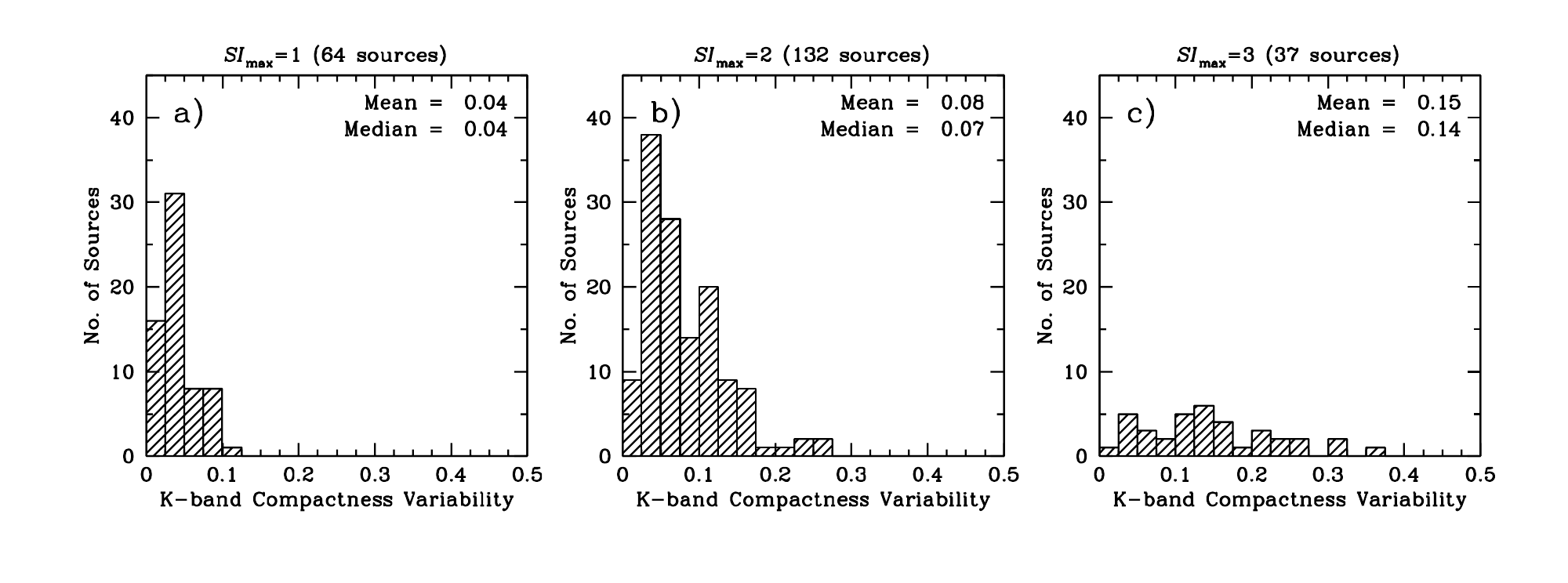}
\figcaption{Distributions of the compactness variability index 
($\sigma_C/\bar{C}$) separated by maximum structure 
index ($SI_{\rm max}$) for the 235 sources K-band sources 
imaged in more than one session.
The three panels represent sources with (a) $SI_{\rm max}=1$, 
(b) $SI_{\rm max}=2$, and (c) $SI_{\rm max}=3$, respectively.  
There are two sources with a $SI_{\rm max}=4$ that are not shown.
\label{FIG:K_COMPACT_VAR_SI}}
\end{figure}

\begin{figure}[hbt]
\epsscale{0.45}
\plotone{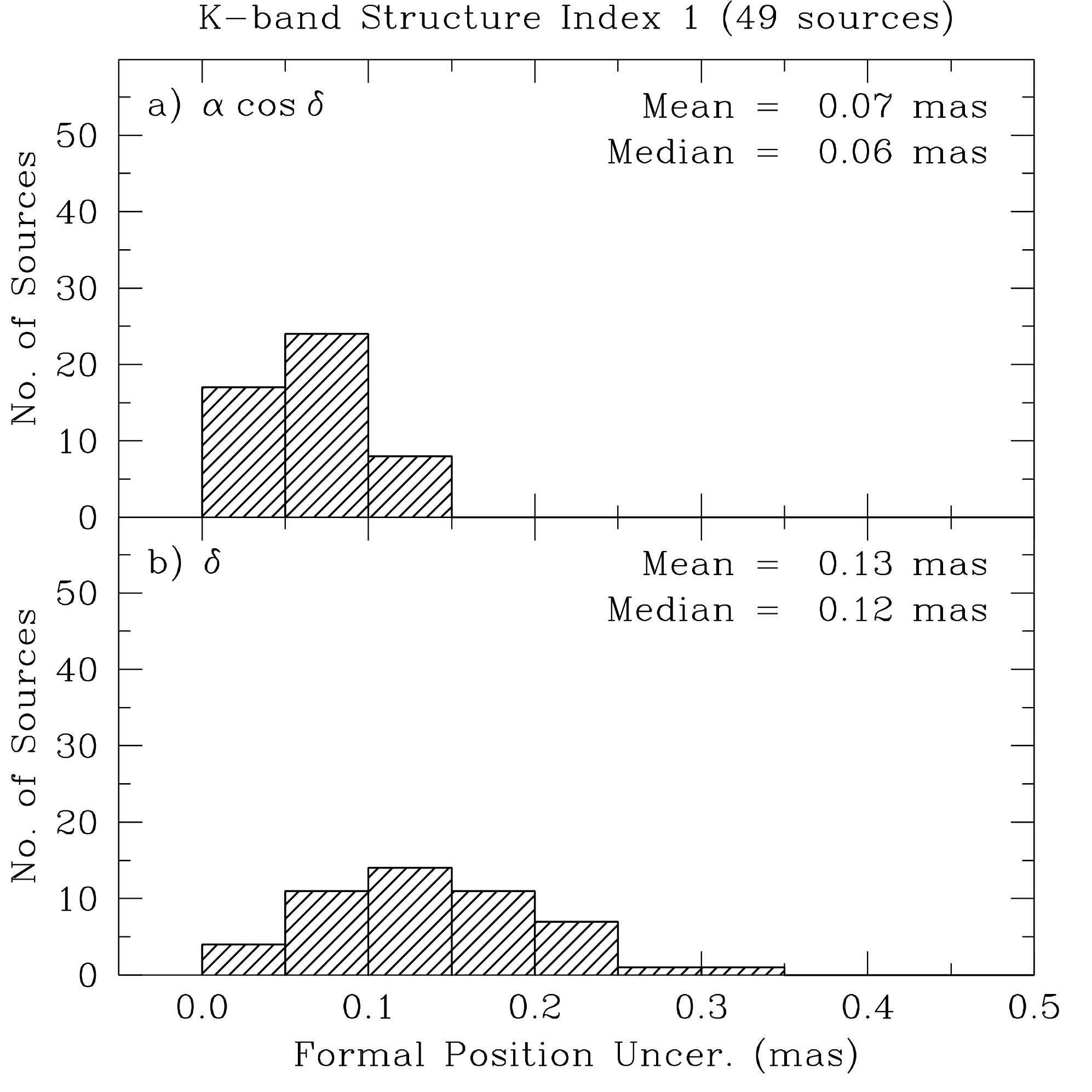}
\plotone{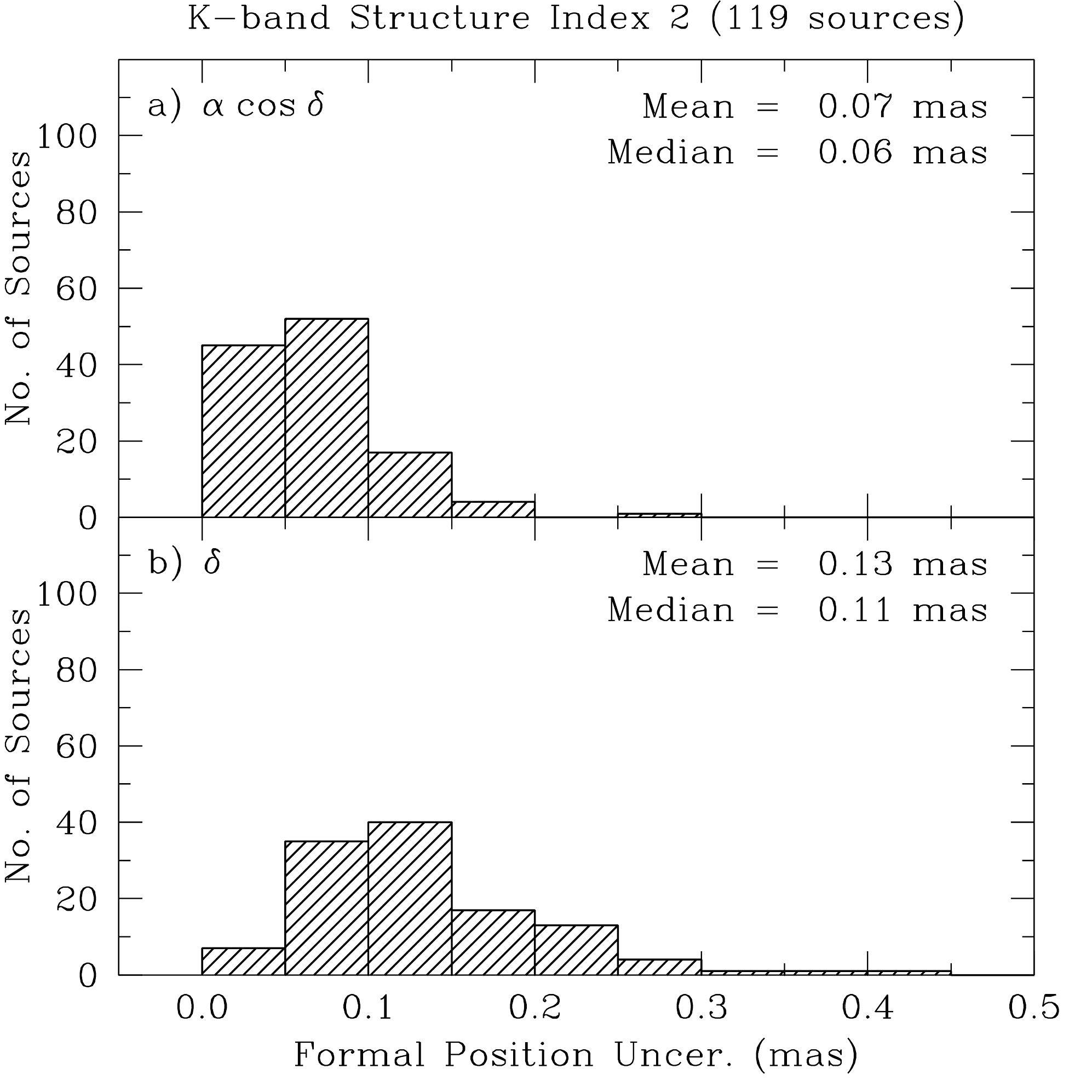}
\plotone{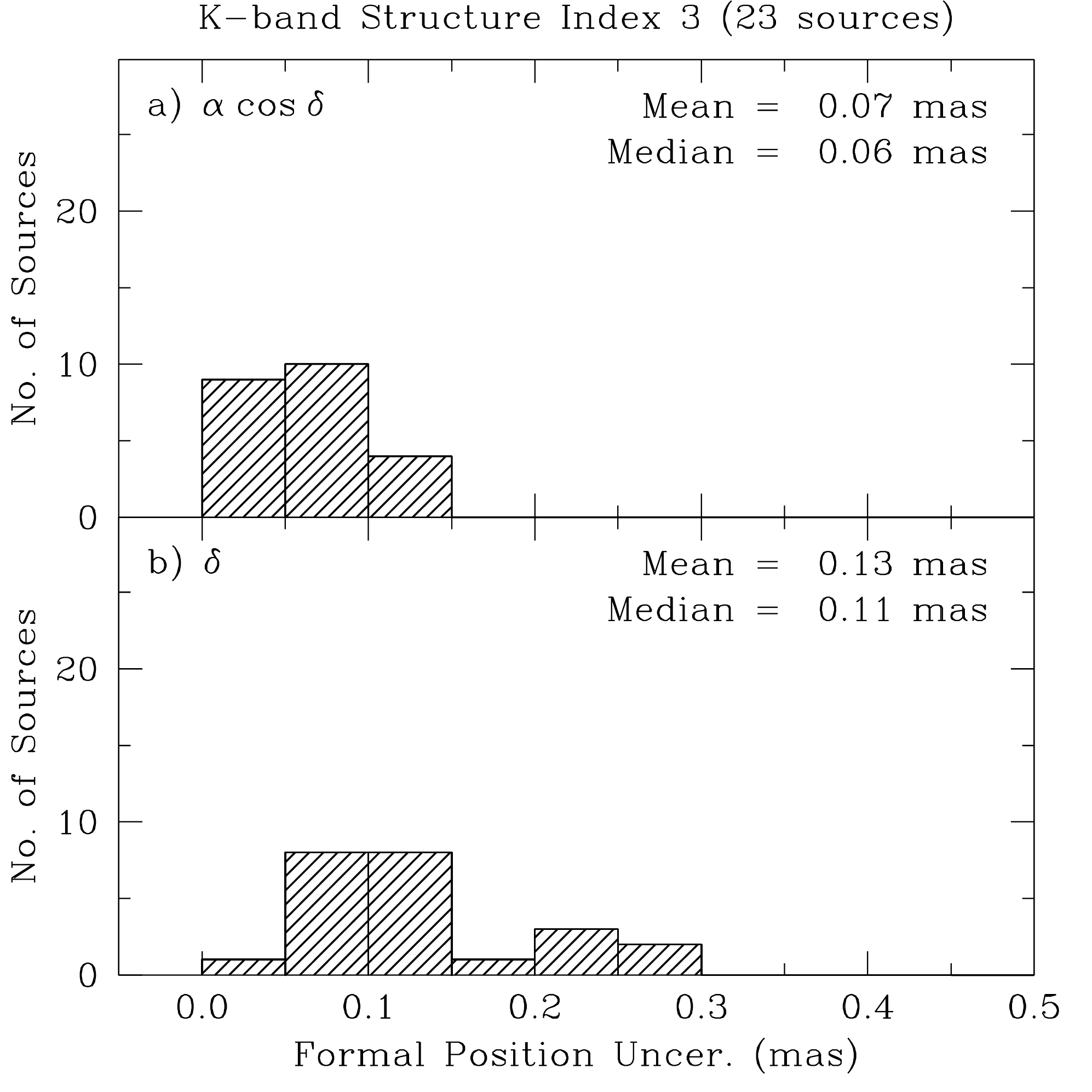}
\figcaption{Distributions of the formal position uncertainties in 
$\alpha \cos \delta$ (a) and in $\delta$ (b) for the 193 sources 
with 100 or more delay observations at K band (24 GHz).  The three 
panels represent sources with a maximum source structure index 
equal to 1, 2 and 3, respectively.  There are two sources with a 
structure index of 4 and with more that 100 delay observations
that are not included in the figure.
\label{FIG:POS_UNCER_HIST_100}}
\end{figure}

\begin{figure}[hbt]
\epsscale{0.5}
\plotone{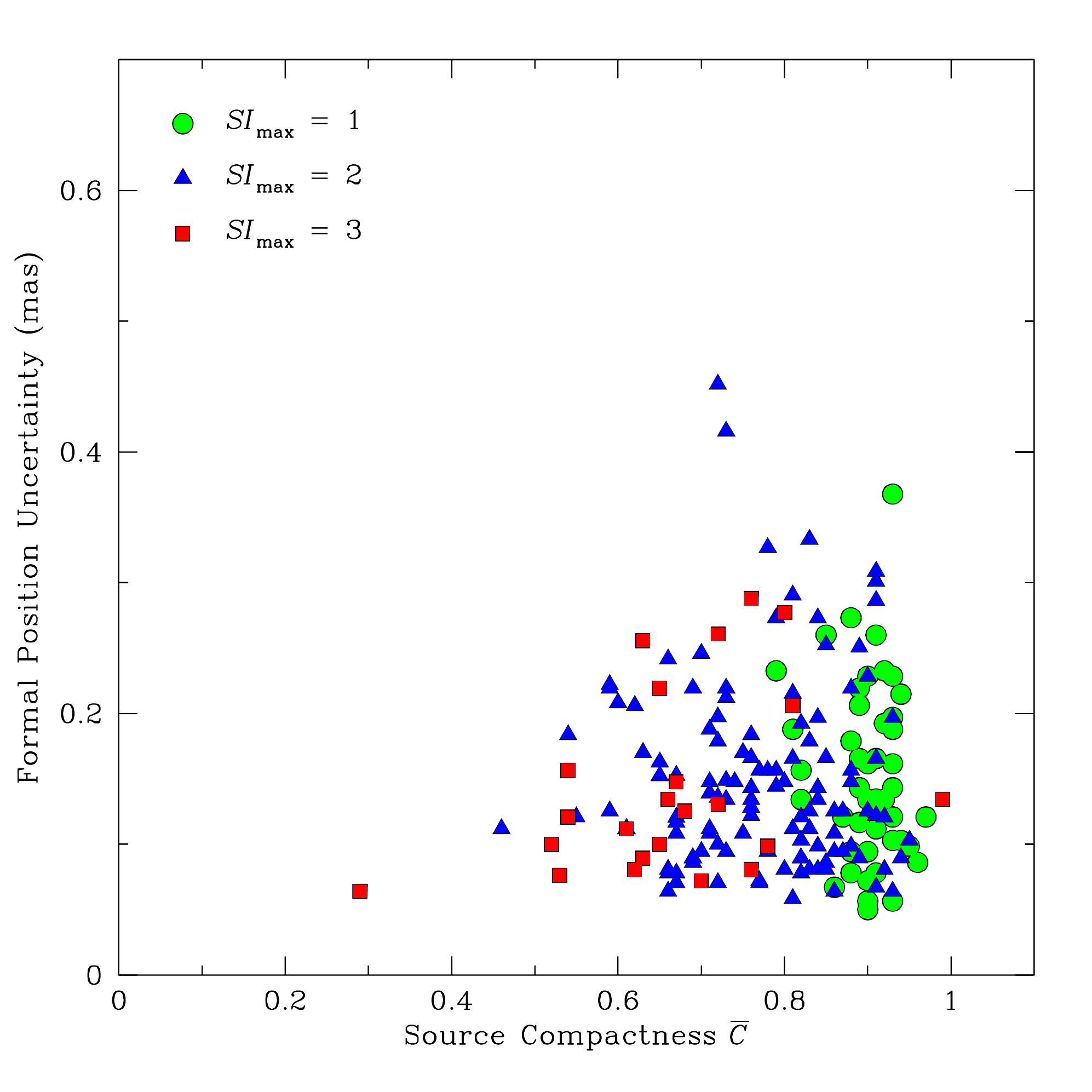}
\figcaption{Comparison of the formal astrometric position uncertainty 
versus the mean source compactness ($\bar C$) for K band sources with 
more than 100 delay measurements.  Point color and type represent the 
maximum structure index for each source over the sessions in which it
was observed.
\label{FIG:POS_COMPACT_SCATTER}}
\end{figure}

\begin{figure}[hbt]
\epsscale{0.5}
\plotone{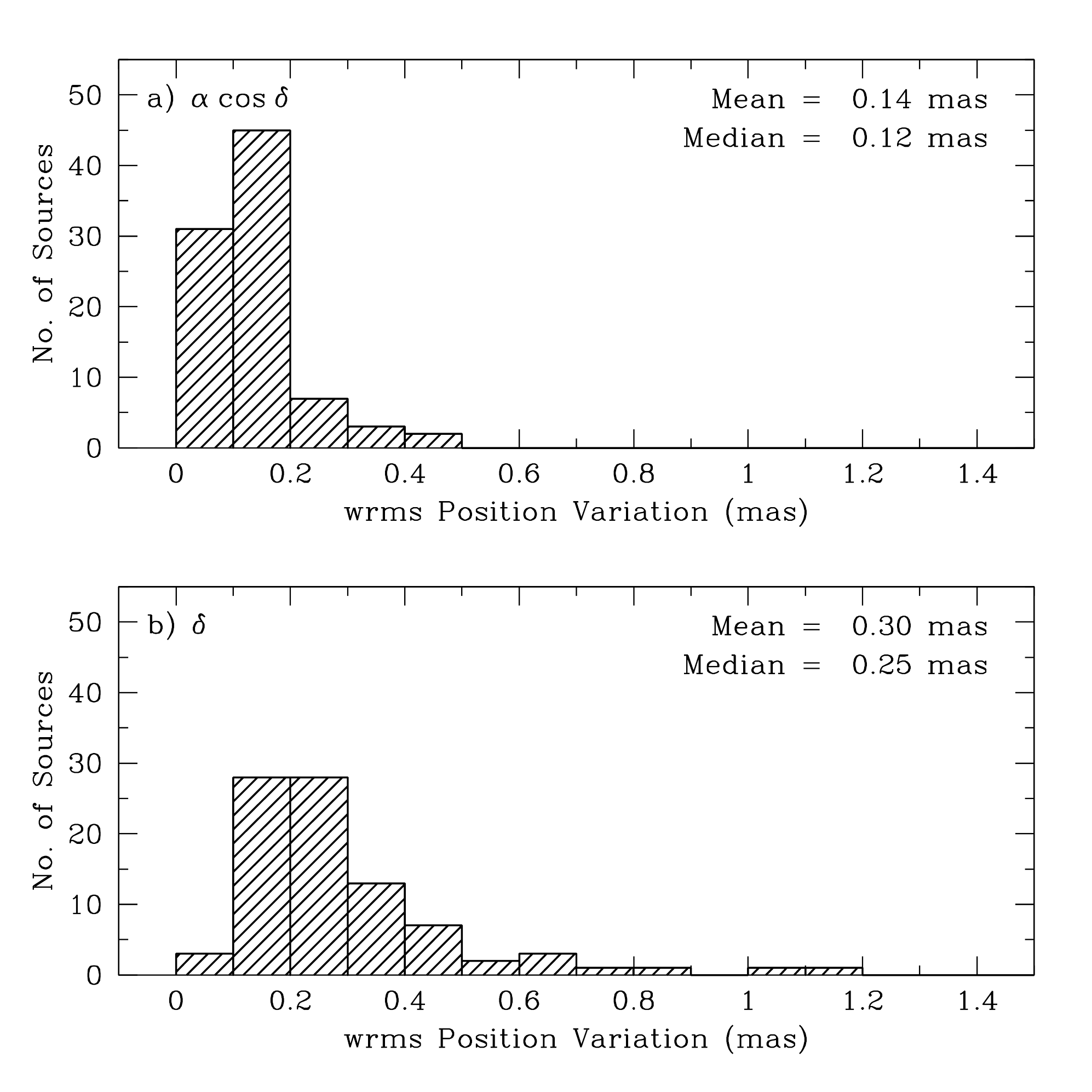}
\figcaption{Distributions of wrms position stability in (a) $\alpha \cos \delta$ 
and in (b) $\delta$ for the 88 sources observed in five or more 
sessions at K-band (24 GHz). \label{FIG:TS_HIST}}
\end{figure}

\begin{figure}[hbt]
\epsscale{0.5}
\plotone{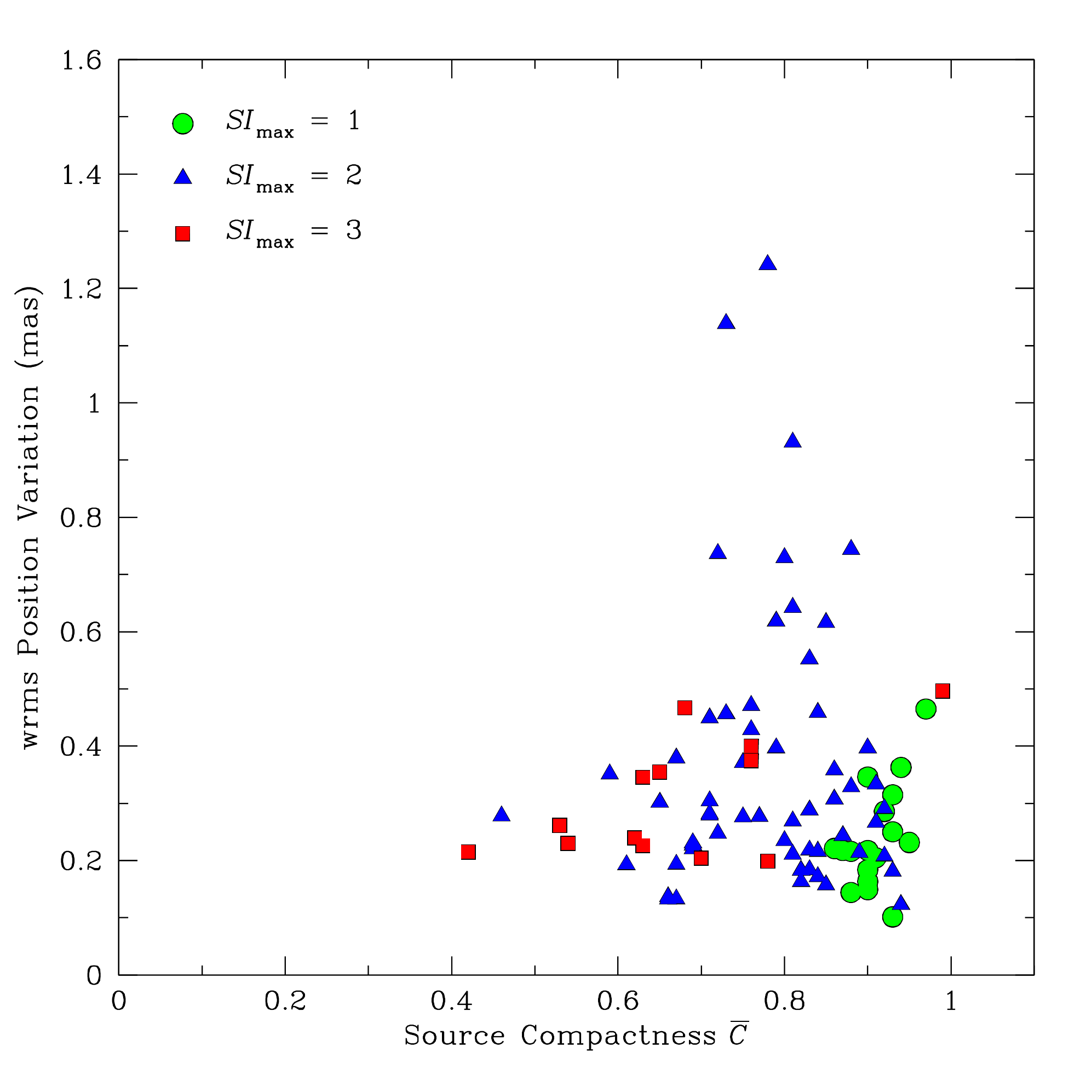}
\figcaption{Comparison of the wrms position stability versus the mean source
compactness ($\bar C$) for sources observed in five or more sessions at 
K band (24 GHz).   Point color and type represent the maximum structure 
index for each source over the sessions in which it was observed.
\label{FIG:TS_COMPACT_SCATTER}}
\end{figure}

\clearpage

\begin{deluxetable}{lcccccccc}
\tablecaption{Summary of VLBA Imaging at K and Q band. \label{TAB:OBS}}
\tablewidth{0pt}
\tablehead{  & & \multicolumn{3}{c}{K-band Sources} & & \multicolumn{3}{c}{Q-band Sources} \\
\cline{3-5}\cline{7-9} 
\colhead{Epoch} & \colhead{Band} & \colhead{Imaged} & \colhead{Overlap} & 
\colhead{New} & & \colhead{Imaged} & \colhead{Overlap} & \colhead{New}
}
\startdata
2002-05-15 & K\tablenotemark{a}/Q\tablenotemark{b}   & 
65  & \nodata   & 65   &  &  65  & \nodata   & 65 \\
2002-08-25 & K\tablenotemark{a}/Q\tablenotemark{b}   & 
65  & 65   & \nodata   &  &  65  & 65  & \nodata \\
2002-12-26 & K\tablenotemark{c}/Q\tablenotemark{d}   & 
67  & 24  & 43            &  &  67  & 24  & 43   \\
2003-05-22\tablenotemark{e} & K\tablenotemark{c}           & 
184 & 62  & 122          & & \nodata   & \nodata   & \nodata  \\
2003-09-13 & K\tablenotemark{c}/Q\tablenotemark{d}   & 
70   & 70  & \nodata            &  & 70 & 46 &  24    \\
2004-02-15 & K\tablenotemark{c}    & 
65   & 65  & \nodata    &  &  \nodata   & \nodata  & \nodata \\
2004-12-14 & K\tablenotemark{f}    & 
101 & 101 & \nodata   &  &  \nodata   & \nodata  & \nodata \\
2005-08-26 & K\tablenotemark{f}      &
92  & 67  & 25             &  &  \nodata   & \nodata  & \nodata \\
2006-07-09 & K\tablenotemark{f}    & 
178 & 163 & 15            &  &  \nodata   & \nodata  & \nodata \\
2007-03-30 & K\tablenotemark{f}     &     
185 & 181 &  4             &  &  \nodata   & \nodata  & \nodata \\
\enddata
\tablenotetext{a}{$\nu=$ 24.25, 24.33, 24.51, 24.65 GHz}
\tablenotetext{b}{$\nu=$ 42.95, 43.03, 43.21, 43.35 GHz} 
\tablenotetext{c}{$\nu=$ 24.21, 24.23, 24.27, 24.34, 24.49, 24.59, 24.66, 24.68 GHz}
\tablenotetext{d}{$\nu=$ 42.91, 42.93, 42.97, 43.04, 43.19, 43.29, 43.36, 43.38 GHz} 
\tablenotetext{e}{K-band only survey of 249 sources.}
\tablenotetext{f}{$\nu=$ 23.72, 23.74, 23.78, 23.85, 24.00, 24.10, 24.17, 24.19 GHz}
\end{deluxetable}

\clearpage
\LongTables

\begin{deluxetable}{lcrrrcccr}
\tabletypesize{\scriptsize}
\tablewidth{0pt}
\tablecaption{Parameters of Naturally Weighted Images at 24~GHz
\label{TAB:KBAND_IMG_PARAM}}
\tablehead{
\colhead{} & \colhead{} & \multicolumn{3}{c}{Beam\tablenotemark{a}} &
\multicolumn{3}{c}{Flux Density}\\
\cline{3-5}\cline{6-8}
\colhead{} & \colhead{Observation}  & \colhead{$a$} & \colhead{$b$} &
\colhead{$\phi$} & \colhead{$S_{\rm peak}$} & 
\colhead{$S_{\rm total}$\tablenotemark{b}} &
\colhead{$S_{\rm rms}$\tablenotemark{c}} &  
\colhead{Contour Levels\tablenotemark{d}} \\
\colhead{Source} & \colhead{Session} & \colhead{(mas)} & \colhead{(mas)} &
\colhead{($^\circ$)} & \colhead{(Jy bm$^{-1}$)} & \colhead{(Jy)} &
\colhead{(mJy bm$^{-1}$)} & \colhead{(mJy bm$^{-1}$)}
}
\startdata
$0006+061$ & 05-22-2003 &    1.01 &    0.29 & $ -11$ &    0.24 &    0.26 &    2.9 &    7.1$\times(1,...,2^{ 5})$ \\
$0007+106$ & 03-30-2007 &    0.72 &    0.30 & $   3$ &    0.42 &    0.51 &    1.5 &    4.5$\times(1,...,2^{ 6})$ \\
$0009+081$ & 03-30-2007 &    0.82 &    0.37 & $  -4$ &    0.31 &    0.45 &    1.4 &    4.1$\times(1,...,2^{ 6})$ \\
$0016+731$ & 12-14-2004 &    0.46 &    0.36 & $  -2$ &    1.82 &    1.98 &    1.4 &    4.1$\times(1,...,2^{ 8})$ \\
$0017+200$ & 03-30-2007 &    0.63 &    0.31 & $  -9$ &    0.43 &    0.59 &    1.8 &    5.4$\times(1,...,2^{ 6})$ \\
$0019+058$ & 03-30-2007 &    0.73 &    0.30 & $  -3$ &    0.87 &    0.94 &    1.3 &    3.9$\times(1,...,2^{ 7})$ \\
$0038-020$ & 05-22-2003 &    1.39 &    0.70 & $  -1$ &    0.29 &    0.35 &    3.6 &   10.9$\times(1,...,2^{ 4})$ \\
$0046+316$ & 03-30-2007 &    0.63 &    0.34 & $   7$ &    0.39 &    0.53 &    1.4 &    3.9$\times(1,...,2^{ 6})$ \\
\enddata
\tablenotetext{a}{The restoring beam is an elliptical Gaussian
with FWHM major axis~$a$ and minor axis~$b$, with major axis in
position angle~$\phi$ (measured north through east).}
\tablenotetext{b}{The total image flux defined by the sum of the CLEAN components.
In some cases this may be slightly less than the peak flux density due
to the presence of components with negative amplitude in the CLEAN
model.}
\tablenotetext{c}{The root mean square (rms) of the residuals of the final hybrid
image.}
\tablenotetext{d}{Contours levels are represented by the geometric
series $1,...,2^n$, \eg\ for $n=5$ the contour levels would be
$\pm1,2,4,8,16,32$.}
\end{deluxetable}

\clearpage

\begin{deluxetable}{lcrrrcccr}
\tabletypesize{\scriptsize}
\tablewidth{0pt}
\tablecaption{Parameters of Naturally Weighted Images at 43~GHz
\label{TAB:QBAND_IMG_PARAM}}
\tablehead{
\colhead{} & \colhead{} & \multicolumn{3}{c}{Beam\tablenotemark{a}} &
\multicolumn{3}{c}{Flux Density}\\
\cline{3-5}\cline{6-8}
\colhead{} & \colhead{Observation}  & \colhead{$a$} & \colhead{$b$} &
\colhead{$\phi$} & \colhead{$S_{\rm peak}$} & 
\colhead{$S_{\rm total}$\tablenotemark{b}} &
\colhead{$S_{\rm rms}$\tablenotemark{c}} &  
\colhead{Contour Levels\tablenotemark{d}} \\
\colhead{Source} & \colhead{Epoch} & \colhead{(mas)} & \colhead{(mas)} &
\colhead{($^\circ$)} & \colhead{(Jy bm$^{-1}$)} & \colhead{(Jy)} &
\colhead{(mJy bm$^{-1}$)} & \colhead{(mJy bm$^{-1}$)}
}
\startdata
$0007+106$ & 09-13-2003 &    0.38 &    0.19 & $   0$ &    3.54 &    3.66 &    5.6 &   16.8$\times(1,...,2^{ 7})$ \\
$0009+081$ & 09-13-2003 &    0.54 &    0.25 & $ -17$ &    0.72 &    0.92 &    4.3 &   12.9$\times(1,...,2^{ 5})$ \\
$0016+731$ & 08-25-2002 &    0.25 &    0.21 & $  27$ &    0.61 &    0.77 &    2.6 &    7.7$\times(1,...,2^{ 6})$ \\
$0017+200$ & 09-13-2003 &    0.40 &    0.20 & $  -3$ &    0.95 &    1.13 &    4.6 &   13.8$\times(1,...,2^{ 6})$ \\
$0046+316$ & 09-13-2003 &    0.61 &    0.36 & $   4$ &    0.74 &    0.80 &    4.3 &   11.9$\times(1,...,2^{ 5})$ \\
$0048-097$ & 09-13-2003 &    0.44 &    0.17 & $   7$ &    1.13 &    1.14 &    7.4 &   22.2$\times(1,...,2^{ 5})$ \\
$0059+581$ & 09-13-2003 &    0.40 &    0.16 & $  33$ &    2.89 &    4.09 &    8.7 &   26.2$\times(1,...,2^{ 6})$ \\
$0119+115$ & 09-13-2003 &    0.61 &    0.25 & $  23$ &    1.23 &    1.45 &    4.5 &   13.4$\times(1,...,2^{ 6})$ \\
\enddata
\tablenotetext{a}{The restoring beam is an elliptical Gaussian
with FWHM major axis~$a$ and minor axis~$b$, with major axis in
position angle~$\phi$ (measured north through east).}
\tablenotetext{b}{The total image flux defined by the sum of the CLEAN components.
In some cases this may be slightly less than the peak flux density due
to the presence of components with negative amplitude in the CLEAN
model.}
\tablenotetext{c}{The root mean square (rms) of the residuals of the final hybrid
image.}
\tablenotetext{d}{Contours levels are represented by the geometric
series $1,...,2^n$, \eg\ for $n=5$ the contour levels would be
$\pm1,2,4,8,16,32$.}
\end{deluxetable}

\clearpage

\begin{deluxetable}{lccccccccccc}
\tabletypesize{\small}
\tablecaption{Source structure indices. \label{TAB:STRUC_INDX}}
\tablewidth{0pt}
\tablehead{  & \multicolumn{3}{c}{X band (8.4 GHz)} & & 
\multicolumn{3}{c}{K band (24 GHz)} & & 
\multicolumn{3}{c}{Q band (43 GHz)} \\
\cline{2-4} \cline{6-8} \cline{10-12}
\colhead{Source} & \colhead{$N_{\rm session}$} & \colhead{$SI_{\rm min}$} &
 \colhead{$SI_{\rm max}$} & & \colhead{$N_{\rm session}$} & 
 \colhead{$SI_{\rm min}$} & \colhead{$SI_{\rm max}$} & & 
 \colhead{$N_{\rm session}$} & \colhead{$SI_{\rm min}$} & 
 \colhead{$SI_{\rm max}$}
}
\startdata
0006+061	&	\nodata	&	\nodata	&	\nodata &	&	1	&	1	&	1	&	&	\nodata	&	\nodata	&	\nodata	\\
0007+106	&	1	&	1	&	1	&	&	6	&	1	&	2	&	&	1	&	1	&	1	\\
0009+081	&	1	&	1	&	1	&	&	6	&	1	&	2	&	&	1	&	1	&	1	\\
0016+731	&	1	&	1	&	1	&	&	4	&	1	&	2	&	&	2	&	1	&	1	\\
0017+200	&	1	&	2	&	2	&	&	6	&	1	&	2	&	&	1	&	1	&	1	\\
0019+058	&	\nodata	&	\nodata	&	\nodata	&	&	3	&	1	&	1	&	&	\nodata	&	\nodata	&	\nodata	\\
0038-020	&	\nodata	&	\nodata	&	\nodata	&	&	1	&	1	&	1	&	&	\nodata	&	\nodata	&	\nodata	\\
0046+316	&	1	&	2	&	2	&	&	6	&	1	&	2	&	&	1	&	1	&	1	\\
\enddata
\end{deluxetable}

\clearpage

\begin{deluxetable}{lccccccc}
\tabletypesize{\small}
\tablecaption{Source compactness vs. source structure index. 
\label{TAB:COMPACT_SI}}
\tablewidth{0pt}
\tablehead{  &  \multicolumn{3}{c}{K Band (24 GHz)} & & 
\multicolumn{3}{c}{Q Band (43 GHz)}   \\
\cline{2-4} \cline{6-8}
\colhead{Struc. Ind.} & \colhead{$N_{\rm sou}$} &
\colhead{Mean} & \colhead{Median} & &  
\colhead{$N_{\rm sou}$} &
\colhead{Mean} & \colhead{Median}  
}
\startdata
1\dotfill  &  84  & 0.89 & 0.90  & &  74  & 0.87 & 0.87 \\
2\dotfill  & 131 & 0.79 & 0.80  & &  48  & 0.74 & 0.75 \\
3\dotfill  &   40 & 0.70 & 0.70  & &   9  & 0.71 & 0.71\\
4\dotfill  & \nodata  & \nodata & \nodata  
& & \nodata  & \nodata & \nodata\\
\enddata
\end{deluxetable}

\begin{deluxetable}{lcccccccccccccc}
\tabletypesize{\scriptsize}
\tablecaption{Source position uncertainties vs. source structure index. 
\label{TAB:POS_UNCER}}
\tablewidth{0pt}
\tablehead{  &  \multicolumn{6}{c}{X band (8.4 GHz)\tablenotemark{a}} & & 
\multicolumn{6}{c}{K Band (24 GHz)\tablenotemark{b}} \\
\cline{2-7} \cline{9-14}
&  & \multicolumn{2}{c}{Mean (mas)}  & & \multicolumn{2}{c}{Median (mas)} & &
& \multicolumn{2}{c}{Mean (mas)}  & & \multicolumn{2}{c}{Median (mas)} \\
\cline{3-4} \cline{6-7} \cline{10-11} \cline{13-14}
\colhead{Struc. Ind.} & \colhead{$N_{\rm sou}$} &
\colhead{$\alpha \cos \delta$} & \colhead{$\delta$} & &  
\colhead{$\alpha \cos \delta$} & \colhead{$\delta$} & & 
\colhead{$N_{\rm sou}$} &
\colhead{$\alpha \cos \delta$} & \colhead{$\delta$} & &
\colhead{$\alpha \cos \delta$} & \colhead{$\delta$} & 
}
\startdata
1\dotfill & 50 & 0.34 & 0.42 & & 0.28 & 0.31 & 
& 49  & 0.27 & 0.33 & & 0.27 & 0.31 \\
2\dotfill & 106 & 0.38 & 0.45 & & 0.29 & 0.32 & 
& 119 & 0.28 & 0.33 & & 0.27 & 0.30 \\
3\dotfill & 72 & 0.44 & 0.54 & & 0.33 & 0.36 & 
& 23 & 0.27 & 0.32 & & 0.27 & 0.30 \\
4\dotfill & 36 & 0.54 & 0.72 & & 0.76 & 1.07 & 
& \nodata & \nodata & \nodata & & \nodata & \nodata \\
\enddata
\tablenotetext{a}{X band ICRF values taken from \cite{FC:00}}
\tablenotetext{b}{K band formal errors used to compute the mean and 
median uncertainties were inflated so as to be comparable to the X-band ICRF 
values (see discussion in the text).}
\end{deluxetable}


\begin{thebibliography}{}

\bibitem[Blandford \& K\" onigl (1979)]{BK:79} 
Blandford, R. D. \& K\" onigl, A. 1979, ApJ, 232, 34

\bibitem[Charlot(1990)]{CHARLOT:90} 
Charlot, P. 1990, \aj, 99, 1309

\bibitem[Charlot(1994)]{CHARLOT:94} 
Charlot, P. 1994, in Proc. of Int. Symp.:
VLBI Technology - Progress and Future Observational Possibilities, 
Kyoto, Japan, 1993 September, ed. T. Sasao, S. Manabe, 
O. Kameya \& M. Inoue (Tokyo: Terra Scientific Publishing
Company), 287 

\bibitem[Charlot(2002)]{CHARLOT:02} 
Charlot, P.\ 2002, in International VLBI Service for Geodesy and Astrometry: 
General Meeting Proc., 2002 May, ed. N. R. Vandenberg \& K. D. Baver, 
NASA/CP-2002-210002, 233 

\bibitem[Charlot et al.(2005)]{CHARLOT:05} 
Charlot, P., Fey, A.~L., Jacobs, C.~S., Ma, C., Sovers, O.~J., 
\& Baudry, A.  2005, Journ{\'e}es 2004--syst{\`e}mes de 
r{\'e}f{\'e}rence spatio-temporels, 21 

\bibitem[Clark(1980)]{CLARK:80}
Clark, B. G. 1980, \aap, 89, 37
   
\bibitem[Fey \& Charlot(1997)]{FC:97} 
Fey, A.~L., \& Charlot, P.\ 1997, \apjs, 111, 95

\bibitem[Fey \& Charlot(2000)]{FC:00} 
Fey, A.~L., \& Charlot, P.\ 2000, \apjs, 128, 17 

\bibitem[Fey et al.(1996)]{FCF:96} 
Fey, A. L., Clegg, A. W., \& Fomalont, E. B. 1996, \apjs, 105, 299
   
\bibitem[Fey et al.(1997)]{FEK:97}
Fey, A. L., Eubanks, T. M., \& Kingham, K. A. 1997, \aj, 114, 2284

\bibitem[Fey et al.(2004)]{FEY:04} 
Fey, A.~L., et al.\ 2004, \aj, 127, 3587

\bibitem[Hovatta et al.(2008)]{HOVATTA:08} 
Hovatta, T., Lehto, H.~J., \& Tornikoski, M.\ 2008, \aap, 488, 897 

\bibitem[Hovatta et al.(2007)]{HOVATTA:07} 
Hovatta, T., Tornikoski, M., Lainela, M., Lehto, H.~J., Valtaoja, E., 
Torniainen, I., Aller, M.~F., \& Aller, H.~D.\ 2007, \aap, 469, 899 

\bibitem[Kovalev et al.(2008)]{KOVALEV:08} 
Kovalev, Y.~Y., Lobanov, A.~P., Pushkarev, A.~B., \& Zensus, J.~A.\ 
2008, \aap, 483, 759

\bibitem[Lanyi et al.(2009)]{LANYI:09}
Lanyi, G. E., et al. 2009, \aj, submitted.  (Paper I)

\bibitem[Lobanov(1998)]{LOBANOV:98} 
Lobanov, A.~P.\ 1998, \aap, 330, 79 

\bibitem[Ma et al.(1998)]{MA:98}
Ma, C., et al.\ 1998, \aj, 116, 516

\bibitem[Napier et al.(1994)]{NAPIER:94}
Napier, P. J., Bagri, D. S., Clark, B. G., Rogers, A. E. E., Romney,
J. D., Thompson, A. R., \& Walker, R. C. 1994, Proceedings of the
IEEE, 82, 658

\bibitem[Pearson \& Readhead(1984)]{PR:84} 
Pearson, T. J., \& Readhead, A. C. S. 1984, \araa, 22, 97

\bibitem[Pearson et al.(1994)]{PSTM:94} 
Pearson, T.~J., Shepherd, M.~C., Taylor, G.~B., \& Myers, S.~T.\ 1994, 
BAAS, 26, 1318 

\bibitem[Rogers(1970)]{ROGERS:70}
Rogers, A.~E.~E. 1970, Radio Science, 5, 1239

\bibitem[Sovers et al.(2002)]{SOVERS:02} 
Sovers, O.~J., Charlot, P., Fey, A.~L., \& Gordon, D.  2002, 
in International VLBI Service for Geodesy and Astrometry: 
General Meeting Proc., 2002 May, ed. N. R. Vandenberg \& K. D. Baver, 
NASA/CP-2002-210002, 243 

\bibitem[Ter\"asranta et al.(2005)]{TERASRANTA:05}
Ter{\"a}sranta, H., Wiren, S., Koivisto, P., Saarinen, V., \& 
Hovatta, T.\ 2005, \aap, 440, 409 

\end{thebibliography}
\end{document}